\documentclass[]{aa} % for a printer version
\usepackage{graphicx}
\usepackage{txfonts}
\usepackage{xcolor}
\usepackage{amsmath}
\usepackage{amsfonts}
\usepackage{natbib}
\usepackage{longtable}
\usepackage{booktabs}
\usepackage[amsmath,thmmarks]{ntheorem}
\theoremseparator{.}
\theorembodyfont{\normalfont} 
{\bfseries}{\itshape}
\usepackage[normalem]{ulem}
\usepackage{hyperref}
\hypersetup{
    colorlinks=true,
    linkcolor=blue,
    filecolor=magenta,      
    urlcolor=cyan,
    citecolor=teal,
    pdftitle={The cosmic waltz of Coma Berenices and Latyshev 2},
    }
\begin{document} 

   \title{The cosmic waltz of Coma Berenices and Latyshev 2 (Group X)\thanks{Full Table \ref{table:list_of_members} is only available in electronic form at the CDS via anonymous ftp to cdsarc.cds.unistra.fr (130.79.128.5) or via https://cdsarc.cds.unistra.fr/cgi-bin/qcat?J/A+A/}}

   \subtitle{Membership, phase-space structure, mass, and energy distributions.}

   \author{J. Olivares\inst{1,2,3}
          \and N.\ Lodieu \inst{2,3}
          \and V.\ J.\ S.\ B\'ejar \inst{2,3}
          \and E.\ L.\ Mart\'in\inst{2,3}
          \and M.\ {\v Z}erjal \inst{2,3}
          \and P.\ A.\ B. Galli\inst{5}
          }

   \institute{
   		 Departamento de Inteligencia Artificial, Universidad Nacional de Educación a Distancia (UNED), c/Juan del Rosal 16, E-28040, Madrid, Spain.  \email{jolivares@dia.uned.es}
   		 \and
         Instituto de Astrof\'isica de Canarias, E-38205 La Laguna, Tenerife, Spain.
         \and
         Departamento de Astrof\'isica, Universidad de La Laguna (ULL), E-38205 La Laguna, Tenerife, Spain
         \and Núcleo de Astrofísica Teórica, Universidade Cidade de São Paulo, R. Galvão Bueno 868, Liberdade, 01506-000, São Paulo, SP, Brazil.
         }
   \date{Received; accepted}

% \abstract{}{}{}{}{} 
% 5 {} token are mandatory
 
  \abstract
   {Open clusters (OCs) are fundamental benchmarks where theories of star formation and stellar evolution can be tested and validated.
Coma Ber and Latyshev 2 (Group X) are the second and third OCs closest to the Sun, making them excellent targets to search for low-mass stars and ultra-cool dwarfs. In addition, this pair will experience a flyby in 10-16 Myr which make it a benchmark to test OCs pair interactions.}
   {We aim at analysing the membership, luminosity, mass, phase-space (i.e., positions and velocities), and energy distributions for Coma Ber and Latyshev 2 and test the hypothesis of the mixing of their populations at the encounter time.}
   {We develop a new phase-space membership methodology and apply it to \textit{Gaia} data. With the recovered members we infer the phase-space, luminosity and mass distributions using publicly available Bayesian inference codes. Then, with a publicly available orbit integration code and members' positions and velocities we integrate their orbits 20 Myr into the future.}
   {In Coma Ber, we identify 302 candidate members distributed in the core and tidal tails. The tails are dynamically cold and asymmetrically populated. The stellar system called Group X is made of two structures: the disrupted OC Latyshev\,2 (186 candidate members) and a loose stellar association called Mecayotl\,1 (146 candidate members), both of them will fly by Coma Ber in $11.3\pm0.5$ Myr and $14.0\pm0.6$ Myr, respectively, and each other in $8.1\pm1.3$ Myr.}
   {We study the dynamical properties of the core and tails of Coma Ber, and also confirm the existence of the OC Latyshev 2 and its neighbour stellar association Mecayotl 1. Although these three systems will experience encounters we find no evidence supporting the mixing of their populations.}


   \maketitle
%
%-------------------------------------------------------------------

\section{Introduction}
\label{intro}

Coma Berenices (hereafter Coma Ber, $85.5\pm0.4$ pc, \citealt{2017A&A...601A..19G}) is the second closest open cluster to the Sun after the Hyades, ($47.03\pm0.20$ pc, \citealt{2019A&A...623A..35L}; $45.5\pm0.42$ pc, \citealt{2017A&A...601A..19G}). Given its proximity, mid-age \citep[between 550 and 800 Myr, e.g.,][]{2020A&A...640A...9M,2018ApJ...862..106T}, and relatively large population \citep[$\sim$200 members, according to][]{2018ApJ...862..106T}, it is an excellent benchmark cluster for studies of stellar evolution and dynamical interactions. Since the pioneering work of \citet{1938LicOB..18..167T}, Coma Ber has been the subject of several studies. Hereafter, we mention only those works done in the \textit{Gaia} \citep{2016A&A...595A...1G} era and refer the reader to the thorough membership review of \citet{2020MNRAS.497.2562C}. 

\citet{2018ApJ...862..106T} identified 192 candidate members within a 5\degr\ radius of the cluster centre using a combination of the Two Microns All Sky Survey \citep[2MASS,][]{2006AJ....131.1163S}, the UKIRT Infrared Deep Sky Survey \citep[UKIDSS,][]{2007MNRAS.379.1599L}, the first USNO Robotic Astrometric Telescope catalogue \citep[URAT1,][]{2015AJ....150..101Z} and \textit{Gaia} Data Release 2 \citep[DR2,][]{2018A&A...616A...1G}. These authors found that the cluster is mass-segregated and elongated along the Galactic plane. In addition, they identified nine sub-stellar members.

\citet{2019A&A...624L..11F}, using \textit{Gaia} DR2, applied the Density Based Spatial Clustering of Applications with Noise algorithm \citep[DBSCAN,][]{10.5555/3001460.3001507} to the space of positions and tangential velocities resulting in the discovery of a kinematically cold group of 177 stars located in the vicinity of Coma Ber. They propose that this new group is younger ($\sim$400 Myr) than Coma Ber, formed in another part of the Galaxy, and will reach a minimum separation of 25 pc with Coma Ber in the following $\sim$13 Myr. 

\citet{2019ApJ...877...12T} identified the tidal tail structures of Coma Ber using \textit{Gaia} DR2 data and found that they have 120 members, extending up to $\sim$50 pc from the cluster centre. In addition, they also identified 218 candidate members of the new group identified by \citet{2019A&A...624L..11F}, which they call Group\,X. They found evidence that this group was composed of two substructures. As mentioned by these authors, groups 10, 81, and 1805 from \citet{2017AJ....153..257O} partially overlap with this new Group X.

\citet{2020A&A...640A...9M} performed a photometric search for new faint objects in the core of Coma Ber. They used UKIDSS, Panoramic Survey Telescope and Rapid Response System \citep[PanSTARRS,][]{2016arXiv161205560C}, Sloan Digital Sky Survey \citep[SDSS,][]{2015ApJS..219...12A} DR12, and AllWISE \citep{2010AJ....140.1868W} data to identify a couple of dozen candidate brown dwarfs within the tidal radius of the cluster. Using spectral types and radial velocities derived from optical spectra collected with the Gran Telescopio de Canarias (GTC), these authors confirmed the membership of four L2$-$L2.5 dwarfs, two of which were already known in the literature \citep{2019ApJ...877...12T}. Given the large Lithium (Li) depletion in the spectra of some of these Coma Ber members, \citet{2020A&A...640A...9M} propose that the Li depletion boundary  must be located at spectral types later than L2.5\@, which restricts the lower limit of the cluster age to 550 Myr. Combining this constraining information with other dating techniques, these authors provide an age estimate of $780\pm230$ Myr for Coma Ber.  

\citet{2020MNRAS.497.2562C} combined their own $BVI_c$ photometry with \textit{Gaia} DR2 parallaxes and rotational periods of K and M members of Coma Ber. They concluded that background giants from the field with short surface rotational periods could mimic main sequence cluster member stars and, thus, contaminate membership lists. The latter may result in older age estimates when using the isochrone fitting technique. These authors then argue that gyrochronology age estimates, as those of \citet[][$590\pm41$Myr]{2009MNRAS.400..451C}, should produce less biased results.

\citet{2021OAst...30..191S} performed an unsupervised blind search for candidate members in Coma Ber and Group\,X using \textit{Gaia} Early Data Release 3 \citep[EDR3,][]{2021A&A...649A...1G}. As a result, they identified $\sim$250 candidate members for both Coma Ber and Group\,X. Also, they fitted isochrones and determined ages of $700\pm70$ Myr and $450\pm100$ Myr, respectively.

\citet{2021ApJ...917...11S} inferred the metallicity of Coma Ber literature members based on high-resolution near-infrared spectra from the SDSS-IV APOGEE survey \citep{2017AJ....154...94M}. Their results show that the Coma Ber members have near solar and homogeneous metallicities ([Fe/H]\,=\,$+$0.04$\pm$0.02 dex).

\citet{2022A&A...657L...3M} used data from the Transiting Exoplanet Survey Satellite (TESS) mission \citep{2015JATIS...1a4003R} to measure the rotation periods of 168 Group\,X candidate members from \citet{2019ApJ...877...12T}. Then, using these periods and the gyrochronology age-mass-rotation relations of \citet{2015MNRAS.450.1787A} and \citet{2008ApJ...687.1264M}, they derived the mean group's age at $300\pm60$ Myr.

\citet{2022arXiv220606254N} performed a search of Group X's members using \textit{Gaia} EDR3. They query the 25 pc volume around the planet-host star TOI-2048, a Group X candidate member according to \citet{2019ApJ...877...12T}, and identify as candidate members of Group X the 208 stars whose tangential velocity differed from that of TOI 2048 by less than 5 $\rm{km\ s^{-1}}$. From this list, only 69 are in common with the candidate members from \citet{2019ApJ...877...12T}. The authors also identified a potential new group, which they named MELANGE-2, containing 81 stars that share the 20 pc radius and 4 $\rm{km\ s^{-1}}$ volume around TIC 224606446. The latter is a non-member of Group X that differs 5 $\rm{km\ s^{-1}}$ from TOI 2048. The rotation periods of this putative new group suggest an age similar to that of Praesepe \citep[$670\pm67$ Myr;][]{2019ApJ...879..100D}. 

\citet{2022arXiv220612170H} recently searched for new clusters by applying DBSCAN on \textit{Gaia} EDR3. They identified 99 candidate members in Coma Ber and 61 in each of the two Group X substructures. These authors also determined isochrone ages of $560\pm65$ Myr, $178\pm20$ Myr, and $141\pm16$ Myr for Coma Ber, Group Xa and Group Xb, respectively. We notice that their age estimates and the numbers of candidate members are at the lower edge of the literature values.

\subsection*{Group X or Latyshev 2?}
In 1977, \citet{1977ATsir.969....7L} reported a new open cluster candidate comprised of seven stars located in the Ursa Majoris constellation. Later on, this candidate open cluster was named Latyshev 2 by \citet{2003stcl.book.....A}. A few years ago, \citet{2016IAUS..314...21M} deemed this candidate open cluster a non-physical one given the large velocity dispersion of its putative members. However, \citet{2018ApJ...863...91F} noted that three stars belonging to Group 10 of \citet[][which contains 29 members]{2017AJ....153..257O} were already reported as candidate members of an unnamed open cluster by \citet{1977ATsir.969....7L}. It turns out that this unnamed open cluster was Latyshev 2. 

As mentioned above, \citet{2019A&A...624L..11F} reported the discovery of a moving group close to the Coma Ber open cluster a few years ago. This group was also found by \citet{2019ApJ...877...12T}, who (re)named it Group X. Given that this group corresponds to Group 10 of \citet{2017AJ....153..257O} and thus to the open cluster candidate reported by  \citet{1977ATsir.969....7L}, hereafter, we refer to it as Latyshev 2 instead of Group X.\\

\noindent Coma Ber and its neighbouring group will experience a flyby in 13 Myr, according to \citet{2019A&A...624L..11F}, or between 10$-$16 Myr, according to \citet{2019ApJ...877...12T}. Given their proximity, relatively young ages and unrelated origin, the possibility of an encounter in this pair makes it an interesting benchmark to study the effects of dynamical interaction(s) and possible merger(s) of stellar populations. 

Several interacting stellar cluster pairs have recently been discovered in the Galaxy \citep[e.g.,][]{2022arXiv220714229Y,2022MNRAS.511L...1P,2022MNRAS.510.5695A}. However, none of these pairs is as close and populated as Coma Ber and Latyshev 2. Moreover, thanks to their proximity to the Sun and large populations, it is possible to obtain high-precision estimates of the pair's parameters and flyby time, and thus, provide constraining information for future studies of the population of OC pairs.

In this work, we aim to analyse the membership, luminosity, mass, phase-space (defined as the joint space of Cartesian positions and velocities), and energy distributions of Coma Ber and Latyshev 2 based on \textit{Gaia} Data Release 3 \citep[DR3,][]{2022arXiv220800211G} and complementary data. These precise and updated distributions will allow us to re-examine the future encounter of Coma Ber and Latyshev 2 and its consequences on the stellar populations of both clusters.

Section \ref{dataset} introduces our \textit{Gaia} DR3 data set, complementary photometry and radial velocity measurements. Then, in Sect. \ref{methods}, we present the methods by which we identify the cluster members and infer the luminosity, mass, phase-space, and energy distributions. In this section, we also describe the trace-forward method by which we will analyse the time of the future encounter between Coma Ber and Latyshev 2. In Sect. \ref{results}, we present our results, and in Sect. \ref{discussion}, we compare them with previous ones from the literature and discuss their implications. Finally, in Sect. \ref{conclusions}, we  present our conclusions.

%--------------------------------------------------------------------
\section{Data}
\label{dataset}

We downloaded\footnote{\url{https://gea.esac.esa.int/archive/}} the astrometry and photometry of $\sim$45 million \textit{Gaia} DR3 sources with the following constraints: \texttt{b}>30\degr, and proper motions within $\rm{-400<\texttt{pmra}/[mas\ yr^{-1}]< +200}$ and $\rm{-400<\texttt{pmdec}/[mas\ yr^{-1}]< +400}$. These proper motion filters are permissive enough to encompass most open clusters and star-forming regions in the solar neighbourhood. We do not apply parallax cuts in the \textit{Gaia} DR3 data to avoid biasing our sample of low-mass stars and brown dwarfs. In addition, we complemented the \textit{Gaia} DR3 data of our candidate members with photometry from 2MASS (\textit{JHK}) and PanSTARRS (\textit{grizy}), as well as radial velocity measurements from the Apache Point Observatory Galactic Evolution Experiment DR17 \citep[APOGEE,][]{2021arXiv211202026A}, the Set of Identifications, Measurements and Bibliography for Astronomical Data \citep[SIMBAD,][]{2000A&AS..143....9W}, and our own radial velocity measurements. The catalogue cross-match and data query with external catalogues were done based on the \textit{Gaia} identifiers using \textit{Astropy} \citep{astropy:2013, astropy:2018}. The next two sections describe our spectroscopic observations and the data processing schema that we use to combine the radial velocity measurements from the different catalogues. 

\subsection{Mid-resolution optical spectroscopy}
We obtained optical spectra of 36 of our bright ($V<15.5$ mag) candidate members (see Sect. \ref{results:membership}) as well as spectrophotometric and radial velocity standards with the Intermediate Dispersion Spectrograph\footnote{\url{https://www.ing.iac.es//astronomy/instruments/ids/}} (IDS, using the RED+2 CCD). IDS is mounted at the Cassegrain focus of the 2.5m Isaac Newton Telescope (INT) on the Roque de Los Muchachos Observatory in La Palma, Spain (P.I. J. Olivares, program C77). We refer the reader to Appendix \ref{appendix:observations} for a detailed description of our INT-IDS observations.

We reduced the spectra using the \textit{spextractor} tool (Bouy et al. in prep). Briefly, it corrects the raw images with the median-combined bias and flat fields. Next, it applies cosmetic corrections (e.g., cosmic rays) and corrects for geometric distortions to align the spectra with the columns and rows of the detector. Then, it extracts the spectra using a rectangular aperture (of 2FWHM width and sky subtraction of a region between  5 and 7 FWHM from the maxima) together with Gaussian and Moffat aperture profiles. Afterwards, it does the wavelength calibration with an interactive approach that allows the user to match the lines of the arc lamps (Copper-Argon and Copper-Neon in the case of INT-IDS) with their corresponding templates. Finally, it corrects for the instrumental response based on the spectra of the spectrophotometric standard stars.

Radial velocities were obtained by cross-correlating the calibrated 1D spectra of our targets with templates created from our radial velocity standards (see Appendix \ref{appendix:observations}). We selected these standards from the catalogue of  \citet{2013A&A...552A..64S} based on the criteria of similarity of spectral type and proximity on the sky. We measured the radial velocities of all our 1D spectra (i.e., targets and standards) using the \textit{iSpec} tool \citep{2019MNRAS.486.2075B,2014A&A...569A.111B} by first removing telluric lines and then applying the cross-correlation function. The templates of the radial velocity standards were created by shifting the 1D spectra with the barycentric correction and the radial velocity reported by \citet{2013A&A...552A..64S}. Zero-point offsets were applied after correcting for the measured radial velocity of the standard star. We notice that our radial velocity measurements were obtained after the membership selection procedure and thus not affect it. The typical uncertainty of our resulting radial velocity measurements is 10 $\rm{km\ s^{-1}}$. Although these measurements are less precise than the median ones from APOGEE ($0.12\,\rm{km\ s^{-1}}$) or \textit{Gaia} ($1.8\,\rm{km\ s^{-1}}$) catalogues, we will use them for those sources with missing radial velocities. The impact that largely heterogeneous and heteroscedastic data have on the cluster parameters will be handled by deconvolution methods (see Sect. \ref{methods:6D_structure}). These types of methods offer an optimal solution for the extraction of the underlying distribution of the population parameters under the presence of heterogenous, heteroscedastic, and incomplete data sets \citep[see, for example,][]{2011AnApS...5.1657B}.  

%We also measured radial velocities of our targets and standards using \textit{iSpec}, but this time doing the cross-correlation with \textit{iSpec}'s internal HARPS masks instead of the templates of our radial velocity standards. This resulted in more precise but less accurate measurements, with typical uncertainties of 2 $\rm{km\ s^{-1}}$. Although the HARPS mask method returns smaller uncertainties, the measurements of our radial velocity standards were incompatible with those from the literature, with deviations beyond the  2$\sigma$ level. For this reason, we preferred the radial velocities computed from the cross-correlation with our templates, as described above.

\subsection{Data processing}

We processed our \textit{Gaia} DR3 data by only applying a parallax zero point correction of $-$0.017 mas. As stated in Sect.\ 7 of \citet{2021A&A...649A...1G}, the current parallax bias correction is only a tentative recipe to correct this systematic. 

The radial velocities were processed as follows. If a candidate member has both \textit{Gaia} DR3 and APOGEE measurements, we use those of APOGEE due to their better precision. We computed zero-point offsets for the different surveys and conclude that given their small values and large uncertainties they can be neglected (see Appendix \ref{appendix:rv_zero-points}). Given that our INT-IDS observations were taken before the release of \textit{Gaia} DR3, 26 of our INT-IDS targets also have DR3 radial velocity measurements. Thus, we used their \textit{Gaia} DR3 radial velocities since they resulted in smaller uncertainties. Some sources, particularly those from APOGEE, have radial velocity uncertainties of a few $\rm{m\ s^{-1}}$ that considerably increase the convergence time for our phase-space modelling methods (see Sect. \ref{methods:6D_structure}). For this reason, we set radial velocity uncertainties smaller than 0.1 $\rm{km \ s^{-1}}$ to this latter value.

\subsection{Initial list of members}
\label{data:initial_list}

As will be explained in Sect. \ref{methods:membership} our membership methodology requires an initial list of members for the cluster under analysis. As this input list, we choose the 214 and 177 candidate members that \citet{2019A&A...624L..11F} identified in Coma Ber and Latyshev 2, respectively. Although those authors identified 214 candidate members in Coma Ber, the cross-match with \textit{Gaia} DR3 excluded one \textit{Gaia} source (\texttt{source\_id 3960475762679374208}) due to its lack of parallax and proper motions. An initial inspection of the 6D space of the input list members revealed that 15 Coma Ber members were outliers in the velocity space. These sources were located more than six standard deviations beyond the candidate members' mean U and W values. Therefore, we decided to exclude them from our input list. In the case of Latyshev 2, we do not find outliers in the initial list of members.

\section{Methods}
\label{methods}
The following sections describe the methodologies we use to determine the different properties of stellar systems under analysis. We start by describing our novel membership methodology. Afterwards, we outline the steps to identify the phase-space structure, mass and energy distributions of Coma Ber and Latyshev 2. Finally, the section concludes by presenting the trace-forward method that we use to estimate the time of encounter between the stellar systems.

\subsection{Membership selection}
\label{methods:membership}
Our novel membership methodology, which we call \textit{Mecayotl}\footnote{Mecayotl means lineage, ancestry or kinship in the mesoamerican Nahuatl language.}, is a generic one and thus, can be applied to open clusters, stellar associations, and moving groups in the solar neighbourhood. It proceeds by iteratively applying the following steps. 

\begin{itemize}
\item In the first step, we use the cluster's input list of members (i.e., their astrometry, radial velocities, uncertainties, and correlations) and the \textit{Kalkayotl}\footnote{\url{https://github.com/olivares-j/Kalkayotl}} code \citep[][see Sect. \ref{methods:6D_structure}]{2020A&A...644A...7O} to determine the cluster's phase-space model. The latter consists of the cluster's central coordinates and size, both in the joint space of physical Cartesian positions and velocities.  
\item In the second step, we take the previous phase-space model and sample $10^6$ synthetic stars whose coordinates we transform to the observed space (i.e., \texttt{ra}, \texttt{dec}, \texttt{parallax}, \texttt{pmra}, \texttt{pmdec}, and \texttt{radial\_velocity}) with the \textit{PyGaia}\footnote{\url{https://github.com/agabrown/PyGaia}} code. 
\item In the third step, we use Gaussian Mixture Models\footnote{Gaussian Mixture Models are flexible statistical models that describe a probability density function as a linear combination of Gaussian distributions, in which the number of Gaussian distributions remains fixed.} (hereafter GMMs) with up to 80 components to independently fit the density distribution of the synthetic cluster's stars and that of an equal size sample of the real field stars randomly selected from the input catalogue. The flexibility of GMMs allows us to model the complex shapes of both the cluster and field stars distributions. We use the GMM algorithms of \citet{2018A&A...617A..15O}, which enable uncertainty deconvolution and the treatment of sources with missing values. The latter is of paramount importance to include sources with missing radial velocities. We select the best number of clusters and field GMM components using the Akaike information criterion \citep[AIC,][]{Akaike1974}. However, we restrict the search to only those models with at least ten stars in each mixture component. This robust restriction avoids spurious components fitted to outliers. 
\item In the fourth step, we use the previous two best GMMs, one for the field and another for the cluster, to compute membership probabilities of the input data set (i.e., all the real stars) as

\begin{equation}
\label{equation:probability}
\begin{split}
& Probability(cluster|data,\mathcal{M}_{cluster},\mathcal{M}_{field})= \\
& \frac{\mathcal{L}(data|\mathcal{M}_{cluster})\cdot \mathcal{P}(cluster)}{\mathcal{L}(data|\mathcal{M}_{cluster})\cdot \mathcal{P}(cluster)+\mathcal{L}(data|\mathcal{M}_{field})\cdot \mathcal{P}(field)},
\end{split}
\end{equation}

where $\mathcal{L}$, $\mathcal{M}$, and $\mathcal{P}$ stand for likelihood, model and prior, respectively. We use uniform prior probabilities for the field and cluster. 
\item In the fifth step, the candidate members are selected based on probability thresholds optimised for the photometric magnitude of the sources (see Sect. \ref{methods:membership:probability_thresholds} below). 
\item In the sixth step, we mask as missing the radial velocities of candidate members laying one standard deviation beyond the mean radial velocity of the cluster. This masking prevents the radial velocities of binary stars from inflating the velocity dispersion of the cluster while conserving good astrometric candidates.
\end{itemize}

Finally, the candidate members resulting from the sixth step are used as the input list for the next iteration of the algorithm, which we iterate until the number of candidate members converges over successive iterations. We use as convergence tolerance the Poisson uncertainty in the number of added members with respect to the previous iteration.

We handle the generation of the cluster's synthetic \textit{Gaia} observables as well as the analysis of the classifier quality with the free open-source code \textit{Amasijo}\footnote{\url{https://github.com/olivares-j/Amasijo}} \citep[see, for example,][ for another application of this code]{2022arXiv220603777C}. The complete membership methodology described here will be publicly available as a free open-source code\footnote{\url{https://github.com/olivares-j/Mecayotl}} (Olivares et al. in prep).

\subsubsection{Optimum probability thresholds}
\label{methods:membership:probability_thresholds}
It is well known that the \textit{Gaia} astrometric uncertainties heavily depend on the apparent magnitude of the sources \citep[see, for example, Fig. 7 of][]{2021A&A...649A...2L}. Therefore, any classifier incorporating \textit{Gaia} astrometric uncertainties will also show a correlation of its quality with the photometric magnitude of the sources to which it is applied. To minimise this correlation's impact and obtain the best classification at different magnitude intervals, we estimate the optimum probability threshold at each \texttt{g} magnitude bin in the full \textit{Gaia} domain (i.e., from 2 to 21 mag). We do this at each iteration of our membership algorithm by generating ten synthetic clusters with $10^3$ sources each. In these clusters, we randomly assign masses from a uniform distribution and obtain their photometry using the \textit{isochrones}\footnote{\url{https://github.com/timothydmorton/isochrones}} code \citep{2015ascl.soft03010M}, with the cluster's observed age and metallicity (see Sect. \ref{intro}). Afterwards, using the photometry and astrometry, we computed observational uncertainties with \textit{PyGaia} and random sample the observed values from these uncertainties. Finally, we applied the fourth step of our membership algorithm to the data sets resulting from combining the synthetic clusters with the real field stars. The obtained membership probabilities together with the true labels of the sources (i.e., cluster or field) allow us to compute classification confusion matrices\footnote{\label{note:confusion_matrix}The confusion matrix of a binary classifier is a four entries matrix with the number of true positives (TP) and true negatives (TN) in the diagonal, and the number of false positives (FP) and false negatives (FN) in the anti-diagonal. The classifier quality can then be evaluated using this matrix and a suitable metric.} and quality indicators, such as the true positive rate (TPR) or the contamination rate (CR), at each magnitude bin and probability threshold. We use magnitude and probability grids with 8 and 40 steps, respectively. The probability grid spans from one to five sigma, with each interval containing ten steps. At each grid step, we select as the optimum probability threshold the one that maximises the Mathews correlation coefficient \citep[MCC,][]{MATTHEWS1975442} of the classifier. This criterion measures the correlation coefficient between the true and observed classifications, and it is known to be the most objective metric to evaluate the quality of a classifier in the presence of unbalanced classes, as is the case here. We also tested other metrics (e.g., Accuracy and F1M), obtaining similar probability thresholds. 

We notice that due to our conservative approach for estimating contamination rates, which assumes as field stars all sources not classified as cluster members by either the literature or a previous iteration of the algorithm, possibly true cluster members will be considered contaminants and thus inflate our estimates of the contamination rate.

\subsubsection{Minimizing contamination}
\label{methods:membership:minimizing_contamination}
During the validation of our membership methodology, we noticed that the contamination rate of successive iterations of the algorithm tended to increase. We inspected the contaminants and found that these resulted from two main sources: radial velocity outliers and faint sources. On the one hand, the radial velocity outliers inflate the velocity dispersion of the cluster and thus result in accreted contaminants in the following iterations of the algorithm. On the other hand, at the faintest end of the cluster photometric sequence, the numbers and uncertainty of the sources increase exponentially. Both these factors result in large chances of confusing field sources as cluster members.

We minimise the contamination rate of our membership methodology and thus prevent possible biases by applying the following two filtering criteria. First, in the sixth step of our membership methods, we mask the radial velocities of outliers as missing. We opt for masking with a rather conservative $1\sigma$ filtering than discarding good astrometric candidate members. Second, we inferred the phase-space cluster model based only on sources within magnitudes bins with contamination rates < 10\%. This effectively discards from the inference of the phase-space cluster model those sources fainter than 20 mag in \texttt{g} band. Although these sources are not used in the inference of the cluster phase-space model, they remain in the list of cluster candidate members.

Due to our conservative approach for estimating the phase-space model, which filters out sources with $G>20$ mag, our membership algorithm may miss some of the faintest cluster members.  

\subsection{Phase-space structure}
\label{methods:6D_structure}
We analyse the phase-space structure of the stellar groups using the \textit{Kalkayotl} code \citep{2020A&A...644A...7O}. In its 6D version (Olivares et al. in prep.), the code simultaneously infers the Cartesian (ICRS) positions and velocities of the stellar system and its members, given the observed \textit{Gaia} astrometry (including its uncertainties and correlations) and (possible missing) radial velocities. The cluster or group-level parameters are modelled with a hierarchical prior, whereas the stellar or source-level parameters are drawn from this prior. Thanks to the hierarchical approach, the influence of the prior is minimised by inferring its parameters at the same time as those of the individual stars. Nonetheless, the prior family (i.e., the type of distribution) must be set in advance. The prior distribution should be selected based on the \textit{a priori} knowledge about the properties of the stellar system. Given that we expect a certain degree of contamination in the list of members and possible substructure in the positions and velocities of these systems, we decided to test the three types of prior families provided by \textit{Kalkayotl}. These are Gaussian, Gaussian Mixture Model (GMM), and Concentric Gaussian Mixture Model (CGMM). They all have the joint 6D space of Cartesian positions and velocities as a domain.  

It is expected that the velocity distributions of these relatively old ($\sim$300 Myr for Latyshev\,2 and 800 Myr for Coma Ber) stellar systems have already been relaxed and thus can be described with a single Gaussian distribution. However, the presence of structures, like tidal tails or halos, requires more flexible models. While the CGMM imposes a common location for all the Gaussian components of the mixture, the GMM is the most flexible of our models with completely free weights (restricted to add to one), means, and covariance matrices. To choose among these, we use the convergence of the inference algorithm and non-negligible components weights as selection criteria. We tested GMM with one to four components, starting with the simplest one. If the model with additional components fails to converge or has weights lower than 10\%, we step down to the previous simple one. 

Our phase-space modelling is conceptually simple but allows constructing complex models based on Gaussian building blocks. It is not the first time that this rather simple multi-component approach has been successfully applied in the literature. For example, \citet{2022A&A...659A..59T} fitted the 3D spatial distribution of open clusters using CGMMs in which, similarly to what we do here, one Gaussian component describes the cluster's core while the other two may be used to describe the halo and tails of the cluster.

We notice the following advantages of our GMM phase-space modelling. First, it allows the inference of the Cartesian positions and velocities for all the cluster members, even those without radial velocity measurements. This latter is accomplished assuming that the cluster members share the global cluster's velocity vector but differ only by the cluster's internal velocity dispersion. We refer the reader to the works of \citet{2002A&A...381..446M,2000A&A...356.1119L,1999A&A...348.1040D} for the conceptual definition of astrometric radial velocities and its applications to nearby open clusters and moving groups. Second, GMMs can be used as classifiers; thus the identified candidate members can be partitioned into the substructures corresponding to each of the Gaussian building blocks of the GMM. This classification is based on the source's multidimensional observed data and the GMM's inferred parameters. Finally, our GMMs live in the joint space of positions and velocities, which allows them to capture correlations among positions and velocities. The latter represents an improvement over previous models living in independent spaces of positions and velocities \citep[e.g.,][]{2019A&A...623A..35L,2021A&A...647A.137J}.  

\subsection{Luminosity and mass distributions}
\label{methods:mass}
We determine the luminosity and mass of each candidate member using the PAdova TRieste Stellar Evolution Code \citep[PARSEC,][]{2020MNRAS.498.3283P,2013MNRAS.434..488M} and BT-Settl \citep{2015A&A...577A..42B,2014IAUS..299..271A} theoretical isochrone models. We utilise a grid of ages spanning the literature values (see Sect. \ref{intro}): Coma Ber $\rm{\in[700Myr-900Myr]}$ and Latyshev 2 $\rm{\in[200Myr-400Myr]}$, both grids with steps of 100 Myr. We infer the luminosity and mass distributions with the free and open code \textit{Sakam}\footnote{\url{https://github.com/olivares-j/Sakam}} \citep{2019A&A...625A.115O}. It infers the joint posterior distribution of luminosity or mass together with $A_v$ and $R_v$ for each candidate member given its distance (see Sect. \ref{methods:6D_structure}) and observed photometry (see Sect. \ref{dataset}). We work with the following prior distributions. For the luminosity and mass, we use a uniform distribution in the logarithmic space and the \citet{2005ASSL..327...41C} distribution, respectively. We use a Gaussian distribution for the total-to-selective extinction: $R_v\sim\mathcal{N}(\mu=3.1,\sigma=0.5)$. In the case of $A_v$, we use a uniform prior, with $A_v\in[0,0.1]$ mag. Appendix \ref{appendix:extinction} provides a detailed explanation to support this decision. Briefly, this is the most probable extinction interval according to the values reported by \cite{2016A&A...596A.109P} for the sky positions of our candidate members. 

We notice the following points. First, in the case of unresolved binaries or multiple systems, the inferred mass for each candidate member will correspond to the system mass distribution. Currently, we have no method to infer the individual masses of possible unresolved binary or multiple systems. Second, none of the theoretical isochrone models that we use here spans the full magnitude interval of our candidate members. Therefore, we infer masses using not only the BT-Settl and PARSEC models independently but also the join of these two, which we call the PB model. For the latter, we select as joining point a mass of 1.4 $M_{\odot}$, which is the one where the two isochrone models reach their maximum agreement \citep[see, for example,][]{2019A&A...623A..35L}.

\subsection{Energy distribution}
\label{methods:energy}
The energy distributions will allow us to identify physical members gravitationally bound to the stellar system as well as possible escapers. In the following, we describe the method to quantify the total energy of each of our candidate members.

We obtain the energy distribution of each candidate member with respect to its parent stellar system by adding its kinetic and potential energies. These energies are computed from the candidate's posterior distributions of mass, positions, and velocities inferred with the methods of the previous sections. Then, we take 1000 samples from these posterior distributions, translate the positions and velocities to the reference frame of the stellar system, and compute the sample's total energy as the sum of its kinetic and potential energies. By taking samples of the posterior distribution, we avoid assuming a particular family distribution and thus the need of Jacobian computations. 

The total energy of each sample is computed as

\begin{equation}
\label{equation:energy}
E=\frac{1}{2}m \cdot v^2 - \frac{G\cdot M\cdot m}{r},
\end{equation}
where $r$ and $v$ are the distance and speed in the reference frame of the stellar system, $M$ is the total mass enclosed within the distance $r$, $m$ is the sample's mass, and $G$ is the gravitational constant. 

We notice the following  assumptions. First, given the low photometric extinction of these two groups ($A_v\lesssim0.75$ mag, see Sect. \ref{results:mass}), we assume no contribution of the dust and gas to the total mass of these systems. Second, our mass estimates for individual candidate members are most likely underestimated in the case of unresolved binary or multiple systems. Thus, we assume that the contribution of binary stars and unresolved systems to the total mass of the stellar group is 20\% of its inferred mass. The estimated value for the fraction of binary systems in open clusters varies from 11\% to 70\% \citep{2010MNRAS.401..577S}, with the most typical value for unresolved binaries being $\sim$20\% \citep[e.g.,][]{2021AJ....162..264J}.  Third, in the energy computation, see Eq. \ref{equation:energy}, we assume that the stellar systems are self-gravitating, and thus we neglect the contribution of the Milky Way's gravitational potential. This assumption is valid for establishing each candidate member's energy relative to its parent group. Due to the negligible size of these systems compared to the Galactic scale, we assume that the Milky Way's gravitational potential can be treated as constant and thus removed from the energy computation. This assumption will be dropped in the next section, where our interest lies in tracing our candidate members' the present-day positions and velocities forward in time. 

\subsection{Trace-forward}
\label{methods:trace-forward}

To determine the time of the fly-by between Coma Ber and Latyshev 2, we trace-forward into the future the present-day phase-space coordinates inferred with \textit{Kalkayotl} (see Sect. \ref{methods:6D_structure}). We trace-forward not only the group-level parameters (i.e., mean positions and velocities of the groups) but also the source-level parameters (i.e., positions and velocities of the candidate members). The uncertainties of these phase-space coordinates are taken into account by tracing forward 100 samples from the posterior distributions of the group- and source-level parameters. We assume that these samples are particles subject only to the Milky Way's gravitational potential, and in the following, we refer to them as sample-particles.

We integrate the orbit of each of these sample-particle 20 Myr into the future using the \textit{Galpy} code \citep{2015ApJS..216...29B}, a time grid with steps of 0.5 Myr and assuming as gravitational potential the \textit{Galpy}'s \textit{MWPotential2014}. The latter assumes a rotational velocity of 220 $\rm{km\ s^{-1}}$ at a solar radius of 8 kpc \citep[for specific details of this gravitational potential, see][]{2015ApJS..216...29B}. 

Finally, for each group-level sample-particle we determine the time of the fly-by as the time at which the distance between the two groups reaches its minimum. We prefer to work with the group-level parameters to estimate the fly-by time rather than the source-level parameters because the former have better precision than the latter. We report the time of encounter with statistics (mean and standard deviation) of the distribution of fly-by times. Once the fly-by time is determined, we use the source-level sample particles to analyse the degree of attachment of the individual stars to their parent group and the rest of the groups and thus test the hypothesis of mixing of stellar populations.

\section{Results}
\label{results}

This section presents the results we obtained after applying the methods of Sect. \ref{methods} to the data introduced in Sec. \ref{dataset}. The comparison of these results with those from the literature will be done in Sect. \ref{discussion}. 

\subsection{Membership}
\label{results:membership}

We applied our novel membership algorithm (see Sect. \ref{methods:membership}) to the $\sim$45 million \textit{Gaia} DR3 sources (see Sect. \ref{dataset}), using as input lists of members those described in Sect. \ref{data:initial_list}.  In Coma Ber, our membership algorithm converged in six iterations to 302 candidate members, while in Latyshev 2, it converged to 332 candidate members in five iterations. Table \ref{table:list_of_members} contains the identifiers and group classification for all our candidate members together with their properties as inferred throughout the rest of this work.

Tables \ref{table:ComaBer_quality} and \ref{table:GroupX_quality} show the quality indicators of the Coma Ber and Latyshev 2 classifiers at the final iteration of the membership algorithm. The first line of the tables (column Strategy All) shows, only for comparison purposes, the properties of the classifier if no magnitude bins  were used for the optimisation. On the other hand, the remaining lines of the tables (Bins 1-9) show the classifier properties measured as a function of the magnitude bins. The edges of these bins were fixed at 2, 6, 8, 10, 12, 14, 16, 18, 20, and 22 magnitudes in the \textit{Gaia} \texttt{g} band. The remaining columns show, for each magnitude bin and from left to right, the mean value of the sources \texttt{g} magnitude, the optimal probability threshold (truncated to two digits), the total number of sources, the entries of the confusion matrix (True Positives, TP, False Positives, FP, True Negatives, TN, and False Negatives, FN), and the quality indicators of True Positive Rate (TPR), Contamination Rate (CR) and MCC  (see Note \ref{note:confusion_matrix} and Sect. \ref{methods:membership:probability_thresholds}). We notice that the values shown correspond to the mean values computed from the ten synthetic clusters used to establish the optimum probability thresholds (see Sect. \ref{methods:membership:probability_thresholds}). As can be observed from these tables, the quality indicators of the membership methodology are excellent, with true positive rates $\gtrsim99\%$ and contamination rates $\lesssim 5\%$. The only exception is at the faintest magnitude bin (\texttt{g}> 20 mag), where the TPR of the Latyshev 2 classifier drops to 95\%, and the CR of the Coma Ber classifier grows up to 44\%.

The astrometric quality of our candidate members is briefly summarised as follows. The parallax fractional error is typically (mode) better than 1\% with only one extreme value of 20\% for a Coma Ber tails member. The mean proper motions uncertainties are $0.16\,\rm{mas\,yr^{-1}}$, $0.08\,\rm{mas\,yr^{-1}}$, $0.09\,\rm{mas\,yr^{-1}}$, and $0.10\,\rm{mas\,yr^{-1}}$ for Coma Ber's tails, core, Latyshev 2, and Mecayotl 1, respectively. With respect to the RUWE parameter \citep[see, for example,][]{2021A&A...649A...2L}, the typical (mode) values for ComaBer's core, tails, Latyshev 2 and Mecayotl 1 are 1.05, 1.06, 1.05, and 1.03, respectively, while their cumulative distributions reach the 1.4 threshold at 85\%, 80\%, 86\% and 83\%, respectively. A RUWE value larger than 1.4 may indicate that the source is non-single or has a problematic astrometric solution\footnote{\url{https://gea.esac.esa.int/archive/documentation/GDR2/Gaia_archive/chap_datamodel/sec_dm_main_tables/ssec_dm_ruwe.html}}. Concerning the astrometric excess noise parameter (\texttt{astrometric\_excess\_noise}, $\epsilon_i$) and its significance (\texttt{astrometric\_excess\_noise\_sig}, $\sigma\epsilon_i$), which are measures of the disagreement between the observations of a source and its best fitting astrometric model \citep[see Sects. 3.6 and 5.1.2 of][]{2012A&A...538A..78L}, the median values in Coma Ber's core are $\epsilon_i=0.13$ mas and $\sigma\epsilon_i=9.36$, in Coma Ber's tail are $\epsilon_i=0.13$ mas and $\sigma\epsilon_i=2.58$, in Mecayotl 1 are $\epsilon_i=0.19$ mas and $\sigma\epsilon_i=3.70$, and in Latyshev 2 are $\epsilon_i=0.12$ mas and $\sigma\epsilon_i=3.46$. However, the \textit{Gaia} DR3 non-single star catalogue \citep[NSS, see, for example,][]{2022arXiv220800211G} classifies only nine, two, and four of our candidate members of Coma Ber, Latyshev 2, and Mecayotl 1, respectively, as potential binaries (i.e., their \texttt{non\_single\_star} entry is non zero). 

\begin{table}[ht!]
\caption{Confusion matrix and quality indicators of the Coma Ber classifier as function of photometric magnitude.}
\label{table:ComaBer_quality}
\centering
\resizebox{\columnwidth}{!}{
\begin{tabular}{ccccccccccc}
\toprule
Strategy &\texttt{g} &  prob &   sources   &     TP &    FP &          TN &   FN &    TPR &    CR &    MCC  \\
\midrule
All      &     - & 0.99 & 45011747.00 & 995.00 & 16.00 & 45010731.00 & 5.00 &  99.50 &  1.58 &  98.96  \\
Bin 1    &  4.00 & 0.68 &     1352.30 & 193.30 &  0.00 &     1159.00 & 0.00 & 100.00 &  0.00 & 100.00  \\
Bin 2    &  7.00 & 0.99 &     9317.10 & 266.00 &  0.00 &     9051.00 & 0.10 &  99.96 &  0.00 &  99.98  \\
Bin 3    &  9.00 & 0.99 &    55735.70 & 163.40 &  0.00 &    55572.00 & 0.30 &  99.82 &  0.00 &  99.91  \\
Bin 4    & 11.00 & 0.99 &   265981.50 & 102.00 &  0.00 &   265879.00 & 0.50 &  99.51 &  0.00 &  99.76  \\
Bin 5    & 13.00 & 0.99 &  1073300.50 &  86.20 &  1.00 &  1073213.00 & 0.30 &  99.67 &  1.16 &  99.25  \\
Bin 6    & 15.00 & 0.99 &  3422830.40 &  91.20 &  1.00 &  3422737.00 & 1.20 &  98.71 &  1.09 &  98.81  \\
Bin 7    & 17.00 & 0.99 &  8554268.70 &  45.80 &  2.00 &  8554220.00 & 0.90 &  98.15 &  4.20 &  96.96  \\
Bin 8    & 19.00 & 0.99 & 18662996.90 &  33.80 &  1.00 & 18662962.00 & 0.10 &  99.70 &  2.92 &  98.38  \\
Bin 9    & 20.72 & 0.99 & 12965963.90 &  14.90 & 11.00 & 12965938.00 & 0.00 & 100.00 & 44.19 &  74.36  \\
\bottomrule
\end{tabular}
}
\end{table}

\begin{table}[ht!]
\caption{Confusion matrix and quality indicators of the Latyshev 2 classifier as function of photometric magnitude.}
\label{table:GroupX_quality}
\centering
\resizebox{\columnwidth}{!}{
\begin{tabular}{ccccccccccc}
\toprule
Strategy &\texttt{g} &  prob &   sources   &     TP &    FP &   TN        &   FN &    TPR &    CR &    MCC  \\
\midrule
All      &     - & 0.99 & 45011716.00 & 991.10 & 13.00 & 45010703.00 & 8.90 &  99.11 &  1.29 &  98.91  \\
Bin 1    &  4.00 & 0.99 &     1479.10 & 315.80 &  0.00 &     1163.00 & 0.30 &  99.90 &  0.00 &  99.94  \\
Bin 2    &  7.00 & 0.99 &     9310.90 & 254.90 &  0.00 &     9056.00 & 0.00 & 100.00 &  0.00 & 100.00  \\
Bin 3    &  9.00 & 0.99 &    55719.00 & 135.90 &  0.00 &    55583.00 & 0.10 &  99.92 &  0.00 &  99.96  \\
Bin 4    & 11.00 & 0.99 &   265964.80 &  78.50 &  0.00 &   265886.00 & 0.30 &  99.61 &  0.00 &  99.80  \\
Bin 5    & 13.00 & 0.99 &  1073278.60 &  68.00 &  1.00 &  1073209.00 & 0.60 &  99.16 &  1.46 &  98.85  \\
Bin 6    & 15.00 & 0.99 &  3422783.90 &  65.60 &  2.00 &  3422716.00 & 0.30 &  99.52 &  2.99 &  98.26  \\
Bin 7    & 17.00 & 0.99 &  8554240.10 &  42.10 &  3.00 &  8554194.00 & 1.00 &  97.67 &  6.72 &  95.45  \\
Bin 8    & 19.00 & 0.99 & 18662979.20 &  24.80 &  1.00 & 18662953.00 & 0.40 &  98.42 &  3.93 &  97.23  \\
Bin 9    & 20.72 & 0.99 & 12965960.40 &  10.90 &  6.00 & 12965943.00 & 0.50 &  95.74 & 37.62 &  77.04  \\
\bottomrule
\end{tabular}
}
\end{table}

\subsection{Phase-space structure}
\label{results:6D_structure}

As explained in Sect. \ref{methods:6D_structure}, we fitted three different models, Gaussian, CGMM, and GMM, to the observed astrometry and radial velocity (when available, see Sect. \ref{dataset}) of our 302 Coma Ber and 332 Latyshev 2 candidate members. The best phase-space models that describe these observables are a CGMM with two components for Coma Ber and a GMM with two components for Latyshev 2. As explained in Sect. \ref{methods:6D_structure}, we selected the best model based on the criteria of the convergence of the inference algorithm and the non-negligible weights of the mixture components. Our attempts to fit models with three and four components returned negligible weights for the third component or failed to converge in the case of four components. 

We observed that one candidate member of Coma Ber (\texttt{source\ id}: 4016309615972923648) and five of Latyshev 2 (\texttt{source\_id}: 1608752435341695872, 1560784144636654976, 857806987370330624, 1560938316782309248, and 1512261012875442304) were located more than three sigma away from the mean UVW velocities of the clusters. To prevent these sources from biasing the internal velocity dispersion of the clusters, we masked their radial velocities as missing. We check the binary status of the previous sources and found that only source 1608752435341695872\footnote{With \texttt{RUWE}=13.0, \texttt{astrometric\_excess\_noise}=3.16, and \texttt{astrometric\_excess\_noise\_sigma}=11568.} is a confirmed binary star according to SIMBAD, while sources 1560938316782309248\footnote{With \texttt{RUWE}=15.65, \texttt{astrometric\_excess\_noise}=2.3, and \texttt{astrometric\_excess\_noise\_sigma}=2102.2.}  and 1560784144636654976\footnote{ With \texttt{RUWE}=6.45, \texttt{astrometric\_excess\_noise}=0.89, and \texttt{astrometric\_excess\_noise\_sigma}=1224.5.} are most likely binary stars. However, the remaining three sources have \texttt{RUWE} values between 1.06 and 1.14 and 
\texttt{astrometric\_excess\_noise} between 0.03 and 0.16, which are not conclusive enough to claim binarity. None of these six stars appears in the \textit{Gaia} DR3 NSS sample (i.e, their \texttt{non\_single\_star} entry is zero), which indicates that according to \textit{Gaia}, they are neither astrometric, spectroscopic or eclipsing binaries. However, due to the low completeness of the NSS sample \citep[see Sect. 3 of][]{2022arXiv220605595G} and the discrepant radial velocities of these sources, we cannot rule out the possibility that they are spectroscopic binaries.

In the fitted CGMM of Coma Ber, we observe that one of the Gaussian components serves to describe the cluster core while the other one describes the cluster tails. In the fitted GMM of Latyshev 2, the two Gaussian components are mutually exclusive in the joint 6D space of positions and velocities (more below). Thus, we identify one of these components with the original candidate open cluster Latyshev 2 and the other with a new stellar association that we call Mecayotl 1. 

\begin{figure}[ht!]
    \centering
     \includegraphics[width=\columnwidth,page=1]{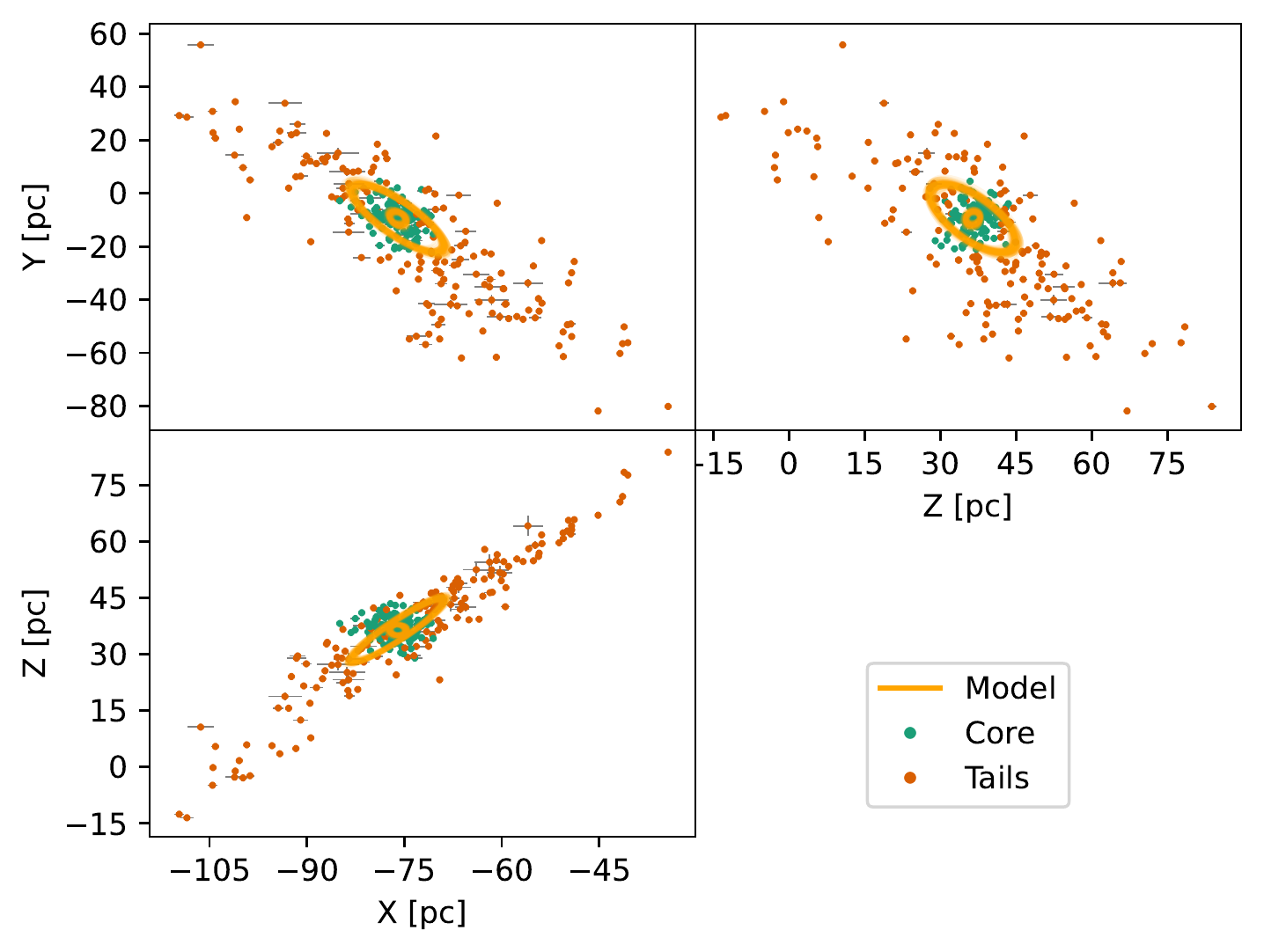}
     \includegraphics[width=\columnwidth,page=2]{Figures/ComaBer_kalkayotl.pdf}
     \caption{Positions (top panel) and velocities (bottom panel) of Coma Ber candidate members. The 100 orange lines show the posterior
samples from the inferred group-level parameters while the dots and error bar depict the mean and standard deviation, respectively, of the inferred positions and velocities.}
\label{fig:ComaBer_kalkayotl}
\end{figure}

\begin{figure}[ht!]
    \centering
     \includegraphics[width=\columnwidth,page=1]{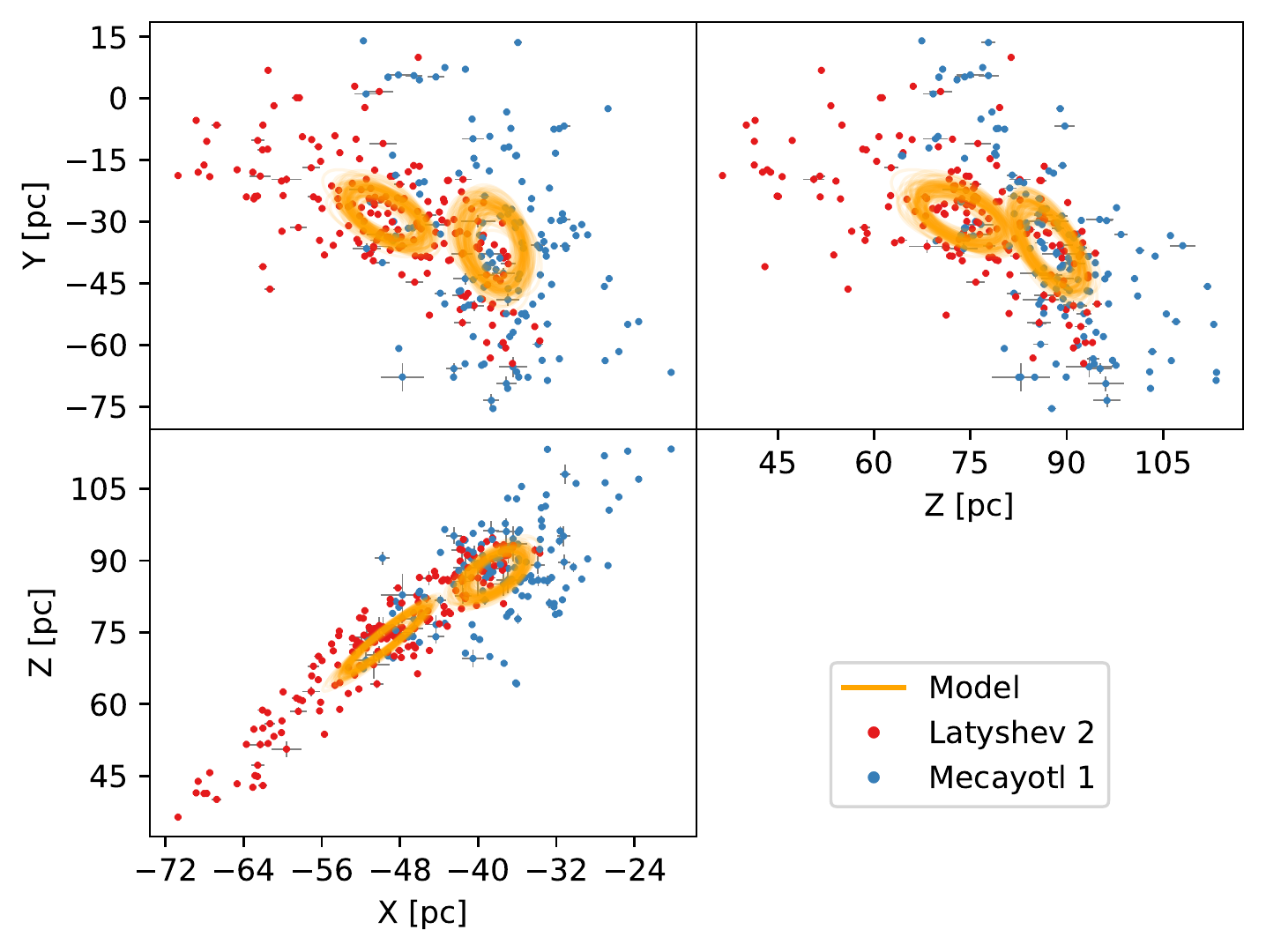}
     \includegraphics[width=\columnwidth,page=2]{Figures/GroupX_kalkayotl.pdf}
     \caption{Positions (top panel) and velocities (bottom panel) of the candidate members of Latyshev 2 and Mecayotl 1. Captions as in Fig. \ref{fig:ComaBer_kalkayotl}.}
\label{fig:GroupX_kalkayotl}
\end{figure}

Figures \ref{fig:ComaBer_kalkayotl} and \ref{fig:GroupX_kalkayotl} show, with orange lines, 100 posterior samples from the inferred group-level parameters of Coma Ber's CGMM and Latyhsev 2 + Mecayotl 1 GMM phase-space models, respectively. These figures show multidimensional projections of the phase-space, with the top panels depicting positions and the bottom ones velocities. The dots and error bars in the figures represent the mean and standard deviation of each candidate member's posterior samples of its phase-space coordinates. The colour code  of the individual sources indicates their probabilistically assigned parent structure, which, as mentioned in Sect. \ref{methods:6D_structure} corresponds to the classification into each of the Gaussian of the GMM. We notice that this classification is done in the 6D space and is based on the model parameters and the source's observed data. This probabilistic classification results in 186 candidate members in Latyshev 2, 146 candidate members in Mecayotl 1, and 136 and 166 candidate members in the core and tails of Coma Ber, respectively.

\begin{figure}[ht]
    \centering
     \includegraphics[width=\columnwidth]{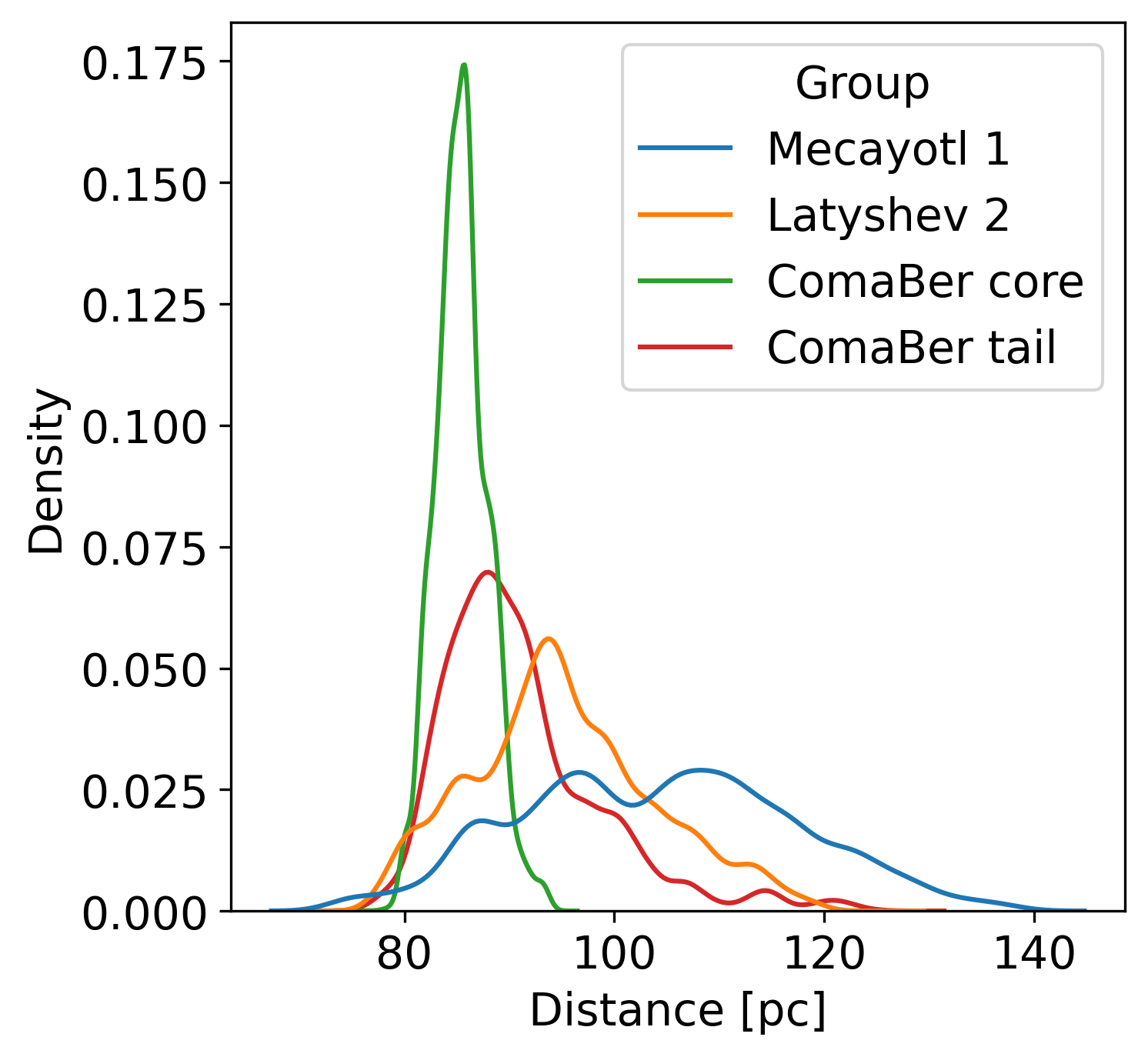}
     \caption{Distance distributions of Coma Ber, Latyshev 2 and Mecayotl 1.}
\label{fig:distances}
\end{figure}

Figure \ref{fig:distances} shows distance distribution of our identified physical groups. These densities were computed by applying a kernel density estimate to the aggregation of 100 posterior samples of each candidate member. As can be observed, the core of Coma Ber is the closest group while the Mecayotl 1 one extends up to 140 pc away.

\begin{table}[ht!]
\caption{Phase-space group-level parameters of the identified substructures.}
\label{table:parameters}
\centering
\resizebox{\columnwidth}{!}{
\begin{tabular}{cccccc}
\toprule
      &             &      Coma Ber core &      Coma Ber tail &         Latyshev 2 &         Mecayotl 1 \\
{} & {} &                    &                    &                    &                    \\
\midrule
$X$ & $\rm{[pc]}$ &  $-76.03 \pm 0.22$ &  $-76.03 \pm 0.22$ &  $-49.53 \pm 0.70$ &  $-38.39 \pm 0.60$ \\
$Y$ & $\rm{[pc]}$ &   $-9.39 \pm 0.55$ &   $-9.39 \pm 0.55$ &  $-28.67 \pm 1.33$ &  $-35.89 \pm 1.81$ \\
$Z$ & $\rm{[pc]}$ &   $36.40 \pm 0.26$ &   $36.40 \pm 0.26$ &   $73.33 \pm 1.15$ &   $87.08 \pm 0.90$ \\
$U$ & $\rm{[km \cdot s^{-1}]}$ &   $-2.27 \pm 0.06$ &   $-2.27 \pm 0.06$ &   $-2.81 \pm 0.06$ &   $-3.15 \pm 0.11$ \\
$V$ & $\rm{[km \cdot s^{-1}]}$ &    $4.65 \pm 0.03$ &    $4.65 \pm 0.03$ &    $7.52 \pm 0.04$ &    $7.25 \pm 0.09$ \\
$W$ & $\rm{[km \cdot s^{-1}]}$ &   $-3.17 \pm 0.03$ &   $-3.17 \pm 0.03$ &   $-4.96 \pm 0.07$ &   $-6.71 \pm 0.21$ \\
$\sigma_X$ & $\rm{[pc]}$ &    $2.62 \pm 0.24$ &   $14.39 \pm 0.79$ &    $8.14 \pm 0.44$ &    $6.24 \pm 0.42$ \\
$\sigma_Y$ & $\rm{[pc]}$ &    $5.36 \pm 0.55$ &   $26.07 \pm 1.44$ &   $13.86 \pm 0.98$ &   $20.27 \pm 1.24$ \\
$\sigma_Z$ & $\rm{[pc]}$ &    $2.98 \pm 0.25$ &   $17.09 \pm 0.93$ &   $13.58 \pm 0.74$ &   $10.15 \pm 0.59$ \\
$\sigma_U$ & $\rm{[km \cdot s^{-1}]}$ &    $0.75 \pm 0.09$ &    $0.51 \pm 0.05$ &    $0.48 \pm 0.04$ &    $1.01 \pm 0.08$ \\
$\sigma_V$ & $\rm{[km \cdot s^{-1}]}$ &    $0.23 \pm 0.02$ &    $1.19 \pm 0.07$ &    $0.37 \pm 0.03$ &    $0.71 \pm 0.07$ \\
$\sigma_W$ & $\rm{[km \cdot s^{-1}]}$ &    $0.43 \pm 0.05$ &    $0.61 \pm 0.05$ &    $0.52 \pm 0.06$ &    $1.38 \pm 0.15$ \\
$weight$ &   &    $0.41 \pm 0.03$ &    $0.59 \pm 0.03$ &    $0.52 \pm 0.03$ &    $0.48 \pm 0.03$ \\
$|\sigma_{XYZ}|$ & $\rm{[pc]}$ &      $6.66\pm0.64$ &     $34.34\pm1.89$ &     $21.04\pm1.29$ &     $23.51\pm1.43$ \\
$|\sigma_{UVW}|$ & $\rm{[km \cdot s^{-1}]}$ &      $0.89\pm0.10$ &      $1.43\pm0.10$ &      $0.80\pm0.08$ &      $1.85\pm0.18$ \\
$\overline{\hat{\vec{e}}_r \cdot \vec{v}}$  & $\rm{[km \cdot s^{-1}]}$ & $0.11 \pm0.39$ & $0.70 \pm0.70$ & $0.04 \pm0.43$ & $-0.14 \pm1.05$ \\
$\overline{ \lVert\hat{\vec{e}}_r \times \vec{v}\rVert}$  & $\rm{[km \cdot s^{-1}]}$ & $0.58\pm 0.43$ & $0.98\pm 0.61$ & $0.67\pm 0.47$ & $1.32\pm 0.74$ \\
$r_t$ & $\rm{[pc]}$ &      $6.32\pm0.22$ &      $5.99\pm0.23$ &      $6.44\pm0.24$ &      $6.02\pm0.25$ \\
\bottomrule
\end{tabular}

}
\end{table}

Table \ref{table:parameters} shows the mean and standard deviation of the inferred group-level parameters of the CGMM of Coma Ber's core and tail and the GMM of Latyshev 2 and Mecayotl 1. For completeness, the bottom five lines of this table show the total position and velocity dispersions of the inferred phase-space models ($\sigma_{XYZ}$ and $\sigma_{UVW}$, respectively), two kinematic proxy indicators, one for expansion and the other for rotation, and the tidal radius ($r_t$). The latter corresponds to the distance from the group's centre at which the gravitational pull exerted by the Galactic potential equals that of the group's potential under the assumption that the latter follows a circular orbit around the centre of the former. We computed this tidal radius using Eq. 1 of \citet{2019ApJ...877...12T} with the Oort's constants values of those authors. The total position and velocity dispersions were computed by adding the dispersion of each coordinate in quadrature. Finally, the proxy indicators for expansion, $\overline{\hat{\vec{e}}_r \cdot \vec{v}}$, and rotation, $\overline{\lVert\hat{\vec{e}}_r\times\vec{v}\rVert}$, are computed as the mean values of the dot and cross products of the unit radial vector, $\hat{\vec{e}}_r$, and the velocity vector, $\vec{v}$; these latter two computed in the cluster's reference frame \citep[see, for example,][]{2019A&A...630A.137G}.

Concerning Coma Ber, the top panel of Fig. \ref{fig:ComaBer_kalkayotl} shows that its tail members reach larger distances, with the farthest candidate member reaching up to 95 pc from the centre, which is roughly two times distance of the latest result from the literature \citep{2019ApJ...877...12T}. In the velocity space (bottom panel of Fig. \ref{fig:ComaBer_kalkayotl}), we observe that the tail members have similar but not identical velocities to those in the core. 

Concerning Latyshev 2 and Mecayotl 1, both panels of Fig. \ref{fig:GroupX_kalkayotl} show that they have distinct positions and velocities. Although their members appear to be entangled, their 6D models are not. Quantitatively, we measure the degree of separation between Latyshev 2 and Mecayotl 1 with the Mahalanobis metric. This metric is the multidimensional extension of the distance to the mean value in a one-dimensional Gaussian distribution when measured in units of the standard deviation. We notice that these distances are not commutative since the internal dispersions of the groups are not the same. Briefly, the Mahalanobis metric can be understood as the multi-dimensional generalisation of measuring the number of standard deviations to the mean value. The Mahalanobis distance that Latyshev 2 has with respect to Mecayotl 1 is 2.03, whereas that of Mecayotl 1 with respect to Latyhsev 2 is 5.5, thus indicating that these groups are mutually exclusive at a level $>2\sigma$. This significance level indicates that our discovery of the two groups is statistically significant, with a credibility $>$95\%. Furthermore, our claim that Latyshev 2 is an open cluster while Mecayotl 1 is a stellar association is based on their velocity dispersion. On the one hand, Latyshev 2 is a compact group in the velocity space with a total velocity dispersion of $\sigma_{UVW}=0.80\pm0.08\ \rm{km\ s^{-1}}$, which is similar to that of Coma Ber's core ($\sigma_{UVW}=0.89\pm0.10\ \rm{km\ s^{-1}}$) and that of the archetypical Pleiades open cluster ($0.8\pm0.1\ \rm{km\ s^{-1}}$ according to \citealt{2017A&A...598A..48G}, and $0.83\pm0.07\ \rm{km\ s^{-1}}$ according to \citealt{2021ApJ...921..117T}, as computed from their 1D radial velocity dispersion of $0.48\pm0.04\ \rm{km\ s^{-1}}$). On the other hand, Mecayotl 1 has a total velocity dispersion of $\sigma_{UVW}=1.85\pm0.18\ \rm{km\ s^{-1}}$, which is more than two times the velocity dispersion of Latyshev 2 or the core of Coma Ber, and similar to the $1.73\pm0.3\ \rm{km\ s^{-1}}$ of the $\beta$Pic stellar association \citep{2020A&A...642A.179M}. Further arguments for these classifications will be provided in Sect. \ref{results:energy}.

\subsection{Luminosity and mass distributions}
\label{results:mass}

To determine the luminosity and mass distributions of the identified physical groups, we first verify that the ages proposed by the literature are indeed compatible with the photometry of our candidate members. Then we compute the luminosity and mass of the individual candidate members and combine them to obtain the corresponding distributions for the physical groups.

\begin{figure*}[ht!]
    \centering
     \includegraphics[width=\textwidth]{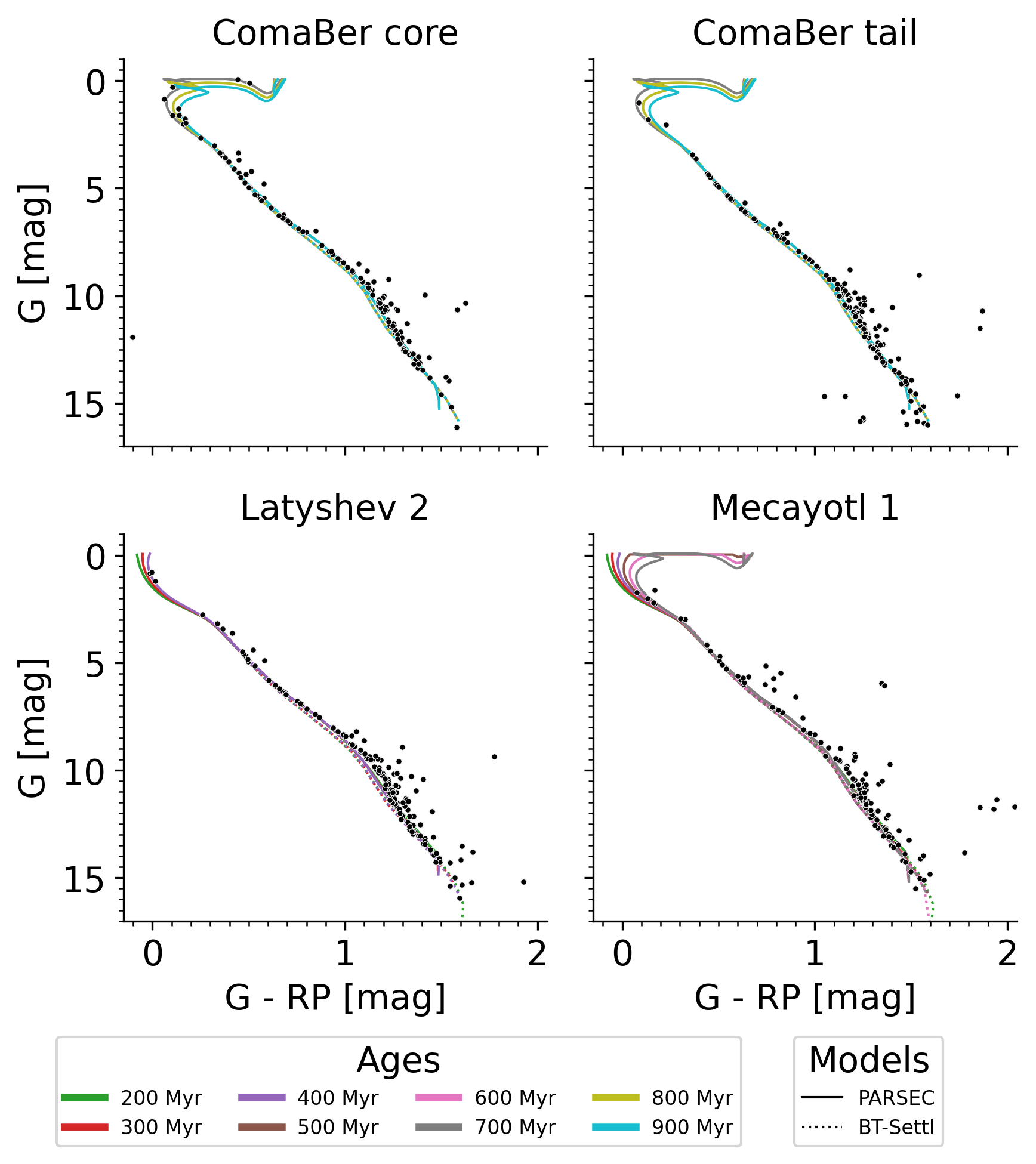}
     \caption{Absolute colour-magnitude diagrams of the candidate members of Coma Ber, Latyshev 2 and Mecayotl 1. The lines depict the theoretical isochrone models with colour-coded ages.}
\label{fig:cmds}
\end{figure*}

Figure \ref{fig:cmds} shows the $G$ vs $G-RP$ absolute colour-magnitude diagrams of our candidate members, together with the theoretical isochrones from the PARSEC  and BT-Settl models at our chosen grids of ages (see Sect. \ref{methods:mass}). In addition, for Mecayotl 1 we also plot the PARSEC isochrones for up to 700 Myr. As can be observed from this figure, the age intervals proposed in the literature are compatible with the empirical isochrones traced by our candidate members. 

In the case of Coma Ber, our brightest candidate members support the ages of 700 to 800 Myr, as \citet{2020A&A...640A...9M} proposed based on lithium measurements and \citet{2019ApJ...877...12T,2018ApJ...862..106T} on isochrone fitting, respectively. In the case of Latyshev 2, the three brightest members originally proposed by \citet{1977ATsir.969....7L} and confirmed here serve as anchors for the isochrone age of the system. These three members seem to indicate that this cluster is young, most likely aged 300 to 400 Myr, which supports the gyrochronology age of $300\pm60$ Myr proposed by \citet{2022A&A...657L...3M} rather than the isochrone fitting ages of \citet[][$450\pm100$ Myr]{2021OAst...30..191S} and \citet[][$141\pm16$ Myr]{2022arXiv220612170H}. On the other hand, the brightest candidate members of Mecayotl 1 show that its isochrone age is between 400 Myr and 600 Myr. Under the lack of constraining evidence, in the following, we assume an age of 400 Myr for this stellar system. As can be observed in Fig. \ref{fig:cmds}, this age corresponds to the closest lower-envelope isochrone to the observed photometry of our candidate members.

As mentioned in Sect. \ref{methods:mass}, we compute the individual luminosity and mass distributions of the candidate members given the theoretical isochrone models and the observed photometry. Then, we combine the resulting posterior samples of all candidate members into the luminosity and mass distributions of the physical groups. Given the excellent recovery rate of our membership algorithm (>95\%, see TPR column in Tables \ref{table:ComaBer_quality} and \ref{table:GroupX_quality}), we do not correct the resulting luminosity and mass distributions for possible missing members. Similarly, given the low contamination rate of our membership algorithm (<7\%, see CR column in Tables \ref{table:ComaBer_quality} and \ref{table:GroupX_quality}) we do not correct the luminosity ans mass distributions for contamination. However, the faintest (\texttt{g}>20 mag) magnitude bin has large contamination rates ($\sim$40\%), which indicates that the luminosity and mass distributions at this magnitude bin can only be taken as upper limits to the true distributions.

\begin{figure}[ht!]
    \centering
     \includegraphics[width=\columnwidth,page=1]{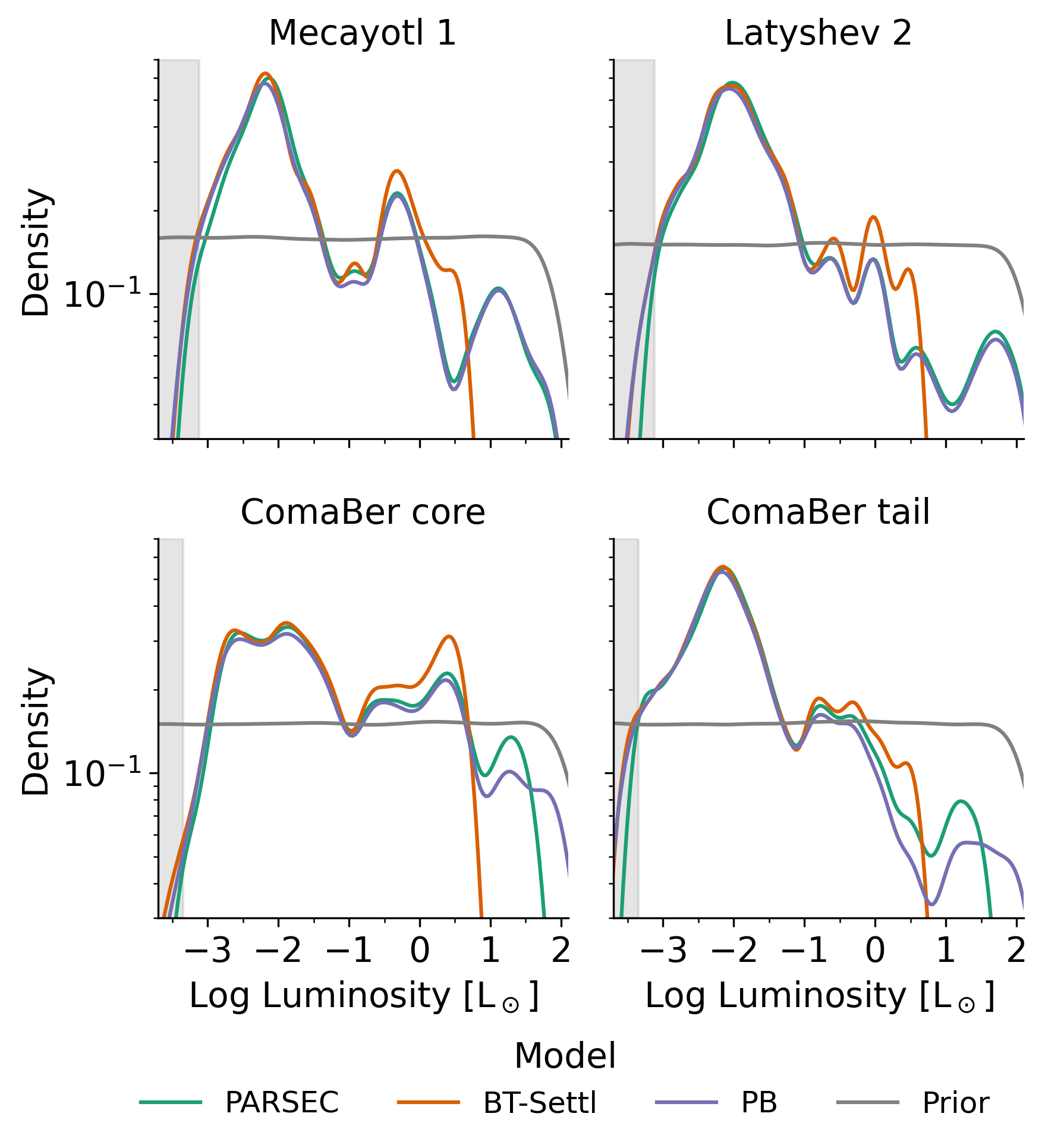}
     \caption{Luminosity distributions inferred with different theoretical models. The grey line and grey region depict the uniform prior and the completeness limit of the \textit{Gaia} data. This latter corresponds to apparent G mag $\sim19$ transformed to luminosity using the mean group distance and BT-Settl model.}
\label{fig:luminosity}
\end{figure}

Figure \ref{fig:luminosity} shows the luminosity distributions of the identified groups as inferred from different models (colour coded). The differences that result from inferring the luminosity and mass distributions using the grid of ages mentioned in Sect. \ref{methods:mass} are negligible. Thus, to avoid overcrowding the plots, we show only the most likely ages: 300 Myr for Latyshev 2, 400 Myr for Mecayotl 1, and 800 Myr for Coma Ber. The lines shown in Figure \ref{fig:luminosity} result from applying a kernel density estimate to 1000 aggregated samples from the posterior luminosity distribution of each of our candidate members. As can be observed, the luminosity distributions resulting from different theoretical models agree well, except at the limit of the BT-Settl model, where there is a concentration of samples from sources more luminous than the BT-Settl limit.

\begin{figure}[ht!]
    \centering
     \includegraphics[width=\columnwidth,page=1]{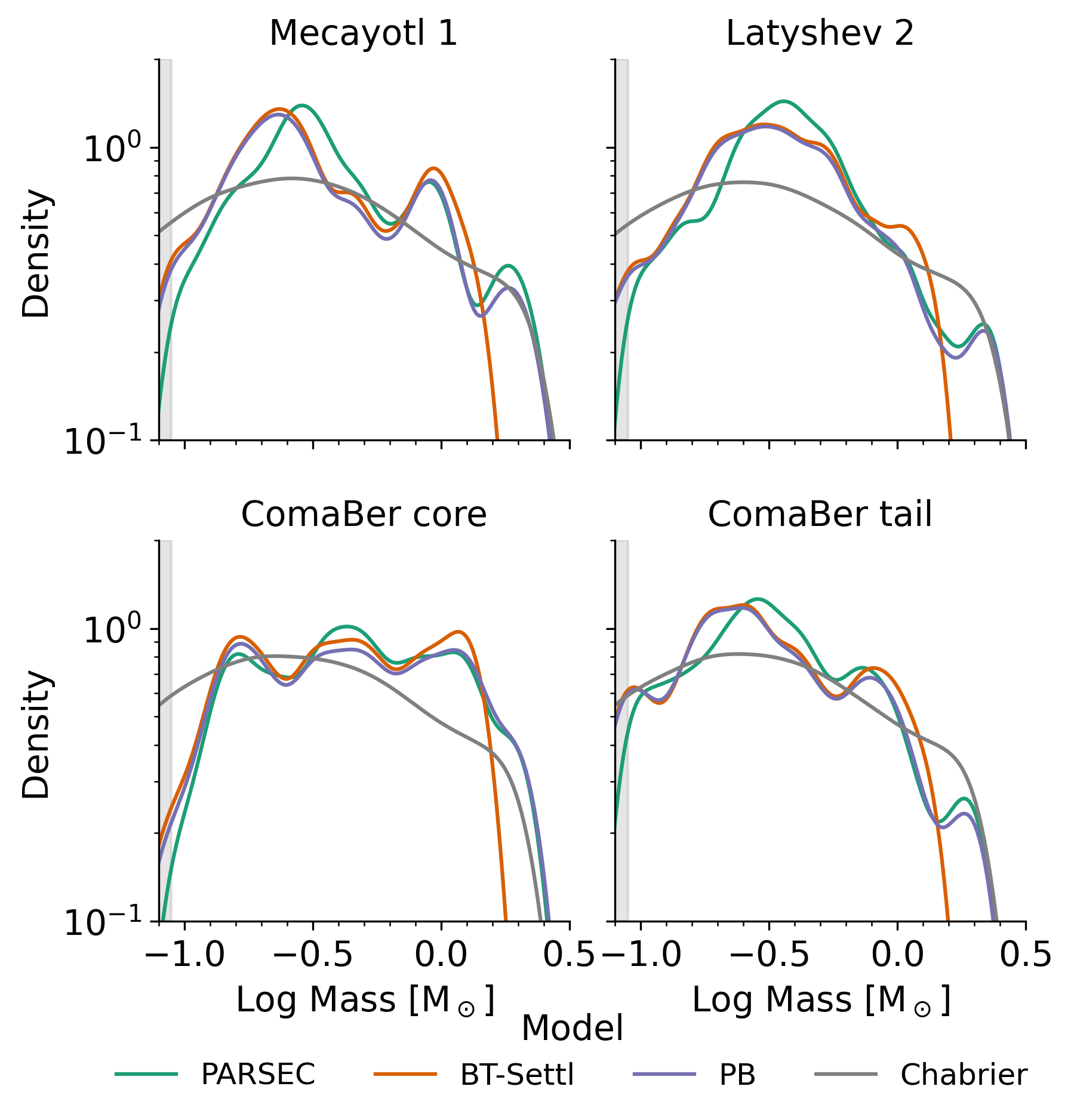}
     \caption{Mass distributions inferred with different theoretical models. Caption as in Fig. \ref{fig:luminosity}. In this case, the grey line depicts the \citet{2005ASSL..327...41C} mass prior.}
\label{fig:mass}
\end{figure}

Similarly to Fig. \ref{fig:luminosity}, Fig. \ref{fig:mass} shows the mass distributions inferred for Coma Ber, Latyshev 2 and Mecayotl 1 using the same models and ages as for the luminosity. The lines also show the kernel density estimate of the aggregated posterior samples. The mass prior (grey lines) corresponds to \citet{2005ASSL..327...41C} mass distribution. As can be observed, the mass distributions resulting from different models have varying degrees of agreement. In the core of Coma Ber mass distributions agree well for all models where as in the rest of the cases the peak of the mass distributions resulting from the PARSEC model are shifted $0.06\,M_\odot$ with respect to those of the BT-Settl and PB models. Given that the BT-Settl models have been specifically created for low-mass stars and brown-dwarfs and the bulk of our candidate members is made of low-mass stars, in the following, we will use the mass estimates obtained from the unified PB model, this is BT-Settl for masses $<1.4 M_\odot$ and PARSEC for masses $>1.4 M_\odot$ (see Sect. \ref{methods:mass}). We select this joint model because it is the only one covering the entire magnitude interval of our candidate members. Concerning the ages, we also observe negligible differences in the mass distributions resulting from using $\pm$100 Myr from the selected ones.

The luminosity distributions of Coma Ber, Latyshev 2 and Mecayotl 1 all show a dip at $\log{L/L_{\odot}}\!\simeq\!-0.89$, which corresponds to the Wielen dip \citep{1974HiA.....3..395W} and to a mass of $\simeq\!0.7M_{\odot}$ \citep{1990MNRAS.244...76K}. This dip has been consistently observed in the solar neighbourhood (<20 pc) and in young (<50 Myr) open clusters \citep[see][and references therein]{2021A&A...655A..45G}, and even in the old (2.5 Gyr) open cluster Ruprecht 147 \citep{2018A&A...617A..15O}. We notice that this dip is less prominent in the luminosity distribution of Latyshev 2 as well as in its mass distribution. A similar situation occurs with the mass distribution of the core of Coma Ber, although in this case the dip is clearly observed in its luminosity distribution.

As mentioned in Sect. \ref{methods:mass}, when inferring the luminosity or mass distributions, the \textit{Sakam} code also delivers posterior distributions of $A_v$ and $R_v$. In the case of $A_v$ the inferred values are all consistent with the chosen prior (i.e., $A_v<0.1$ mag, see Sect. \ref{methods:mass}). Concerning $R_v$, the posterior distributions are also consistent with the prior (i.e., $R_v=3.1$) in spite of the different theoretical isochrones used. Furthermore, it shows no apparent correlation with $A_v$, the luminosity, or the mass of the candidate members. 

\subsection{Energy distribution}
\label{results:energy}

\begin{figure}[ht!]
    \centering
     \includegraphics[width=\columnwidth,page=1]{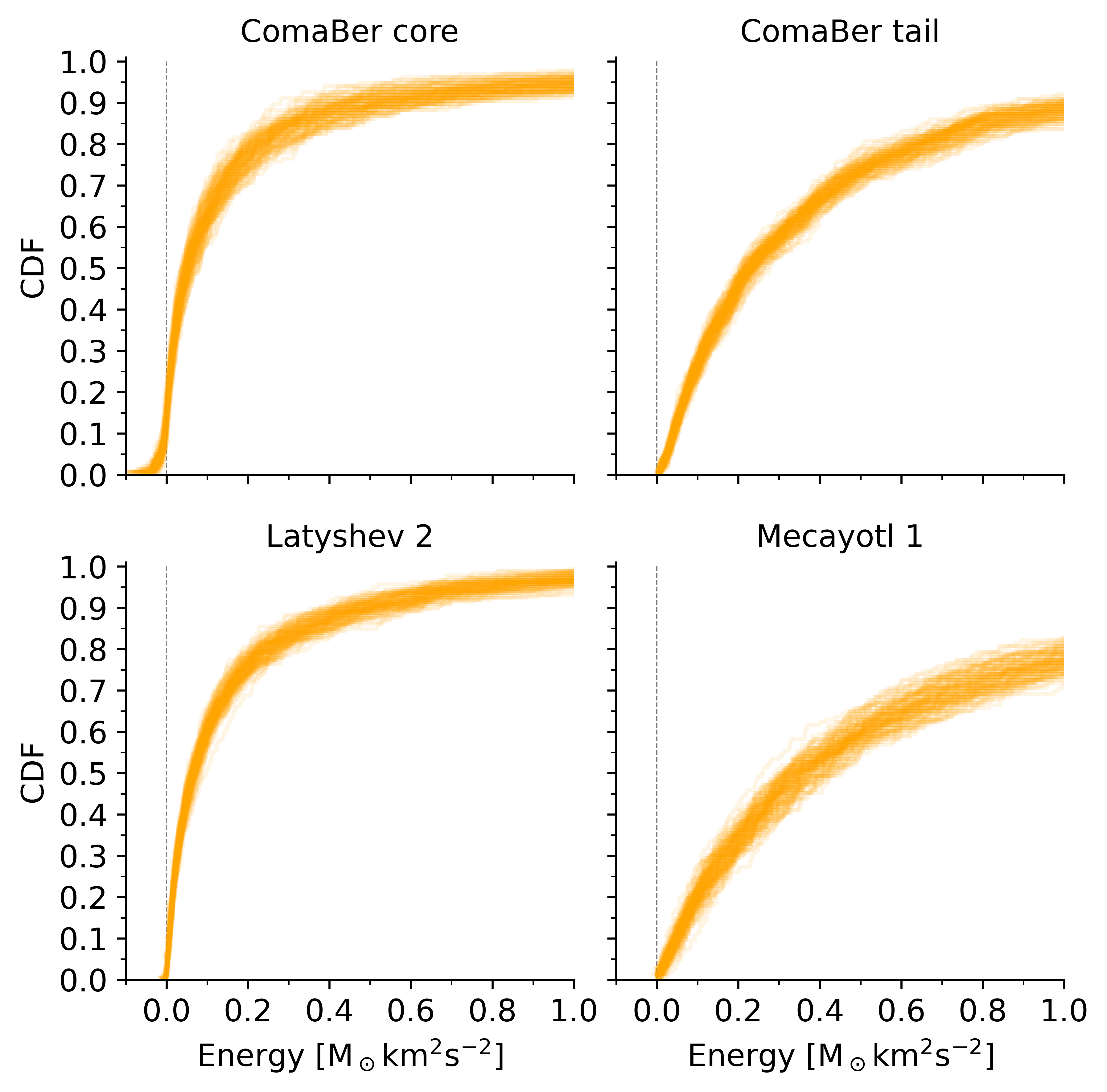}
     \caption{Cumulative energy distribution functions. Each orange line shows the result obtained from a sample of the posterior distributions of mass, position and velocity. For visual aid, the vertical grey dashed lines show the zero energy.}
\label{fig:energy_cdfs}
\end{figure}

We now compute the energy distributions of our candidate members with the methods described in Sect. \ref{methods:energy}. Figure \ref{fig:energy_cdfs} shows each group's cumulative energy distribution function (CEDF). In this figure, each orange line depicts the CEDF resulting from taking one sample of the posterior distributions of position, velocity and mass. We choose to plot the cumulative distribution function rather than the probability distribution function because it allows direct reading of the fraction of sources within a given energy value. For example, it allows us to directly read that only 20-30\% of the sources in the core of Coma Ber have negative energies.

We observe that the CEDF of Latyshev 2 resembles that of Coma Ber's core while the one of Mecayotl 1 resembles that of Coma Ber's tail. On the one hand, the CEDFs of Latyhsev 2 and the core of Coma Ber have steep slopes, with 80\% of its sources having energies smaller than $0.3\ \rm{M_{\odot}\ km^2\ s^{-2}}$ and 95\% having energies smaller than $1.0\ \rm{M_{\odot}\ km^2\ s^{-2}}$.  On the other hand, Mecayotl 1 has a large energy dispersion, with only 40\% of its sources having energies smaller than $0.3\ \rm{M_{\odot}\ km^2\ s^{-2}}$ and 75\% having energies smaller than $1.0\ \rm{M_{\odot}\ km^2\ s^{-2}}$. This energy dispersion is even larger than that of the tails of Coma Ber, where 85\% of the sources have energies smaller than $1.0\ \rm{M_{\odot}\ km^2\ s^{-2}}$. The previous results provide additional evidence supporting our classification of Latyshev 2 as an open cluster and Mecayotl 1 as a stellar association. Nonetheless, we notice that the low fraction of bound sources in these open clusters indicate that they are disrupted, as in the case of Latyshev 2, or in the process of disruption, as in the case of the core of Coma Ber. We will come back to these points in Sect. \ref{discussion}, where we will attempt to link the properties of these systems with their possible origin.

\begin{figure}[ht!]
    \centering
     \includegraphics[width=\columnwidth]{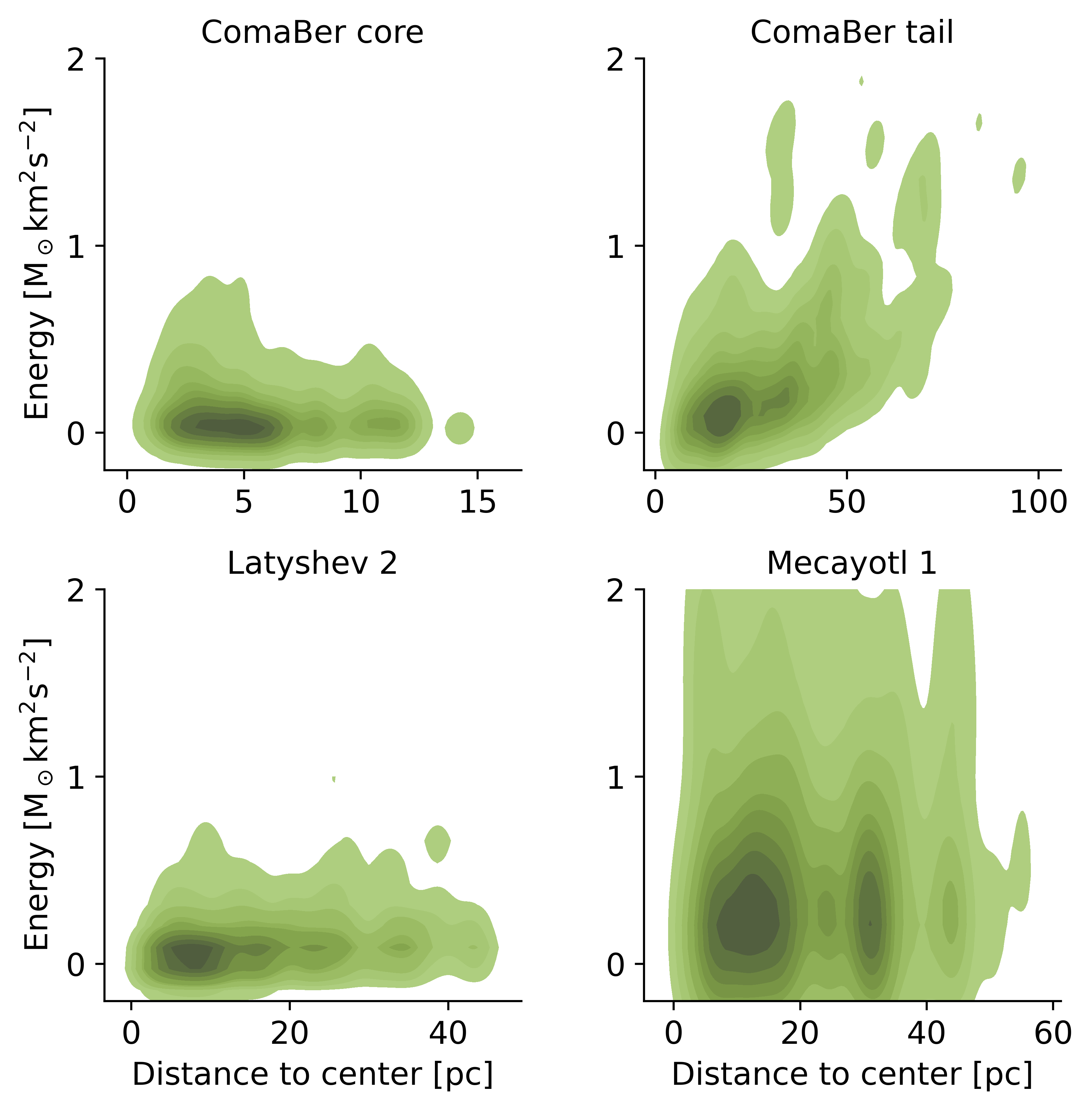}
     \caption{Density of candidate members as a function of energy and distance to the group centre.}
\label{fig:HvsR_diagram}
\end{figure}

Figure \ref{fig:HvsR_diagram} shows the 2D kernel density estimate of energy versus distance to the group centre. This figure will help us to explore possible correlations between energy and position.

Concerning Coma Ber, we observe that the core candidate members show no clear correlation between the energy and distance to the centre, with only a slow decay in the density of sources as a function of distance to the centre. On the contrary, the tidal tail candidate members show a clear correlation between energy and distance, with more distant sources resulting in larger energies. Interestingly, the candidate members do not uniformly distribute along the distance-energy diagram but agglomerate at a distance of 10-15 pc.

Concerning Latyshev 2, similarly to  the core of Coma Ber, it shows no clear correlation between distance to the centre and energy, with only a smooth decay in the density of sources with distance to the centre, which we interpret as further evidence of its similarity to the core of Coma Ber. On the other hand, Mecayotl 1 shows large dispersion in energy with over-densities in the distance to the centre. We spot at least two clear over-densities and hints of a possible third one. The two most prominent ones are uniformly spread at distance intervals of 5-15 pc and 30-35 pc, whereas the third and less prominent one is located at 45 pc. Furthermore, we observe that these three over-densities have energy dispersion larger than that of Coma Ber's tails.  We will discuss the implications of these over-densities and their dispersion in Sect. \ref{discussion}.

\subsection{Trace-forward}
\label{results:trace-forward}

As mentioned in Sect. \ref{methods:trace-forward}, we use the source- and group-level parameters inferred with \textit{Kalkayotl} to trace-forward in time the current positions and velocities of our candidate members. In the following, we will first obtain the time of flyby between Coma Ber and the other two groups using the time-integration of their group-level parameters. Then, we will estimate the candidate members' binding energies with respect to the other groups using the source-level parameters. These binding energies with respect to all the groups will allow us to investigate possible population mixing.

\subsubsection{Flyby time}

\begin{figure*}[ht!]
    \centering
     \resizebox{\hsize}{!}{\includegraphics{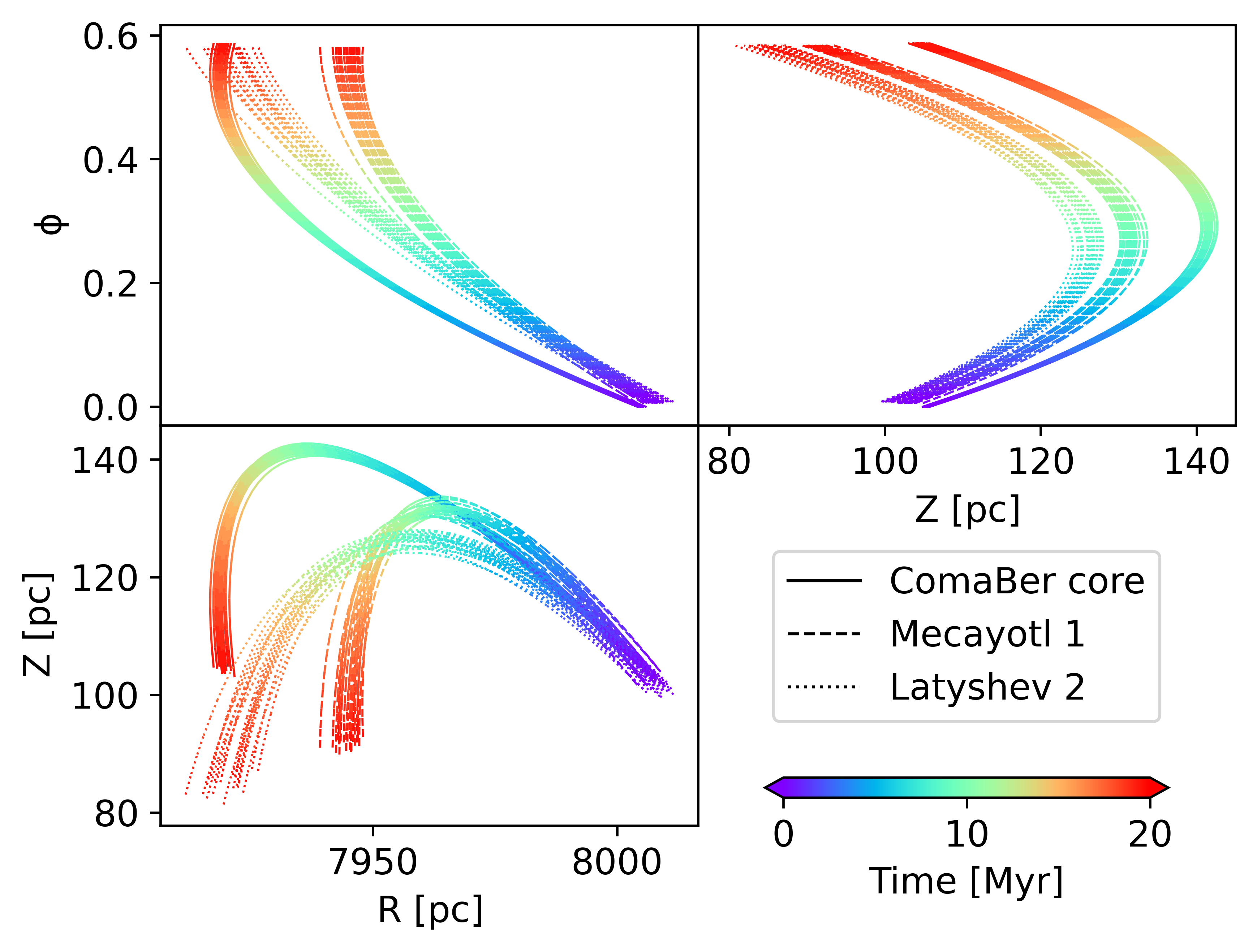}}
     \caption{Galactic coordinates ($R,Z,\phi$) of Coma Ber, Mecayotl 1 and Latyshev 2 as functions of time. To improve visibility, only the orbits of 20 samples of each stellar system are shown.}
\label{fig:orbit}
\end{figure*}

We integrate forward in time (see Sect. \ref{methods:trace-forward}) 100 sample-particles drawn from the posterior distribution of the group central phase-space coordinates. Although we give these central coordinates to \textit{Galpy} in the ICRS frame, it internally transforms them into a Galactic frame \citep[$R$, $Z$, and $\phi$; see][for the definition of these Galactic coordinates]{2015ApJS..216...29B}. Figure \ref{fig:orbit} shows the time evolution (colour coded) of the central Galactic coordinates of Coma Ber, Mecayotl 1 and Latyshev 2. The figure shows that these latter two will experience an abrupt divergence in the following 10 Myr. This divergence will mainly occur in their galactocentric distance, $R$. 

To obtain the time of flyby, we compute for each sample-particle and, as a function of time, the euclidean distances among the three groups. We notice that the core and tails of Coma Ber have, by construction, the same central position; thus, there is no need to treat them as different groups. The closest distance between Coma Ber and Latyshev 2, $28.4\pm1.5$ pc, will occur in $11.3\pm0.5$ Myr, between Coma Ber and Mecayotl 1, $23.6\pm2.0$ pc, will occur in $14.0\pm0.6$ Myr, and that between Mecayotl 1 and Latyshev 2, $12.5\pm2.0$ pc, will occur in $8.1\pm1.3$ Myr.

\subsubsection{Energies as a function of time}
We now compute the total energy of each sample-particle with respect to its parent physical group and with respect to the rest of the groups. While the former helps us to analyse the future state of these systems, the latter helps us to spot sources that may be captured by a group different from its parent one. The presence or absence of these sources will allow us to test the hypothesis of population mixing. 

\begin{figure}[ht!]
    \centering
     \includegraphics[width=\columnwidth,page=1]{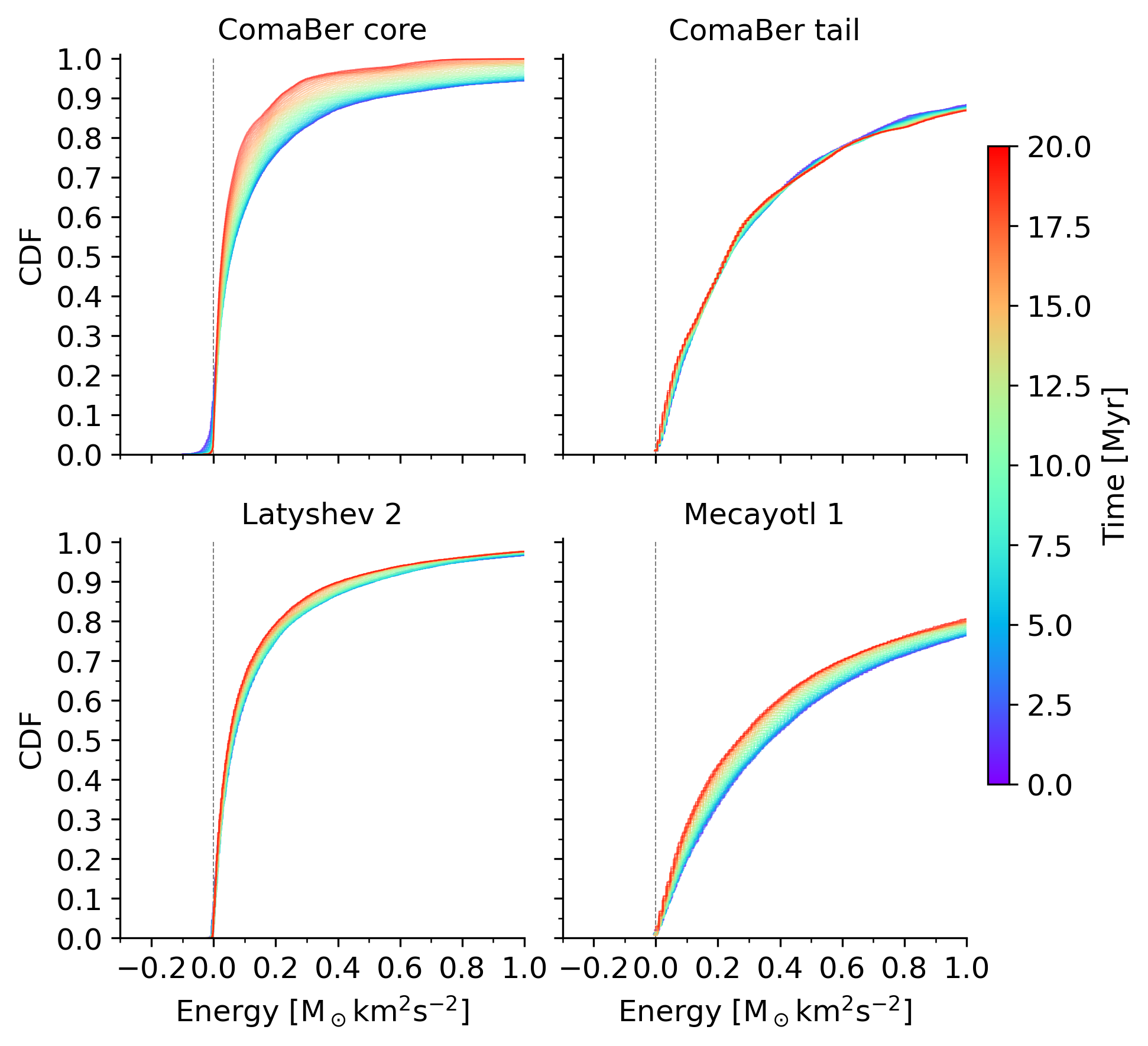}
     \caption{Time evolution of each physical group's cumulative energy distribution function. For visual aid, the vertical grey dashed lines show the zero energy. }
\label{fig:energy_cdfs_time}
\end{figure}

In Fig. \ref{fig:energy_cdfs_time}, we show the CEDF of each group as a function of time. In this figure, the energy of the sources was computed with respect to their parent group. As can be observed, the most significant change in energy in the following 20 Myr will occur in the core of Coma Ber followed by that of Mecayotl 1. In the case of Coma Ber's core, this change points towards a more steep CEDF function, which indicates that, as time goes by, the group's sources will homogenize their energies, and as a consequence, the fraction of gravitationally bound stars (energies $<$ 0) will diminish. In the case of Mecayotl 1, we  observe that its CEDF has a similar but less pronounced behaviour as that of Coma Ber's core given that it has virtually no energetically bound sources. The slope of its CEDF will increase for near-zero energy sources and will remain similar for higher energy sources, which will result in an overall increase in the fraction of sources for any given energy value.

We use the present-day values of the energy of each sample-particle with respect to the other groups as an additional validation to our membership classification. We find that none of our candidate members has present-day energies lower than those with respect to its parent group. It shows that our statistical classification methodology is robust. 

We observe that at all time stamps of our orbit integration, the energies of each sample-particle with respect to the other groups are higher than those with respect to its parent group. Thus indicating that at the flyby distances the gravitational potentials of the groups are not strong enough to capture members from another group. Therefore, we conclude that, under our current assumptions and observational uncertainties, we find no evidence of population mixing. 

\section{Discussion}
\label{discussion}

In this Sect. we compare the results presented in the previous sections with the most recent ones from the literature and the relevant theories that describe them. Table \ref{table:properties} summarises the general properties of the stellar systems analysed here. The distance and total mass values correspond to the mean and standard deviation computed from the sample of candidate members.

\begin{table}[ht!]
\caption{Number of members and mean properties of the stellar systems.}
\label{table:properties}
\centering
%\resizebox{\columnwidth}{!}{
\begin{tabular}{c|c|c|c|c}
\toprule
Name & Members      & Distance   & Age     & Mass \\
{}   &      {}      &  [pc]      & [Myr]   & [$\rm{M_{\odot}}$] \\
\midrule
Coma Ber core & 136 & $ 85\pm3$  & 800     & $ 87\pm  4$\\
Coma Ber tail & 166 & $ 91\pm8$  & 800     & $ 74\pm  4$\\
Latyshev 2    & 186 & $ 95\pm9$  & 300     & $ 92\pm  5$\\
Mecayotl 1    & 146 & $104\pm13$ & 400-600 & $ 75\pm  5$\\
\bottomrule
\end{tabular}
%}
\end{table}

\subsection{Physical groups}
\label{discussion:groups}
We found that Coma Ber is composed of two subgroups: the core and the tails, with the tails being more populous than the core (166 vs 136) but with a lower total mass ($74\pm4\ M_{\odot}$ vs $87\pm4\ M_{\odot}$, see also Fig. \ref{fig:mass}). This behaviour is expected when the less massive stars gain energy, for example thanks to dynamical relaxation (e.g., two-body encounters) or tidal effects (e.g., a constant tidal field or tidal shocks), and thus, move towards larger cluster radii until eventually they get expelled through the Lagrangian points $L_1$ or $L_2$, and end up in the leading or trailing tidal tails, respectively \citep[see, for example,][and references therein]{2012MNRAS.420.2700K}.

We found that the moving group simultaneously discovered by \citet{2019A&A...624L..11F} and \citet{2019ApJ...877...12T}, and called Group X by the latter, is composed of two independent physical groups. One corresponds to the candidate open cluster Latyshev 2, while the other is a sparse and kinematically loose stellar association that we call Mecayotl 1.

Latyshev 2 has a compact velocity dispersion ($0.80\pm0.08\ \rm{km\ s^{-1}}$) similar to that of the Pleiades \citep[$0.8\pm0.1\ \rm{km\ s^{-1}}$,][]{2017A&A...598A..48G} and the core of Coma Ber ($0.89\pm0.1\ \rm{km\ s^{-1}}$). Furthermore, its CEDF and energy-distance distribution (see Figs. \ref{fig:energy_cdfs} and \ref{fig:HvsR_diagram}) resemble those of Coma Ber's core. The previous results allow us to confirm that Latyshev 2 is an open cluster as originally proposed by \citet{1977ATsir.969....7L}. We nonetheless notice that its large spatial dispersion ($\sigma_{XYZ}=21.04\pm1.29$ pc) and negligible fraction of energy-bound sources ($\lesssim 5$\%, see Fig. \ref{fig:energy_cdfs}) indicate that it is in a disrupted state, which may have resulted from past encounters with molecular clouds or other stellar systems.

Mecayotl 1 has a large velocity dispersion ($1.85\pm0.18\ \rm{km\ s^{-1}}$), similar to that of the $\beta$Pic stellar association \citep[$1.73\pm0.3\ \rm{km\ s^{-1}}$][]{2020A&A...642A.179M}, and even larger than that of the tidal tails of Coma Ber ($1.43\pm0.1\ \rm{km\ s^{-1}}$). Furthermore, its CEDF and energy-distance distribution (see Figs. \ref{fig:energy_cdfs} and \ref{fig:HvsR_diagram}) have larger dispersions than those of the tails of Coma Ber. Concerning its spatial dispersion, $23.5\pm1.43$ pc, we observe that it is three times larger than the core of Coma Ber ($6.66\pm0.64$ pc), slightly larger than that of Latyshev 2 ($21.04\pm1.29$ pc), but smaller than that of Coma Ber's tails ($34.34\pm1.89$ pc). Therefore, we conclude that Mecayotl 1 resembles a stellar association rather than an open cluster. We notice that its large (space, velocity, and energy) dispersions may be primordial, the result of dynamical encounters that disrupted the system, or artefacts of averaging unidentified and unrelated substructures.

Concerning this latter point, the energy vs distance diagram of Mecayotl 1 shows over-densities (see Fig. \ref{fig:HvsR_diagram}) that could be related to possible unidentified substructures. However, the present observational uncertainties do not provide evidence supporting these substructures. Therefore we conclude that Mecayotl 1 is one single physical group.

\subsection{Membership}
\label{discussion:membership}
\begin{figure}[ht!]
    \centering
     \includegraphics[width=\columnwidth,page=1]{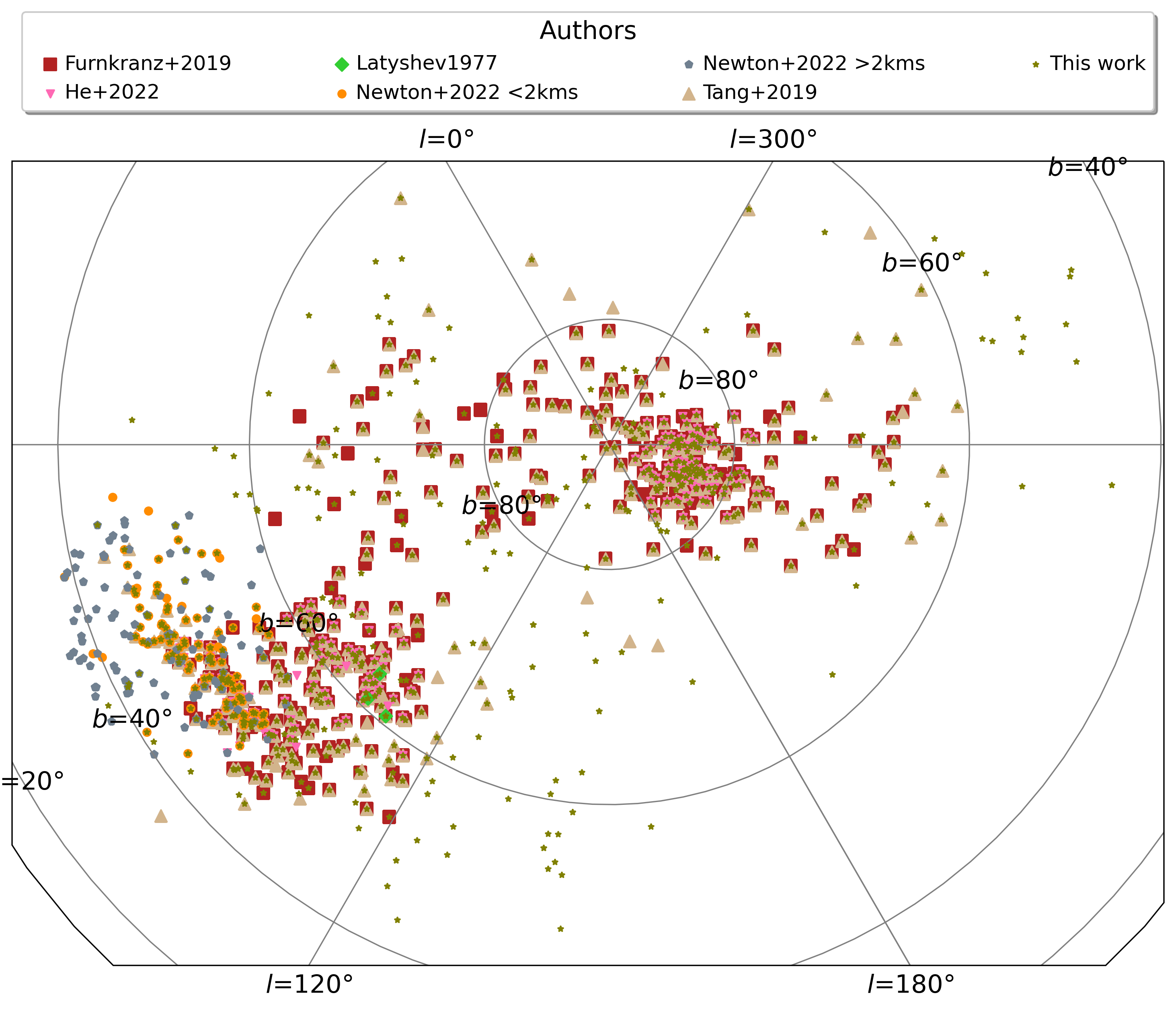}
     \caption{Galactic coordinates of our candidate members (olive stars) and those from the literature. The candidate members of Latyshev\,2 and Mecayotl 1 are located to the left bottom of the image, while those of Coma Ber are at the centre right.}
\label{fig:sky}
\end{figure}

\begin{figure*}[ht!]
    \centering
     \resizebox{\hsize}{!}{\includegraphics{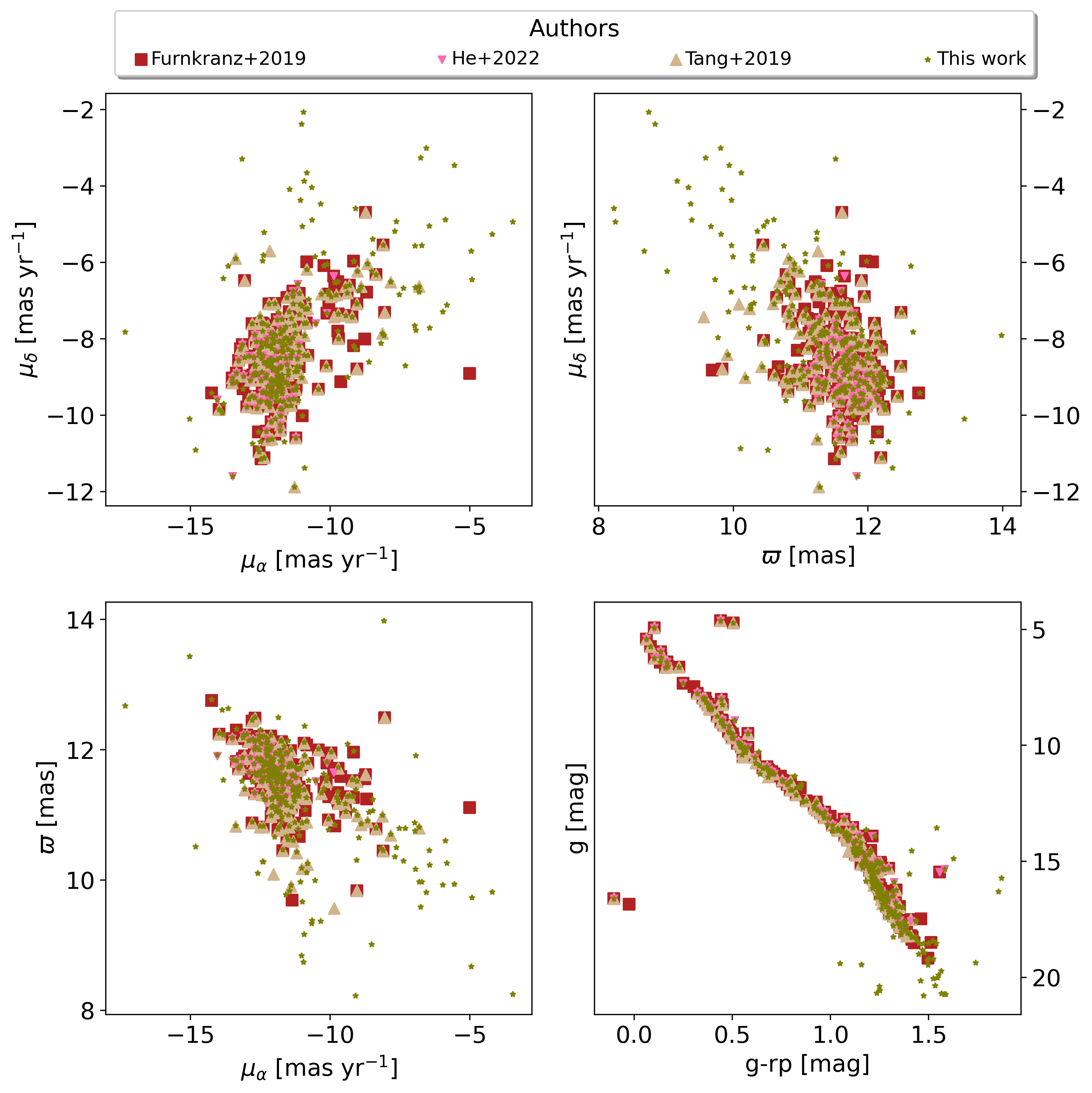}}
     \caption{\textit{Gaia} astrometry and photometry of the Coma Ber candidate members from the literature and this work.}
\label{fig:ComaBer_observables}
\end{figure*}

\begin{figure*}[ht!]
    \centering
     \includegraphics[width=\textwidth,page=1]{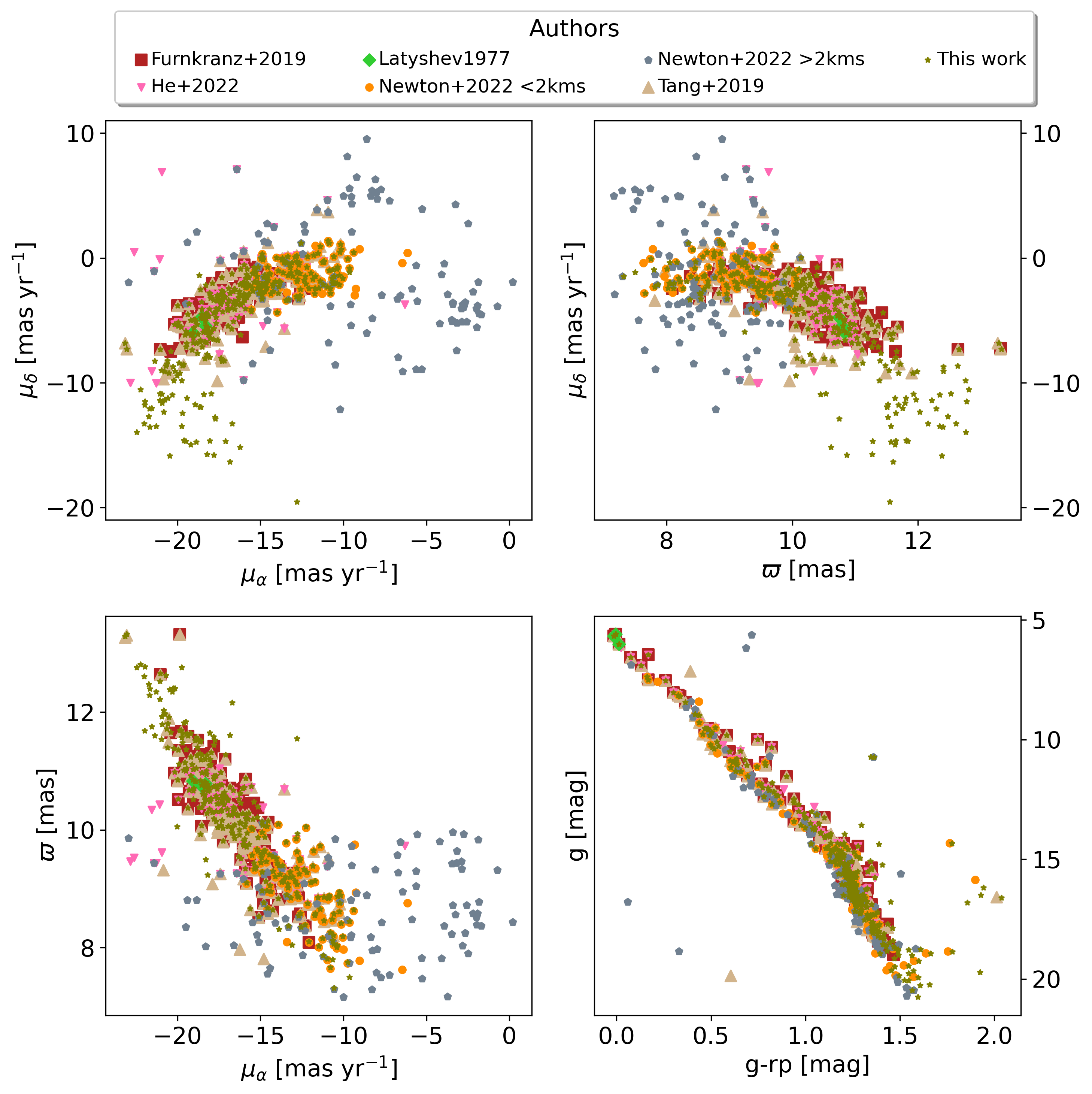}
     \caption{\textit{Gaia} astrometry and photometry of Latyshev 2 (+Mecayotl 1) candidate members from the literature and this work.}
\label{fig:GroupX_observables}
\end{figure*}

We now compare our membership lists with previous ones from the literature, in particular with the most recent ones obtained based on \textit{Gaia} data. Figures \ref{fig:sky} to \ref{fig:GroupX_observables} compare the astrometry and photometry of our candidate members with those previously known in the literature (i.e., \citealt{2019ApJ...877...12T}, \citealt{2019A&A...624L..11F}, \citealt{2022arXiv220612170H}, and \citealt{2022arXiv220606254N}; these latter authors split their sample into those having velocity offsets with respect to TOI-2048 larger and smaller than $2\ \rm{km\ s^{-1}}$). As can be observed, our candidate members extend farther in the sky and further in photometric magnitudes than in previous searches from the literature. In particular, Fig. \ref{fig:sky} shows that our search for candidate members extends to half the northern Galactic hemisphere (\texttt{b}>30\degr\@ without parallax cuts). The latter represents a considerable improvement with respect to previous works that restrict their searches in space volume, for example, to radii of 85 pc \citep{2019ApJ...877...12T}, 75 pc \citep{2019A&A...624L..11F} or 25 pc \citep{2022arXiv220606254N}.

We compare our Latyshev 2 list of candidate members with those proposed by \citet{1977ATsir.969....7L}. We recovered only three\footnote{The \citet{1977ATsir.969....7L} sources that we recover as candidate members are 81 UMa, HD 119765, and 84 UMa, with the following \textit{Gaia} \texttt{source\_id}: 1562168842092340352, 1559292485314843008 and 1561439694084279552, respectively. We notice that 83 UMa and 86 UMa could be ejected members that, due to their discrepant distance and large negative radial velocities, were not identified as members by our membership algorithm. } out of the seven candidate members originally proposed in that work. These three sources are shown as green diamonds in Figs. \ref{fig:sky} and \ref{fig:GroupX_observables}. The remaining four candidate members were discarded  due to their small parallaxes (<7 mas) and highly discrepant proper motions. 

Table \ref{table:ComaBer_literature} shows the number of Coma Ber's candidate members from the literature that our membership algorithm recovers and rejects. Similarly, Table \ref{table:GroupX_literature} shows the number of recovered and rejected candidate members from the literature, but this time for what is known as Group X. In this latter case, we aggregated our lists of candidate members for Latyshev 2 and Mecayotl 1 into a single one corresponding to what is known as Group X in the literature works.
 
\begin{table}[ht!]
\caption{Number of Coma Ber members from the literature that we recover and reject.}
\label{table:ComaBer_literature}
\centering
\resizebox{\columnwidth}{!}{
\begin{tabular}{ccccc}
\toprule
Work & Members & In dataset & Recovered & Rejected \\
\midrule
\text{\citet{2019A&A...624L..11F}} & 214 & 213 & 200 & 13 \\
\text{\citet{2019ApJ...877...12T}} & 197 & 196 & 189 &  7 \\
\text{\cite{2022arXiv220612170H}}  & 99  & 99  & 98  &  1 \\
\bottomrule
\end{tabular}
}
\end{table}

\begin{table}[ht!]
\caption{Number of Group X (Latyshev 2 + Mecayotl 1) members from the literature that we recover and reject.}
\label{table:GroupX_literature}
\centering
\resizebox{\columnwidth}{!}{
\begin{tabular}{ccccc}
\toprule
Work & Members & In dataset & Recovered & Rejected \\
\midrule
\text{\citet{2019A&A...624L..11F}}                                   & 177 & 177 & 172 &  5 \\
\text{\citet{2019ApJ...877...12T}}                                   & 218 & 218 & 197 & 21 \\
\text{\citet{2022arXiv220606254N}}($\Delta v_{tan}<2\rm{km s^{-1}}$) & 105 & 105 &  93 & 12 \\
\text{\citet{2022arXiv220606254N}}($\Delta v_{tan}>2\rm{km s^{-1}}$) & 103 & 103 &  11 & 92 \\
\text{\cite{2022arXiv220612170H}}                                    & 122 & 122 & 105 & 17 \\
\bottomrule
\end{tabular}
}
\end{table}

As can be observed in Tables \ref{table:ComaBer_literature} and \ref{table:GroupX_literature}, our novel membership methodology recovers between 42\% to 305\%, and between 52\% and 275\% more candidate members in Coma Ber and Group X (Latyshev 2 + Mecayotl 1), respectively, than the previous works from the literature. In particular, we increase the number of Coma Ber and Group X (Latyshev 2 + Mecayotl 1) candidate members by 42\% and 87\% with respect to those of \citet{2019A&A...624L..11F}, by 53\% and 52\% with respect to \citet{2019ApJ...877...12T}, and by 305\% and 275\% with respect to \citet{2022arXiv220612170H}, respectively. The work of \citet{2022arXiv220606254N} focused only on Group X, and split their list of candidate members into those having velocity offsets larger and smaller than 2 $\rm{km\ s^{-1}}$ with respect to TOI-2048. Comparing our list of Group X (Latyshev 2 + Mecayotl 1) candidate members with those of the previous authors, we increase their numbers by 60\% and reject 11\% of their $\Delta v_{tan}<2\rm{km s^{-1}}$ sample and 90\% of their $\Delta v_{tan}>2\rm{km s^{-1}}$ sample.

In Coma Ber, our new candidate members are primarily located in the farthest regions of the tidal tails and in the faintest magnitudes of the cluster photometric sequence (see Fig. \ref{fig:ComaBer_observables}). In the case of Group X (Latyshev 2 + Mecayotl 1) our new candidate members are primarily located in the regions of large parallaxes ($\varpi>11$ mas) and faintest magnitudes (see Fig. \ref{fig:GroupX_observables}).  

Due to the large recovery rates ($>95$\%) and low contamination rates ($\leq5\%$) of our classifiers (see Sect. \ref{results:membership}), we expect that our lists of candidate members have larger purity than those from the literature. In particular, our recovery and contamination rates are better than those reported by \citet[][90\% of recovery rate and 5-6\% of contamination]{2019ApJ...877...12T}. Unfortunately, \citet{2019A&A...624L..11F} only reported an estimated contamination rate of a "few per cent", and the most recent works of \citet{2022arXiv220606254N} and \citet{2022arXiv220612170H} do not report recovery or contamination rates. 

To provide a quantitative comparison of our classifiers' quality with respect to those from the literature, we use the numbers of recovered and rejected candidate members from those works together with our new candidate members. We compute the literature classifier's recovery, contamination, and missing rates as the fractions of recovered, rejected and missed sources, with respect to our lists of candidate members, respectively, and assuming that these latter are complete. These quality indicators are shown in Table \ref{table:quality_literature}. As previously mentioned, the classifier's quality indicators of \citet{2019ApJ...877...12T} agree well with the ones they report. The 4\% to 10\% contamination rates that we measure for \citet{2019A&A...624L..11F} classifiers are larger than their estimated "few per cent" contamination rate. In the case of \cite{2022arXiv220606254N}, we observe that the quality of their classifier for the sample with $\Delta v_{tan} < 2\rm{km s^{-1}}$ is similar to that of \citet{2019A&A...624L..11F}, whereas the sample with $\Delta v_{tan} > 2\rm{km s^{-1}}$ has worse quality indicators than those of a random classifier.

\begin{table}[ht!]
\caption{Recovery, contamination, and missing rates of the literature classifiers.}
\label{table:quality_literature}
\centering
\resizebox{\columnwidth}{!}{
\begin{tabular}{lcccccc}
\toprule
Work   & \multicolumn{2}{c}{Recovery} & \multicolumn{2}{c}{Contamination} & \multicolumn{2}{c}{Missing} \\
{}     & \multicolumn{2}{c}{[\%]}     & \multicolumn{2}{c}{[\%]}          & \multicolumn{2}{c}{[\%]}    \\
{}     &  Group X    & Coma Ber    &  Group X         &    Coma Ber & Group X      &  Coma Ber \\
\midrule
\text{\citet{2019A&A...624L..11F}}                                     & 90.4 & 96.4 & 9.6  & 3.6 & 46.7 & 29.5 \\
\text{\citet{2019ApJ...877...12T}}                                     & 97.2 & 93.9 & 2.8  & 6.1 & 34.3 & 35.1 \\
\text{\citet{2022arXiv220606254N}}($\Delta v_{tan} < 2\rm{km s^{-1}}$) & 88.6 &      & 11.4 &     & 68.4 &      \\
\text{\citet{2022arXiv220606254N}}($\Delta v_{tan} > 2\rm{km s^{-1}}$) & 10.6 &      & 89.3 &     & 68.9 &      \\
\text{\citet{2022arXiv220612170H}}                                     & 86.0 & 98.9 & 13.9 & 1.0 & 68.4 & 67.5 \\
\bottomrule
\end{tabular}
}
\end{table}

From our previous estimates, we notice the following points. First, the \textit{StarGo} classifier of \citet{2019ApJ...877...12T} misses only one-third of our candidate members. This classifier performs better in the more sparse case of Group X (Latyshev 2 + Mecayotl 1) than in the case of Coma Ber. Second, the classifier of \citet{2019A&A...624L..11F} performs better in Coma Ber than in Group X (Latyshev 2 +Mecayotl 1), which is not surprising given that this classifier was designed to identify extended structures like tidal tails and stellar streams \citep[see][]{2019A&A...622L..13M}. Third, the classifier of \cite{2022arXiv220612170H} works better in Coma Ber than in Latyshev 2 and Mecayotl 1, which comes as no surprise since it was tuned for more compact open clusters. Fourth, the DBSCAN classifier of \cite{2022arXiv220612170H} and the \textit{FindFriends} query method used by \citet{2022arXiv220606254N} miss up to two-thirds of our candidate members.

We highlight that the different performances that the literature classifier's have when applied to stellar systems in distinct dynamical states point towards intrinsic differences in the machine-learning algorithms they use and their fine-tuned and manually chosen parameters, particularly in the number of neurons used by the self-organizing map of \citet{2019ApJ...877...12T} and the $\epsilon$ and $minPts$ DBSCAN parameters used by \citet{2019A&A...624L..11F} or \citet{2022arXiv220612170H}. Contrary to the previous methods, our novel membership methodology offers clear physical and statistical interpretability of both the membership algorithm and its resulting phase-space model. Furthermore, our new method returns similar contamination and recovery rates (see Tables \ref{table:ComaBer_quality} and \ref{table:GroupX_quality}) in spite of the dynamical state of the stellar system to which it is applied. The latter are examples of the advantages that physical forward modelling offers over classical machine-learning algorithms imported from other domains. In spite of the previous advantages, our methodology still faces difficulties, particularly at recovering the tail's most distant and low-mass members (see discussion in Sect. \ref{discussion:mass}). Future improvements of the tail's phase-space model are warranted.

\subsection{Phase-space structure}
\label{discussion:6D_structure}

Our comprehensive phase-space modelling allows us, on the one hand, to reveal that Latyshev 2 and Mecayotl 1 are two physically distinct stellar systems and, on the other hand, to extend the tidal tails of Coma Ber up to distances of 95 pc from the cluster centre. We now discuss the position and velocity distributions of our candidate members.

\begin{figure}[ht!]
    \centering
     \includegraphics[width=\columnwidth,page=1]{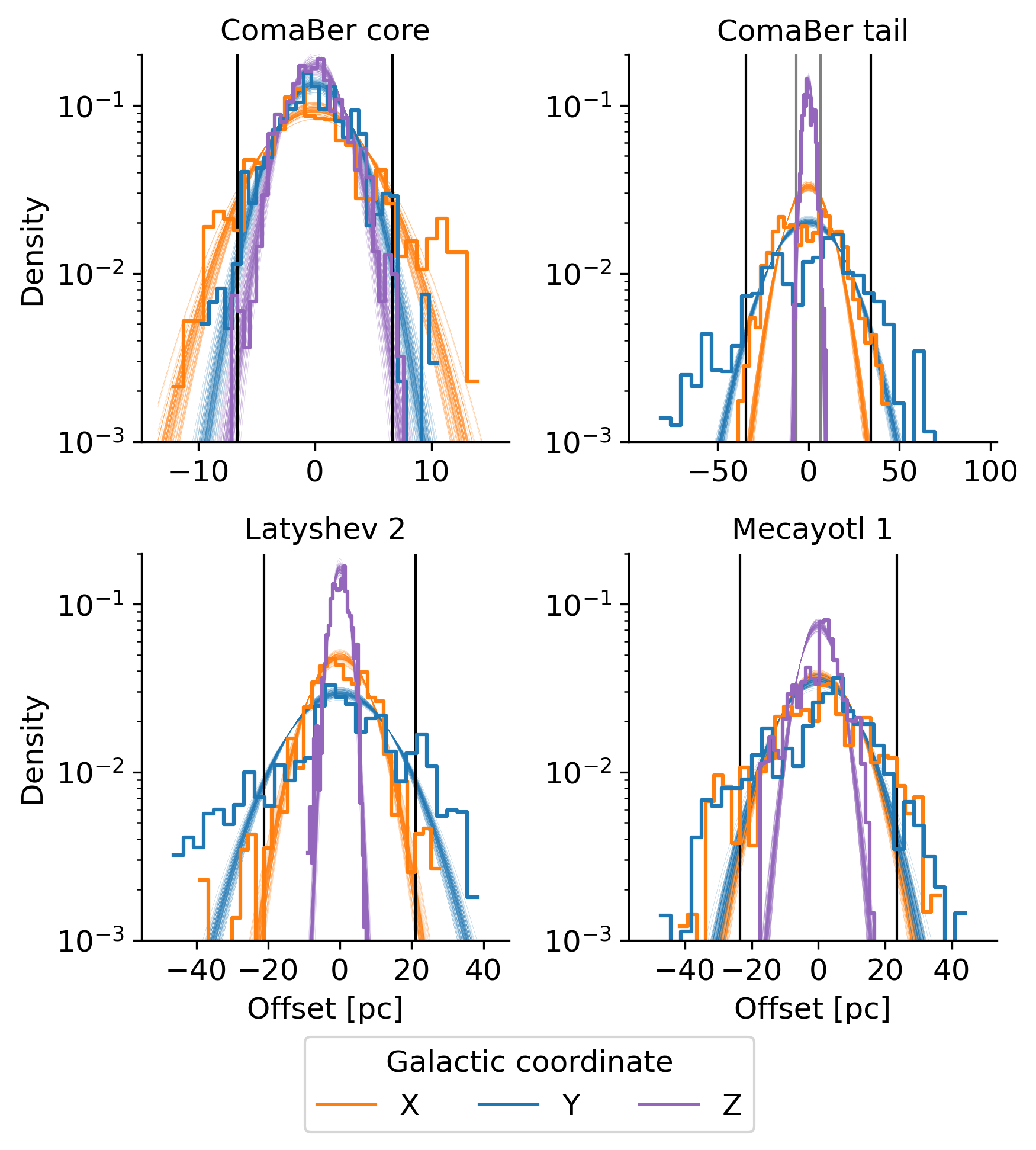}
     \includegraphics[width=\columnwidth,page=1]{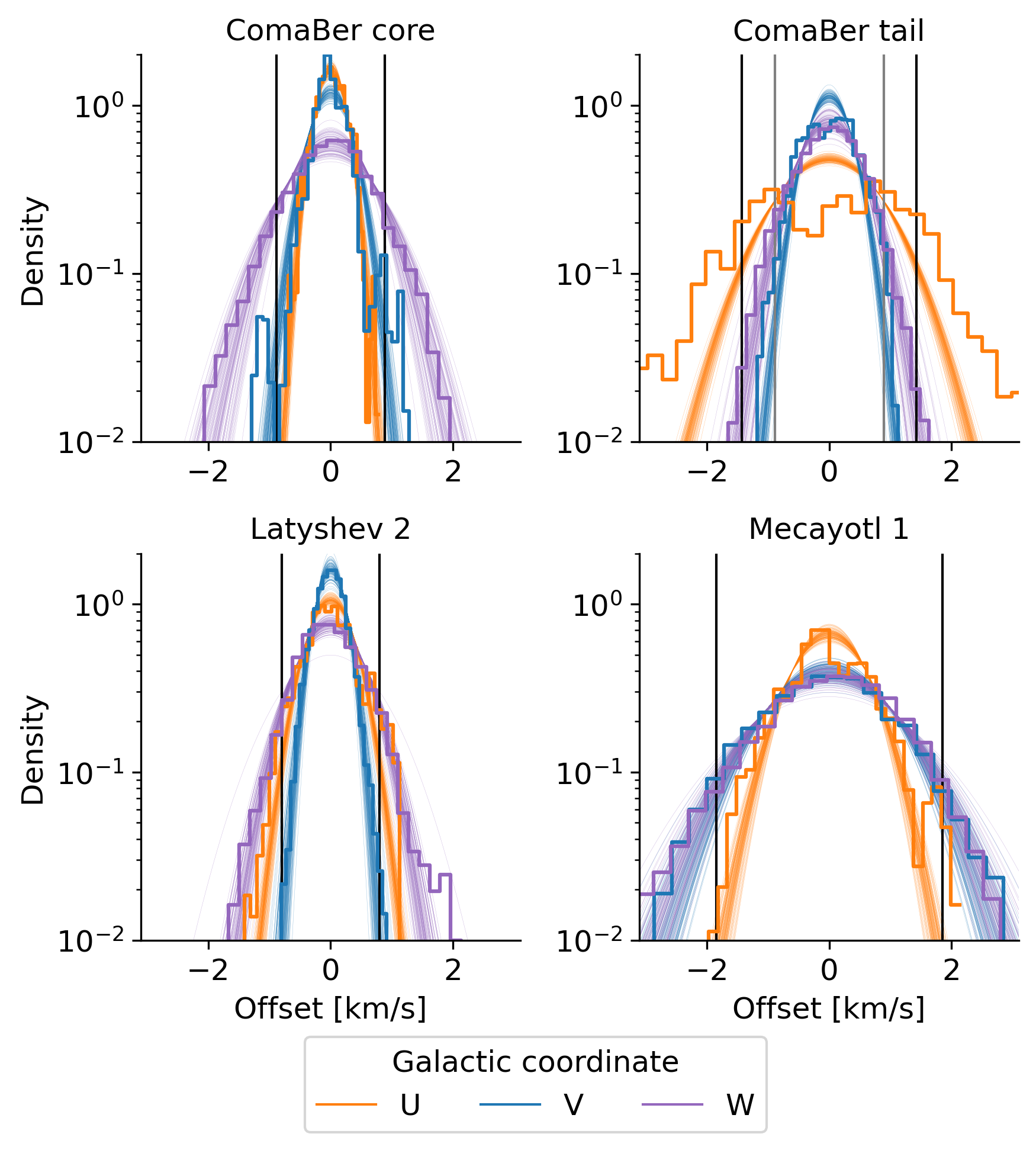}
     \caption{Distribution of position (top) and velocity (bottom) offsets with respect to the group center in Galactic heliocentric coordinates. The histograms and smooth lines show samples from the posterior distribution of the source- and group-level parameters, respectively.}
\label{fig:offsets}
\end{figure}

In Fig. \ref{fig:offsets}, we show the distributions of offsets, both in distance and in velocity, that our candidate members have with respect to the centre of their parent group (histogram). These offsets are shown in Galactic heliocentric coordinates, for which we follow the \textit{PyGaia} reference frame definitions and transformations. The figure also shows the groups Gaussian models obtained in Sect. \ref{results:6D_structure} after being transformed from the ICRS to the Galactic reference frame. The uncertainties in both the individual stars and the groups' parameters were incorporated by taking 100 samples from their posterior distributions. This figure allows us to confirm that although our model building blocks are Gaussian, the inferred position and velocity distributions of the individual sources are not necessarily Gaussian, as can be observed by the long tails in both positions and velocities of Coma Ber's tails. For comparison purposes, the figures depict the group's total dispersions ($\sigma_{XYZ}$ and $\sigma_{UVW}$ from Table \ref{table:parameters}) with black lines. In the case of Coma Ber's tail, we also add the dispersion of Coma Ber's core in grey lines. 

The 3D spatial distribution of Latyshev 2, Mecayotl 1, and the core and tails of Coma Ber show smaller dispersion in the Galactic $Z$ direction than in the other two. This phenomenon has been observed in several open clusters and galactic structures \citep[see, for example,][]{2022arXiv220604567M,2019A&A...622L..13M,2014AJ....147..146K}, which indicates that it results from the interaction with the Galactic potential. Furthermore, the tidal tails in the Galactic $Z$ direction have the same typical size as the cluster core, thus indicating that the escaped stars remain collimated in this direction. On the other hand, the Galactic $X$ and $Y$ distributions show the following properties. In the case of Latyshev 2, the $X$ and $Y$ distributions appear Gaussian, although with wider tails. Similarly, the $X$ and $Y$ distributions of Mecayotl 1 have wider tails than a Gaussian distribution, but in this case, they present the over-densities described in Sect. \ref{results:energy}, particularly seen in the $Y$ distribution. In Coma Ber, the core has two over-densities in the $Y$ distribution roughly corresponding with the spatial dispersion and tidal radius of the core ($6.66\pm0.64$ pc and $6.32\pm0.22$ pc, respectively, see Table \ref{table:parameters}). In the $X$ distribution, these over-densities are shifted out at -9 pc and 11 pc. We interpret these over-densities as the result of the accumulation of escaping stars at the Lagrangian points, where they are most likely to escape. The tails of Coma Ber also have two over-densities in both $X$ and $Y$ located at -14 pc and 9 pc, which we interpret as the accumulation of escaped stars still remaining in the cluster's vicinity due to their low escaping velocities. We notice that, as pointed out by \citet{2012MNRAS.420.2700K}, the tidal radius ($r_t$, see Table \ref{table:parameters} and Sect. \ref{results:6D_structure}) is time-dependent for clusters with eccentric orbits and can significantly change between peri- and apogalacticon. Therefore, there can be previously escaped stars within the current tidal radius and future bound stars outside of it. This breathing of the tidal radius will result in the accumulation of stars in the Lagrangian regions. 

The 3D velocity distributions of Latyshev 2, Mecayotl 1, and the tails of Coma Bar have fewer features than their 3D spatial counterparts. On the other hand, the velocity distribution of Coma Ber's core have symmetric over-densities located at the wings of the $V$ distribution. Given that these over-densities are only present in the Galactic plane and in the $V$ direction, we interpret them as resulting from either potential escapers or misclassified sources actually belonging to the tails. The Coma Ber's tail velocity dispersion in the $U$ direction lacks the central peak present in the other directions ($V$ and $W$) and thus resembles a flat-top Gaussian. We interpret the lack of this central peak as resulting from escapers gradually accelerated in this direction and thus effectively moving them away from the central zero-velocity peak. Although significant only at the one-sigma level, the observed expansion of the tails ($0.7\pm0.7\ \rm{km\ s^{-1}}$; see Table \ref{table:parameters}) also supports this scenario. Furthermore, while the tidal tail velocity dispersion in the $U$ and $V$ directions increases with respect to those of the core, the velocity dispersion in the $W$ direction remains similar. This effect indicates that the main Galactic forces acting upon the tails remain in the Galactic plane.

Our results show that the stellar systems analysed here have non-zero rotation velocities (see Table \ref{table:parameters}), with similar values to those observed in open clusters and star-forming regions. For example, the rotation velocities of Coma Ber's core and Latyshev 2 are consistent with the Praesepe value of $0.4\ \rm{km\ s^{-1}}$ reported by \citet{2020AN....341..638L}, and the rotation velocity of Mecayotl 1, $1.32\pm0.74\ \rm{km\ s^{-1}}$, is also consistent with the $1.5\pm0.1\ \rm{km\ s^{-1}}$ reported by \citet{2019A&A...630A.137G} for the Taurus complex. In addition, the total rotation of Coma Ber ($1.13\pm0.75\ \rm{km\ s^{-1}}$, computed by adding in quadrature the core and tail values) is similar to that of Mecayotl 1 and compatible with that of the Taurus complex. Although our rotational velocities are non-zero and compatible with those from other stellar systems in the solar neighbourhood, their significance still not reaches the two-sigma level (i.e., 95\%) needed to discard the null hypothesis of no rotational velocity. Future work is still needed both in the modelisation and the observational sides.

\subsubsection{Tail overdensity discussion}
\label{discussion:tail_densities}
\begin{figure}[ht!]
    \centering
     \includegraphics[width=\columnwidth,page=1]{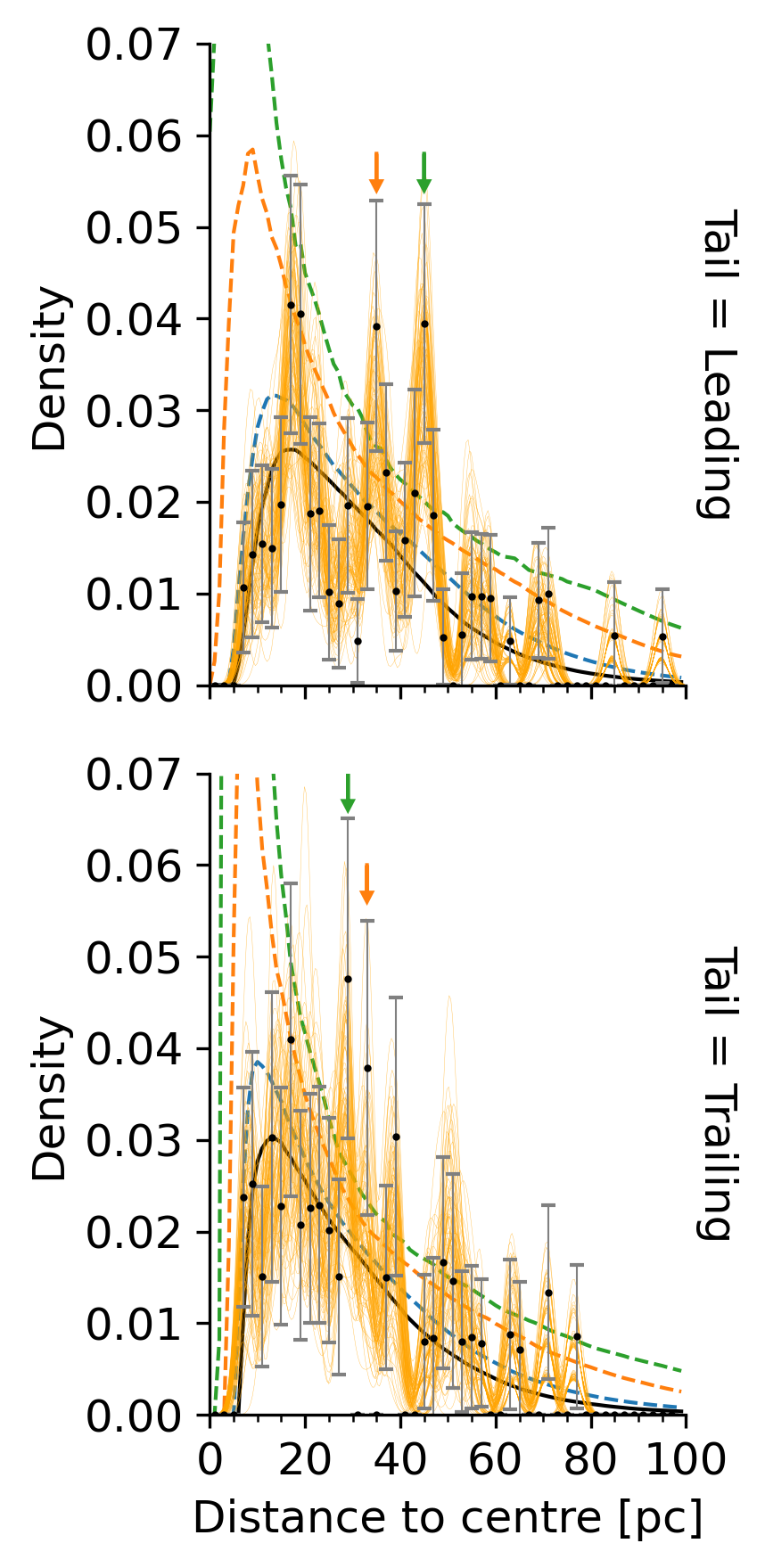}
     \caption{Distance-to-center distribution of Coma Ber's leading and trailing tail members. The orange lines depict the kernel density estimate of samples from the radial distance while the black dots and grey bars depict radial bins mean and standard deviation of 100 bootstrap realisations from the tail's members. A Gamma distribution was fitted to these bootstrap realisation and its median value (black solid lines) and one-sigma (blue dashed lines), two-sigma (orange dashed lines) and three-sigma (green dashed lines) confidence intervals are also plotted. The vertical arrows indicate the radial bin with densities exceeding the two-sigma (orange arrows) and three-sigma (green arrows) levels.}
\label{fig:tidal_tails}
\end{figure}

We investigate the radial distribution of Coma Ber's candidate members in the two tails. To identify the candidate members in the leading and trailing tails we use the inner product of their radial vector (with respect to the cluster centre) with the vector of the cluster's orbit (see Sect. \ref{results:trace-forward}). If the resulting inner product is positive we associate the candidate to the leading tail, otherwise to the trailing tail. We identify 101 and 65 candidate members in the leading and trailing tails, respectively. This represents a  leading-to-trailing tail number ratio of $1.55\pm0.34$\footnote{Given the large TPR and low FPR of our classifier (see Sect. \ref{results:membership}) we expect the uncertainties in the number of sources in the leading and trailing tails to be Poisson dominated.}. A similarly large ratio of 1.76 was also reported for the Hyades open cluster \citep{2021A&A...647A.137J}.

Figure \ref{fig:tidal_tails} shows the distribution of the distance to the cluster centre for the Coma Ber's leading and trailing tails. In this figure, the orange solid lines depicts the kernel density estimate of samples from each candidate member radial distance to the cluster centre, as computed from samples of their posterior distributions of the Cartesian spatial coordinates ($X$, $Y$ and $Z$). The black dots and grey bars depict the mean and standard deviations obtained after binning the mean radial distance of the candidate members in bins of 2 pc width. This width represents the maximum uncertainty in the radial distance determination (the typical one being 0.5 pc). As can be observed, the extension of the tails differ, with the leading one reaching up to 95 pc while the trailing one only 80 pc. The radial distributions of both tails show over-densities. We investigate the statistical significance of these over-densities by comparing them to a model of the radial distribution. For such, we use a Gamma distribution whose parameters we infer from 100 bootstrap realisations of each tail candidate members. This figure also depicts, at each radial distance, the median value of the model (black solid lines) as well as one-sigma (blue dashed lines), two-sigma (orange dashed lines) and three-sigma (green dashed lines) confidence intervals. As can be observed, only two radial bins in each tail have densities (median minus standard deviation) exceeding the two-sigma (orange arrows) and three-sigma (green arrows) levels of the Gamma distribution. In the leading tail these correspond to the over-densities located at $35\pm1$ pc and $45\pm1$ pc, while in the trailing tail, these correspond to the over-densities at $29\pm1$ pc and $33\pm1$ pc. We notice that these over-densities are still significantly detected for bin-widths of up to $\sim$4 pc and up to $\sim$3 pc in the leading and trailing tails, respectively.

Clumps in the tidal tails are expected to appear due to gravitational shocks (e.g., due to encounters with molecular clouds or passages through the Galactic disk) or by the epicyclic motion of stars escaping from a cluster in a circular galactic orbit and in a static tidal field \citep[see][and references therein]{2008MNRAS.387.1248K}. The latter are known as epicyclic over-densities and correspond to the places where escaped stars spend more time in their epicyclic motion away from the cluster \citep{2008MNRAS.387.1248K}. These over-densities are expected to depend on several factors, like the mass of the cluster, the eccentricity of its orbit, or the velocity of the escaped stars. Nonetheless, the location of these over-densities can be roughly approximated to be $y_C\simeq3\pi r_t$ \citep[see Eq. 4 of][]{2012MNRAS.420.2700K} with $r_t$ is the tidal radius. As mentioned before, this tidal radius is expected to vary as a function of time for clusters in excentric orbits (see for example, Eq. 2 of the aforementioned authors).

Assuming that the observed over-densities correspond to the epicyclic ones, their distances would result in tidal radii of 
$4\pm1$ pc and $5\pm1$ pc for the leading tail, and of $3\pm1$ pc and $4\pm1$ pc for the trailing tail. These values are compatible within their uncertainties and smaller than the present-day value of the tidal radius, $6.32\pm0.21$ pc (see Table \ref{table:parameters}). A possible explanation for these small tidal radii is that their associated over-densities were produced at perigalacticon and thus the cluster has an excentric orbit. Future work will be needed to clarify if the over-densities found here are indeed of epicyclic origin.

\subsection{Luminosity and mass distributions}
\label{discussion:mass}

Concerning the luminosity and mass distributions (see Figs. \ref{fig:luminosity} and \ref{fig:mass}), we observe that in the case of Coma Ber, they show clear differences between the core and the tails, in particular at the low-mass domain. We interpret these differences as resulting from the ejection and evaporation of low-mass stars. We notice that the fact that our luminosity and mass distributions at the faint magnitude bin represent only upper limits to the true distributions do not impact the previous conclusion. On the other hand, Latyshev 2 and Mecayotl 1 have similar luminosity and mass distributions with peak masses of $\sim\!0.33\,M_{\odot}$ ($0.36\,M_\odot$ for PARSEC and $0.30\,M_\odot$ for BT-Settl and PB) and $\sim\!0.25M_{\odot}$ ($0.28\,M_\odot$ for PARSEC and $0.22\,M_\odot$ for BT-Settl and PB), respectively. This relative shift towards higher masses in the peak of Latyshev 2 further supports our older age estimate for this system with respect to that of Mecayotl 1. 

The mass distribution of Coma Ber's tails also shows a lack of low-mass stars, $\lesssim 0.2\,M_\odot$, (see Fig. \ref{fig:mass}) for what is expected from the initial mass distribution of \citet{2005ASSL..327...41C}. Given that this lack is well within the \textit{Gaia} completeness limits, we conclude that our membership algorithm still misses low-mass stars in the most distant edges of the tidal tails. Identifying this distant population of low-mass stars and brown dwarfs will be a significant future challenge given the low contrast and large uncertainties of these faint sources. Future steps will be taken to improve our current forward model of the cluster's tails.

%Nonetheless, we notice that our extinction values were inferred based on the assumption that the observed photometry of the sources corresponds to that of single stars. Thus, the presence of unresolved binaries may result in an overestimation of both the extinction value and the mass of the assumed single star. However, the mass overestimate compensates for the possible diminished mass resulting from assuming that the binary is a single star. In any case, we stress the fact that the lack of agreement between the theoretical isochrone models and observations, particularly at faint magnitudes, may lead to extinction overestimates and thus to higher masses. Thus, we conclude that the extinction values inferred here can only be taken as upper limits to the true ones.

%Finally, we notice that the extinction values of the core of Coma Ber (for the BT-Settl, PB, and \textit{Gaia} DR3 distributions) are lower  than those of Coma Ber's tail, Latyshev 2, and Mecayotl 1. We interpret this low extinction as resulting from the core's closer distance and from the gas expulsion that it possibly suffered at its early ages. We hypothesise that the foreground (up to 80 pc) contribution to the extinction amounts up to $A_v\!\simeq\!0.2$ mag, which is the one observed at Coma Ber's core, while the remaining $A_v\!\simeq\!0.5-0.7$ mag may come from the background region (90 to 140 pc, see Fig. \ref{fig:distances}) where Mecatyotl 1, Latyhsev 2 and the tails of Coma Ber dwell.  

\subsection{Energy distribution}
\label{discussion:energy}

In Sect. \ref{results:energy}, we presented the energy distributions of Coma Ber, Latyshev 2 and Mecayotl 1 and described their observed properties. To the best of our knowledge, this is the first time that such observationally derived energy distributions are presented for open clusters in the literature. We now analyse these distributions and compare them with each other and theoretical models.

%In the case of Coma Ber, as mentioned before, only a small fraction (<30\%) of the sources have negative energies, while the majority ($\sim$80\%) have small but positive values, and almost all (>95\%) have energy values lower than 1 $\rm{M_{\odot}\ \ km^2 \ \ s^{-2}}$. The same behaviour is observed for the energy distribution of Latyshev 2, except for the fact that even less sources ($\sim$10\%) have negative energies. On the contrary, the energy distributions of Mecayotl 1 and the tails of Coma Ber have virtually no sources (<3\%) with negative energies, and only 70\% to 80\% of these sources have energies lower than 1 $\rm{M_{\odot}\ \ km^2 \ \ s^{-2}}$.

Classically, the phase-space distribution of open clusters has been described with King's profile \citep{1962AJ.....67..471K} and model \citep{1966AJ.....71...64K}. These were derived based on either the radial density profiles of globular clusters \citep{1962AJ.....67..471K} or the assumption that their velocity distribution is Gaussian and reaches a zero value at the cluster surface \citep{1966AJ.....71...64K}. \citet{2010MNRAS.407.2241K} provides a clear disambiguation between the latter two. The literature has shown that \citet{1962AJ.....67..471K} profile faces difficulties when fitting the radial density profiles of open clusters \citep[e.g.,][]{2022AJ....164...54Z,2022A&A...659A..59T,2018A&A...612A..70O}, particularly when these have halos or tails. \citet{2010MNRAS.407.2241K} analysed the profiles of dynamically evolved simulated clusters and found that although \citet{1962AJ.....67..471K} profile is a reasonable tool to measure the Jacobi radius it fails to capture the distribution of escaped stars in the form of the tidal debris. Concerning \citet{1966AJ.....71...64K} model, the literature has shown that it fails to describe the observed properties of globular clusters \citep[see, for example,][]{2019MNRAS.487..147C} and several works have created improved models of these stellar systems \citep[see, for example][and references therein]{2022arXiv221206847W}. However, little attention has been paid to the modelling of open clusters despite the challenges they pose \citep[see, Sect. 9.1 of][]{2008IAUS..246..131K}, and King's profile and model continue to be widely used by the open cluster community. 

\begin{figure}[ht!]
    \centering
     \includegraphics[width=\columnwidth]{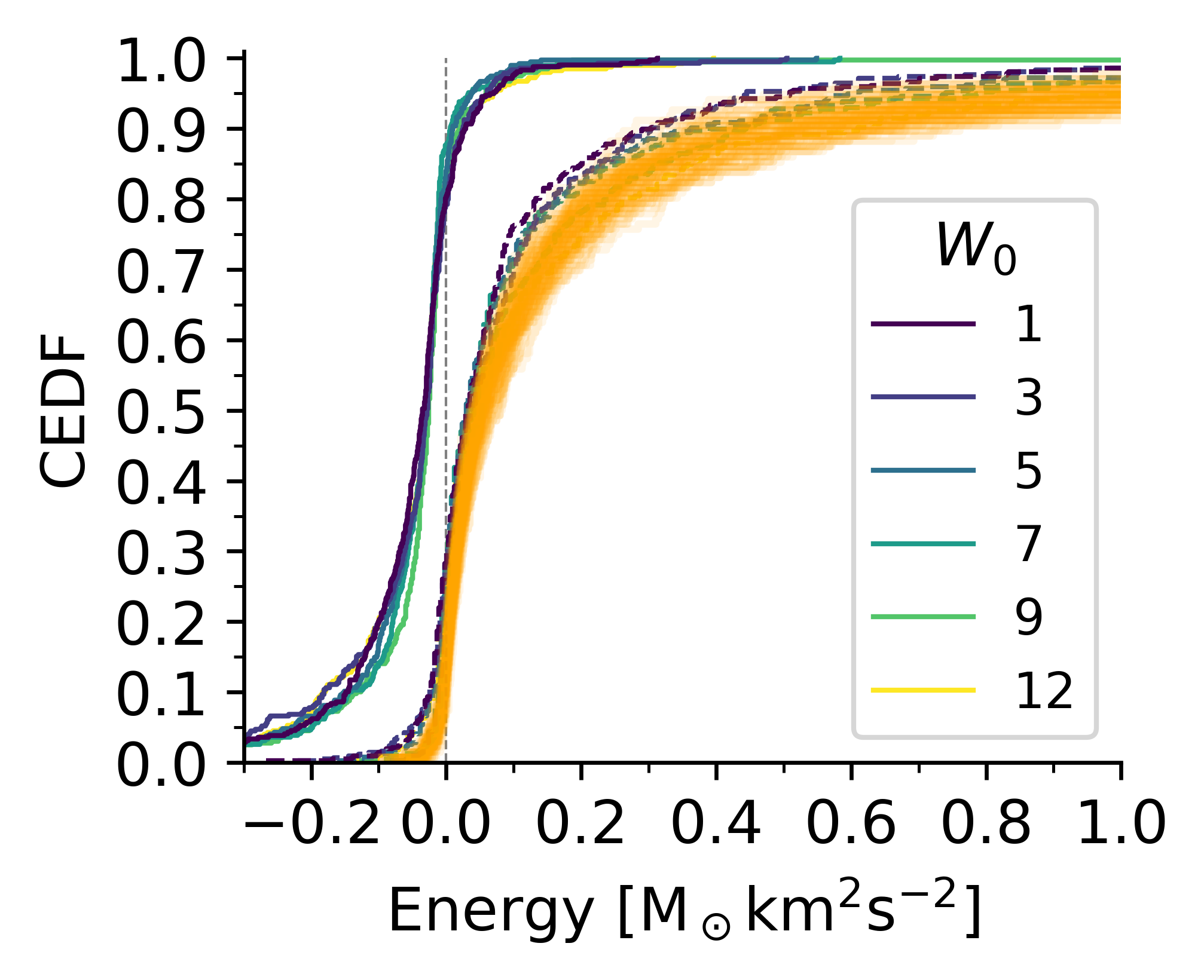}
     \caption{CEDFs of synthetic clusters with positions and velocities following the \citet{1966AJ.....71...64K} profile and mimicking the properties of Coma Ber (see text). The orange lines depict the empirical CEDFs of Coma Ber's core (see Fig. \ref{fig:energy_cdfs})}
\label{fig:king_cdfs}
\end{figure}

To compare our observational energy distribution with that expected from the \citet{1966AJ.....71...64K} model, we simulated six stellar clusters using the \textit{McLuster} code \citep{2011MNRAS.417.2300K}. Each synthetic cluster has a total mass of 160 $M_{\odot}$ and between 280 and 330 stars, with phase-space coordinates and masses randomly sampled from the \citet{1966AJ.....71...64K} and \citet{2001MNRAS.322..231K} distributions, respectively. These clusters have half-mass radii of 1 pc and concentration parameters, $W_0$, varying between 1 and 12. We choose the total mass and half-mass radii of these synthetic clusters to resemble those of Come Ber; the chosen $W_0$ values cover the full spectrum of this parameter. We compute the CEDFs of these synthetic clusters in the same way as our observed ones. Figure  \ref{fig:king_cdfs} shows the empirical CEDFs of ComaBer's core (orange lines) together with the CEDFs of these synthetic clusters colour-coded with the value of their $W_0$ parameter. The solid lines show the original CEDFs, and the dashed lines show the CEDFs that result from subtracting a total mass of 80 $M_{\odot}$. The latter is similar to the total mass estimated for the tidal tails (see Table \ref{table:properties}). As can be observed, the original CEDFs of these simulated clusters show that 90\% of its stars have negative energies, which is in clear disagreement with our observational CEDF of Coma Ber's core. On the contrary, the CEDFs resulting from subtracting 50\% of the cluster mass produce CEDFs resembling the observed ones. Therefore, we conclude that the \citet{1966AJ.....71...64K} model fails to capture the energy distribution of open clusters in late dynamical stages in which they have lost considerable fractions of their initial mass, as already shown by dynamical simulations \citep{2010MNRAS.407.2241K}. Future work will be needed to analyse the energy distributions of other open clusters and stellar associations at different stages of their dynamical evolution and to develop more realistic models of their phase-space and energy distributions.

We observe that the main characteristic that distinguishes our observational CEDFs is their slope. On the one hand, low positive slope values result in dispersed energy distribution, as in the case of Mecayotl 1 and the tails of Coma Ber. On the other hand, large positive slopes result in concentrated energy distributions, as in the core of Coma Ber and Latyshev 2, where most of the sources have small but positive energies. The CEDFs of these two open clusters agree with previous results from the literature, in which energetically unbound stars remain within the cluster for several dynamical time-scales, and are preferentially located at the cluster's edge (as mentioned in Sect. \ref{discussion:6D_structure}), where the dynamical time-scale is of the order of the orbital time-scale of the cluster \citep{2012MNRAS.420.2700K}.

\subsection{Trace-forward}
\label{discussion:trace-forward}
In Sect. \ref{results:trace-forward}, we estimated the times at which Mecayotl 1 and Latyshev 2 will have their closest distance to Coma Ber and between themselves. These flyby times have a precision between 5\% and 16\%, which represents a major improvement with respect to the 37\% precision of \citet{2019ApJ...877...12T}. The latter reported a single flyby time at 10-16 Myr, while  \citet{2019A&A...624L..11F} also reported a single flyby time of 13 Myr but with no associated uncertainty.

%Our trace-forward analysis also shows that the closest distances of Mecayotl 1 and Latyshev 2 with respect to Coma Ber will be $23.6\pm2.0$ pc and $28.4\pm1.5$ pc, respectively, and $12.5\pm2.0$ pc between them. As already mentioned by \citet{2019A&A...624L..11F}, given the cluster's sizes and minimum encounter distances, the members of these groups will occupy the same space volume. Nonetheless, at any time, including the time of encounters, the candidate members of each group will have lower energy with respect to its parent group than with respect to any other group. In other words, the candidate members of one group will not be captured by another group. 

We stress the fact that our flyby times and test for population mixing are computed under the assumption that the clusters and their stars move as free particles in the gravitational potential of the Milky Way. However, this assumption neglects both the gravitational potential of the clusters as well as possible encounters between the cluster's members. Numerical N-body simulations will be needed to overcome this assumption. However, these simulations are beyond the scope of the present work and will be carried out in a following paper of this series.

In an attempt to disentangle the possible origin of Mecayotl 1 and Latyshev 2, we also performed a trace-back orbit integration down to -50 Myr and found no other past encounters among these three systems. In addition, the distances of Mecayotl 1 and Latyshev 2 to Coma Ber rapidly diverge in the first -10 Myr, thus confirming that Latyshev 2 and Mecayotl 1 are unrelated to Coma Ber. Furthermore, the distance between Latyshev 2 and Mecayotl 1 also diverges in time but at a slower rate, reaching separations of 70 pc in the past 50 Myr. This result suggests that if the current approach between Latyshev 2 and Mecayotl 1 is the origin of the large energy dispersion observed in the latter (see Fig. \ref{fig:HvsR_diagram}), then numerical N-body trace-back simulations will be needed to clarify this scenario. However, the unravelling of the origin of these two stellar systems still awaits the arrival of more precise radial velocity measurements, precise age determinations, and numerical trace-back simulations. 

\section{Conclusions}
\label{conclusions}

Using public data from 2MASS, APOGEE, \textit{Gaia} and PanSTARRS,  in combination with our own INT-IDS observations, we studied the well-known Coma Ber open cluster and its neighbour moving group called Group X by \citet{2019ApJ...877...12T}. Our novel forward model membership method confirmed, with more than 95\% confidence level, that Group X is composed of two independent physical structures as \citet{2019ApJ...877...12T} originally suggested. These structures correspond to the disrupted open cluster Latyshev 2 (an open cluster proposed by \citealt{1977ATsir.969....7L}) and a newly identified stellar association that we call Mecayotl 1. 

Our novel membership method unravels $\sim$50\% more candidate members than recent studies from the literature. Moreover, thanks to carefully crafted synthetic data sets that mimic the real stellar systems and have the uncertainty properties of the \textit{Gaia} data, we measure recovery and contamination rates $\gtrsim$90\% and $\sim$5\%, respectively. 

Based on the most complete samples of candidate members for Coma Ber, Latyshev 2, and Mecayotl 1 to date, we analyse their phase space, mass, and energy distributions and integrate their Galactic orbits 20 Myr into the future and 50 Myr into the past. From the previous set of results, we draw the following conclusions.

The tidal tails of Coma Ber are dynamically cold and formed through mass segregation and evaporation. The leading tail is more populous (101 sources) and elongated (95 pc) than the trailing one (65 sources, 80 pc). Although we identify tidal tails' members at up to 95 pc from the cluster centre, their mass distribution indicates that our membership method still misses low-mass stars within the \textit{Gaia} completeness limit. 

Latyshev 2 is an open cluster, as originally proposed by \citet{1977ATsir.969....7L}, that is in a disrupted stage. Its mass ($92\pm5\ \rm{M_\odot}$), velocity dispersion ($0.80\pm0.08\ \rm{km\ s^{-1}}$) and energy distribution (see Fig. \ref{fig:energy_cdfs}) resemble those of Coma Ber's core, although with a larger population (186 candidate members). Its relatively young age ($\sim$400 Myr) and proximity to the Sun make it an excellent target for studies of cluster evolution and searches for ultra-cool dwarfs and exoplanets. 

Mecayotl 1 is a relatively populous (146 candidate members), relatively young (400-600 Myr), and sparse stellar association with a large velocity dispersion ($1.85\pm0.18 \ \rm{km \ s^{-1}}$), a total mass of $75\pm5\ \rm{M_\odot}$, and an energy distribution with more dispersion than that of the tails of Coma Ber.

Our trace-forward analysis shows that Latyshev 2 and Mecayotl 1 will experience flyby encounters with Coma Ber in $11.3\pm0.5$ Myr and $14.0\pm0.6$ Myr, respectively, and between them in $8.1\pm1.3$ Myr. The precision in the time of these encounters (5\% and 16\%) represents a considerable improvement with respect to the 37\% precision of \citet{2019ApJ...877...12T}. Although the encounters' distances will be smaller to the groups' size, we find no evidence for the mixing of the groups' populations. On the other hand, our trace-back analysis shows that these two stellar systems are most likely unrelated in origin.

Despite our previous results, the dynamical analysis of Coma Ber, Latyshev 2 and Mecayotl 1 still requires further methodological improvements, like better phase-space models for the tidal tails, trace-back and trace-forward numerical N-body simulations, together with precise and accurate absolute age determinations.

Finally, we notice that thanks to their proximity to the Sun, their high Galactic latitude, and their different dynamical stages, the three stellar systems analysed here will be fundamental for the search and characterization of ultra-cool dwarfs in the \textit{Euclid} \citep{2011arXiv1110.3193L} era.

\begin{acknowledgements}
JO acknowledge financial support from “Ayudas para contratos postdoctorales de investigación UNED 2021”.

JO, NL, VB, and ELM acknowledge  support  from  the Agencia  Estatal  de  Investigaci\'on  del  Ministerio  de  Ciencia  e Innovaci\'on (AEI-MCINN) under grant PID2019-109522GB-C53\@. 

NL and M{\v Z} acknowledge support from the Consejer{\'{\i}}a de Econom{\'{\i}}a, Conocimiento y Empleo del Gobierno de Canarias and the European Regional Development Fund (ERDF) under grant with reference PROID2020010052\@.

P.A.B. Galli acknowledges financial support from São Paulo Research Foundation (FAPESP) under grants 2020/12518-8 and 2021/11778-9\@.

% ERC
Co-funded by the European Union (ERC, SUBSTELLAR, project number 101054354). Views and opinions expressed are however those of the author(s) only and do not necessarily reflect those of the European Union or the European Research Council. Neither the European Union nor the granting authority can be held responsible for them.\@

% INT
We acknowledge Lucía Suarez, Olga Zamora, George Hume, and the rest of the INT staff for their support during the observations, as well as their kind help afterwards. 
This article is based on observations made in the Observatorios de Canarias del IAC with the Isaac Newton telescope, under Spanish Service Time, operated on the island of La Palma by the Isaac Newton Group of Telescopes in the Observatorio Roque de Los Muchachos.

% INT 
The Isaac Newton Telescope is operated on the island of La Palma by the Isaac Newton Group of Telescopes in the Spanish Observatorio del Roque de los Muchachos of the Instituto de Astrofísica de Canarias.

% PyGaia
We acknowledge Anthony Brown, the Gaia Project Scientist Support Team and the Gaia Data Processing and Analysis Consortium (DPAC) for providing the \textit{PyGaia} code.

% Gaia
This work has made use of data from the European Space Agency (ESA) mission
{\it Gaia} (\url{https://www.cosmos.esa.int/gaia}), processed by the {\it Gaia}
Data Processing and Analysis Consortium (DPAC,
\url{https://www.cosmos.esa.int/web/gaia/dpac/consortium}). Funding for the DPAC
has been provided by national institutions, in particular the institutions
participating in the {\it Gaia} Multilateral Agreement.

% IAC Supercomputer
This research made use of computing time available on the high-performance computing systems at the Instituto de Astrofísica de Canarias. The author thankfully acknowledges the technical expertise and assistance provided by the Spanish Supercomputing Network (Red Española de Supercomputacion), as well as the computer resources used: the Deimos-Diva Supercomputer, located at the Instituto de Astrofísica de Canarias.

% 2MASS
This publication makes use of data products from the Two Micron All Sky Survey, which is a joint project of the University of Massachusetts and the Infrared Processing and Analysis Center/California Institute of Technology, funded by the National Aeronautics and Space Administration and the National Science Foundation.

% PanSTARRS
The Pan-STARRS1 Surveys (PS1) and the PS1 public science archive have been made possible through contributions by the Institute for Astronomy, the University of Hawaii, the Pan-STARRS Project Office, the Max-Planck Society and its participating institutes, the Max Planck Institute for Astronomy, Heidelberg and the Max Planck Institute for Extraterrestrial Physics, Garching, The Johns Hopkins University, Durham University, the University of Edinburgh, the Queen's University Belfast, the Harvard-Smithsonian Center for Astrophysics, the Las Cumbres Observatory Global Telescope Network Incorporated, the National Central University of Taiwan, the Space Telescope Science Institute, the National Aeronautics and Space Administration under Grant No. NNX08AR22G issued through the Planetary Science Division of the NASA Science Mission Directorate, the National Science Foundation Grant No. AST-1238877, the University of Maryland, Eotvos Lorand University (ELTE), the Los Alamos National Laboratory, and the Gordon and Betty Moore Foundation.

% APOGEE
Funding for the Sloan Digital Sky 
Survey IV has been provided by the 
Alfred P. Sloan Foundation, the U.S. 
Department of Energy Office of 
Science, and the Participating 
Institutions. 

SDSS-IV acknowledges support and 
resources from the Center for High 
Performance Computing  at the 
University of Utah. The SDSS 
website is www.sdss.org.

SDSS-IV is managed by the 
Astrophysical Research Consortium 
for the Participating Institutions 
of the SDSS Collaboration including 
the Brazilian Participation Group, 
the Carnegie Institution for Science, 
Carnegie Mellon University, Center for 
Astrophysics | Harvard \& 
Smithsonian, the Chilean Participation 
Group, the French Participation Group, 
Instituto de Astrof\'isica de 
Canarias, The Johns Hopkins 
University, Kavli Institute for the 
Physics and Mathematics of the 
Universe (IPMU) / University of 
Tokyo, the Korean Participation Group, 
Lawrence Berkeley National Laboratory, 
Leibniz Institut f\"ur Astrophysik 
Potsdam (AIP),  Max-Planck-Institut 
f\"ur Astronomie (MPIA Heidelberg), 
Max-Planck-Institut f\"ur 
Astrophysik (MPA Garching), 
Max-Planck-Institut f\"ur 
Extraterrestrische Physik (MPE), 
National Astronomical Observatories of 
China, New Mexico State University, 
New York University, University of 
Notre Dame, Observat\'ario 
Nacional / MCTI, The Ohio State 
University, Pennsylvania State 
University, Shanghai 
Astronomical Observatory, United 
Kingdom Participation Group, 
Universidad Nacional Aut\'onoma 
de M\'exico, University of Arizona, 
University of Colorado Boulder, 
University of Oxford, University of 
Portsmouth, University of Utah, 
University of Virginia, University 
of Washington, University of 
Wisconsin, Vanderbilt University, 
and Yale University.

% SIMBAD 
This research has made use of the SIMBAD database,
operated at CDS, Strasbourg, France.

% Diego
We thank Diego Godoy-Rivera for the useful discussions that took place during the Cool Stars 21 and greatly improved the quality of this work.
\end{acknowledgements}

\bibliographystyle{aa} 
\bibliography{cbgx}

\begin{appendix}
\section{INT-IDS observations}
\label{appendix:observations}

We use INT-IDS spectrograph to acquire the spectra of 36 targets and four radial velocity standards in visitor mode on the nights of April 12th to April 18th 2022. We selected our IDS targets from our list of candidate members (see Sect. \ref{results:membership}) based on two criteria: poor quality (uncertainties > 10 $\rm{km\ s^{-1}}$) or non-existing radial velocity measurements in \textit{Gaia} EDR3, APOGEE or SIMBAD, and $V$ band magnitude brighter than 15.5 mag. The spectra were collected with the R1200R grating centred at 6500{\AA}, covering the 5300–7100{\AA} wavelength range and yielding a spectral resolving power of 6731. We used the IDS default detector, RED+2 CCD, which has pixels of 15.0 micron, a plate scale of 66.7 pixels/mm and 0.44\arcsec/pixel, and 2200 unvignetted pixels.

Tables \ref{table:log_obs} and \ref{table:log_sds} show the identifiers and characteristics of our 36 targets and four radial velocity standards, respectively. As can be observed, our targets have radial velocity uncertainties with values ranging from 5 to 40 $\rm{km\ s^{-1}}$, with a typical value of $\sim$10 $\rm{km\ s^{-1}}$. We notice that the radial velocities that we measured for our standard stars depart from the values reported by \citet[][shown in the last two columns of Table \ref{table:log_sds}]{2013A&A...552A..64S}. For this reason, we computed radial velocity zero-points by median combining the standards of each night assuming \citet{2013A&A...552A..64S} values.  These zero-points were used to correct our target's radial velocity, and its uncertainty was added in quadrature. 

\begin{table}[ht!]
\caption{Observing log of our 36 radial velocity targets.}
\label{table:log_obs}
\centering
\resizebox{\columnwidth}{!}{
\begin{tabular}{lcccccccccc}
\toprule
        Gaia DR3 ID &     RAJ2000 &     DEJ2000 &       V & Exposures &      SNR & Airmass & Slit width &        MJD &                      rv &                rv error \\
                    &       [hms] &       [dms] &   [mag] & \multicolumn{3}{l}{[seconds]} &   [arcsec] &     [days] & $\rm{[km\cdot s^{-1}]}$ & $\rm{[km\cdot s^{-1}]}$ \\
\midrule
1562168842092340352 & 13:34:07.27 & +55:20:54.3 &  $5.61$ &      1x10 & $250.00$ &  $1.52$ &     $1.20$ & $59682.89$ &                $-13.75$ &                 $33.59$ \\
4020493738752834048 & 11:56:23.80 & +30:38:28.2 & $10.52$ &     2x100 & $100.00$ &  $1.06$ &     $2.00$ & $59682.92$ &                  $4.75$ &                  $7.45$ \\
 858266239632493440 & 11:32:52.59 & +58:54:39.6 & $10.95$ &     2x100 & $100.00$ &  $1.17$ &     $2.00$ & $59682.94$ &                $-40.43$ &                  $6.49$ \\
 858266239633259648 & 11:32:52.64 & +58:54:39.2 & $10.97$ &     2x100 & $100.00$ &  $1.16$ &     $2.00$ & $59682.95$ &                 $15.08$ &                  $6.43$ \\
3958058555085345280 & 12:49:00.40 & +25:21:35.4 & $11.95$ &     1x600 & $100.00$ &  $1.04$ &     $2.00$ & $59682.97$ &                $-10.14$ &                  $6.09$ \\
1608114095830830592 & 14:16:01.29 & +52:47:10.8 & $10.49$ &     1x150 & $100.00$ &  $1.19$ &     $2.00$ & $59683.00$ &                 $-1.49$ &                  $7.88$ \\
4002733774106104192 & 12:07:41.75 & +24:12:59.2 & $12.30$ &     1x600 &  $90.00$ &  $1.01$ &     $2.00$ & $59683.01$ &                 $12.54$ &                  $6.60$ \\
1662855584317183744 & 13:39:27.23 & +61:03:42.2 & $12.22$ &     1x600 &  $90.00$ &  $1.19$ &     $2.00$ & $59683.03$ &                $-16.86$ &                  $5.80$ \\
1608114100122827904 & 14:16:01.11 & +52:47:10.6 & $10.17$ &     1x100 & $100.00$ &  $1.11$ &     $2.00$ & $59683.05$ &                 $-4.46$ &                  $7.91$ \\
1607476280298633984 & 14:48:02.81 & +56:09:33.1 & $10.01$ &      1x90 & $100.00$ &  $1.15$ &     $2.00$ & $59683.06$ &                 $-8.97$ &                 $11.04$ \\
1600815335778533888 & 15:12:57.88 & +56:02:47.2 &  $7.53$ &      1x10 & $100.00$ &  $1.16$ &     $2.00$ & $59683.07$ &                 $-6.64$ &                 $23.02$ \\
1600815335778533760 & 15:12:57.51 & +56:02:45.7 & $11.11$ &     1x250 & $100.00$ &  $1.16$ &     $2.00$ & $59683.07$ &                 $-0.21$ &                  $7.09$ \\
1614257105945926656 & 15:04:17.58 & +59:32:06.2 &  $7.41$ &      1x10 & $110.00$ &  $1.18$ &     $2.00$ & $59683.07$ &                $-17.53$ &                 $30.89$ \\
1615954442661853056 & 15:04:25.72 & +59:52:50.8 &  $9.67$ &      1x60 &  $96.00$ &  $1.18$ &     $2.00$ & $59683.08$ &                 $-6.33$ &                  $9.99$ \\
1404937579808394112 & 15:51:09.36 & +52:54:26.1 &  $9.03$ &      1x40 & $100.00$ &  $1.15$ &     $2.00$ & $59683.08$ &                  $3.56$ &                 $17.68$ \\
1606191329163112576 & 14:50:05.62 & +53:38:08.6 & $13.83$ &    2x1100 &  $80.00$ &  $1.10$ &     $2.00$ & $59683.10$ &                $-12.41$ &                  $7.09$ \\
1575260142928002688 & 12:09:47.58 & +57:00:24.6 & $14.01$ &    2x1250 &  $80.00$ &  $1.46$ &     $2.00$ & $59683.14$ &                $-16.50$ &                 $12.56$ \\
1643889047394003840 & 15:17:12.74 & +63:58:49.8 & $14.25$ &    2x1500 &  $80.00$ &  $1.27$ &     $2.00$ & $59683.18$ &                $-11.05$ &                  $7.89$ \\
1498322916287022976 & 14:06:16.44 & +41:35:55.3 & $14.29$ &    2x1600 &  $80.00$ &  $1.34$ &     $2.00$ & $59683.20$ &                 $-5.25$ &                  $8.39$ \\
3981006148735077760 & 11:34:40.89 & +23:20:03.1 & $13.80$ &     2x100 &  $80.00$ &  $1.11$ &     $2.00$ & $59683.88$ &                  $3.18$ &                  $7.97$ \\
3959503725681020032 & 12:42:38.84 & +25:09:37.3 & $14.50$ &    2x1800 &  $75.00$ &  $1.08$ &     $2.00$ & $59683.94$ &                  $6.70$ &                  $9.09$ \\
4006473213152260736 & 11:59:05.20 & +26:44:34.5 & $14.40$ &    2x1800 &  $80.00$ &  $1.00$ &     $2.00$ & $59683.99$ &                  $7.83$ &                  $9.92$ \\
1503770755884281344 & 13:45:13.16 & +46:18:40.0 & $14.23$ &    2x1500 &  $80.00$ &  $1.07$ &     $2.00$ & $59684.01$ &                $-28.29$ &                 $39.23$ \\
3960449099522357376 & 12:31:27.72 & +25:23:39.7 & $14.90$ &    3x1500 &  $70.00$ &  $1.07$ &     $2.00$ & $59684.06$ &                 $14.30$ &                 $11.08$ \\
1604441971804002176 & 14:25:53.46 & +50:24:59.3 & $14.50$ &    2x1800 &  $75.00$ &  $1.16$ &     $2.00$ & $59684.15$ &                 $18.30$ &                 $10.13$ \\
1608751851226140544 & 14:18:50.19 & +54:49:43.5 & $14.55$ &    2x1800 &  $75.00$ &  $1.37$ &     $2.00$ & $59684.21$ &                  $8.18$ &                  $9.00$ \\
3960080316450597376 & 12:26:00.25 & +24:09:20.8 & $14.55$ &    2x1800 &  $75.00$ &  $1.24$ &     $2.00$ & $59684.88$ &                $-10.34$ &                 $10.82$ \\
3960936801648700928 & 12:44:29.98 & +24:56:02.5 & $14.68$ &    4x1000 &  $72.00$ &  $1.07$ &     $2.00$ & $59684.94$ &               $-110.26$ &                 $13.50$ \\
4011987813721043456 & 12:24:43.54 & +30:17:50.1 & $14.81$ &    3x1500 &  $72.00$ &  $1.23$ &     $1.20$ & $59685.88$ &                 $-1.71$ &                  $9.50$ \\
3958947235358426496 & 12:30:57.37 & +22:46:14.9 & $15.04$ &    3x1500 &  $65.00$ &  $1.06$ &     $1.20$ & $59685.94$ &                 $11.90$ &                 $10.46$ \\
3960536716855239936 & 12:37:56.29 & +25:51:45.2 & $14.97$ &    3x1500 &  $66.00$ &  $1.00$ &     $1.20$ & $59686.00$ &                 $-0.40$ &                  $9.59$ \\
1659300798800682496 & 13:47:31.92 & +58:21:03.4 & $14.97$ &    3x1500 &  $66.00$ &  $1.19$ &     $1.20$ & $59686.10$ &                  $4.43$ &                 $14.60$ \\
1473286418047384064 & 13:26:51.67 & +34:54:25.2 & $14.88$ &    3x1500 &  $70.00$ &  $1.24$ &     $2.00$ & $59686.15$ &                $-37.89$ &                  $6.68$ \\
4009851290829830528 & 12:16:00.84 & +28:05:47.8 & $15.15$ &    3x1500 &  $61.00$ &  $1.06$ &     $2.00$ & $59686.92$ &               $-153.91$ &                 $10.82$ \\
3953836670952932224 & 12:23:55.51 & +23:24:51.9 & $15.04$ &    3x1500 &  $65.00$ &  $1.01$ &     $2.00$ & $59687.00$ &                $124.21$ &                  $6.97$ \\
1524852688757168000 & 13:13:58.11 & +40:27:08.0 & $15.04$ &    3x1500 &  $65.00$ &  $1.03$ &     $2.00$ & $59687.04$ &               $-148.77$ &                  $5.06$ \\
\bottomrule
\end{tabular}

}
\end{table}

\begin{table}[ht!]
\caption{Observing log of the four radial velocity standards.}
\label{table:log_sds}
\centering
\resizebox{\columnwidth}{!}{
\begin{tabular}{lcccccccccccc}
\toprule
Hipparcos ID &     RAJ2000 &     DEJ2000 &      V & Exposures &      SNR & Airmass & Slit width &        MJD &                      rv &                rv error &           rv literature &     rv error literature \\
             &       [hms] &       [dms] &  [mag] & \multicolumn{3}{l}{[seconds]} &   [arcsec] &     [days] & $\rm{[km\cdot s^{-1}]}$ & $\rm{[km\cdot s^{-1}]}$ & $\rm{[km\cdot s^{-1}]}$ & $\rm{[km\cdot s^{-1}]}$ \\
\midrule
    HIP73005 & 14:55:11.04 & +53:40:49.2 & $7.77$ &      1x50 & $200.00$ &  $1.23$ &     $1.20$ & $59682.21$ &                $-13.16$ &                  $4.83$ &                $-14.69$ &                  $0.03$ \\
    HIP69526 & 14:13:57.08 & +30:13:01.9 & $8.03$ &      1x50 & $200.00$ &  $1.32$ &     $1.20$ & $59682.22$ &                $-12.96$ &                  $3.31$ &                $-16.12$ &                  $0.02$ \\
    HIP57563 & 11:47:56.36 & +27:20:26.2 & $7.48$ &      1x50 & $200.00$ &  $1.09$ &     $1.20$ & $59682.91$ &                 $-9.23$ &                  $6.52$ &                $-21.86$ &                  $0.01$ \\
    HIP73005 & 14:55:11.04 & +53:40:49.2 & $7.77$ &      1x50 & $200.00$ &  $1.37$ &     $2.00$ & $59683.25$ &                 $-7.42$ &                  $7.42$ &                $-14.69$ &                  $0.03$ \\
    HIP57563 & 11:47:56.36 & +27:20:26.2 & $7.48$ &      1x50 & $200.00$ &  $1.24$ &     $1.20$ & $59683.86$ &                 $-4.89$ &                  $8.25$ &                $-21.86$ &                  $0.01$ \\
    HIP48714 & 09:56:08.67 & +62:47:18.5 & $8.99$ &     1x120 & $200.00$ &  $1.22$ &     $1.20$ & $59683.87$ &                 $35.24$ &                  $5.87$ &                 $15.38$ &                  $0.01$ \\
    HIP73005 & 14:55:11.04 & +53:40:49.2 & $7.77$ &      1x50 & $200.00$ &  $1.33$ &     $2.00$ & $59684.23$ &                $-11.46$ &                 $10.97$ &                $-14.69$ &                  $0.03$ \\
    HIP57563 & 11:47:56.36 & +27:20:26.2 & $7.48$ &      1x50 & $200.00$ &  $1.21$ &     $2.00$ & $59684.87$ &                 $-7.34$ &                 $11.49$ &                $-21.86$ &                  $0.01$ \\
    HIP48714 & 09:56:08.67 & +62:47:18.5 & $8.99$ &     1x120 & $200.00$ &  $1.21$ &     $2.00$ & $59684.87$ &                 $35.81$ &                  $7.59$ &                 $15.38$ &                  $0.01$ \\
    HIP73005 & 14:55:11.04 & +53:40:49.2 & $7.77$ &      1x50 & $200.00$ &  $1.39$ &     $1.20$ & $59685.24$ &                 $-8.96$ &                  $9.66$ &                $-14.69$ &                  $0.03$ \\
    HIP48714 & 09:56:08.67 & +62:47:18.5 & $8.99$ &     1x120 & $200.00$ &  $1.22$ &     $1.20$ & $59685.86$ &                 $35.23$ &                  $6.92$ &                 $15.38$ &                  $0.01$ \\
    HIP48714 & 09:56:08.67 & +62:47:18.5 & $8.99$ &     1x120 & $200.00$ &  $1.21$ &     $1.20$ & $59686.87$ &                 $35.29$ &                  $5.69$ &                 $15.38$ &                  $0.01$ \\
\bottomrule
\end{tabular}

}
\end{table}

\subsection{Radial velocity comparison}
\begin{figure}[ht!]
\vspace{-0.3cm}
    \centering
    
     \includegraphics[width=\columnwidth,page=2]{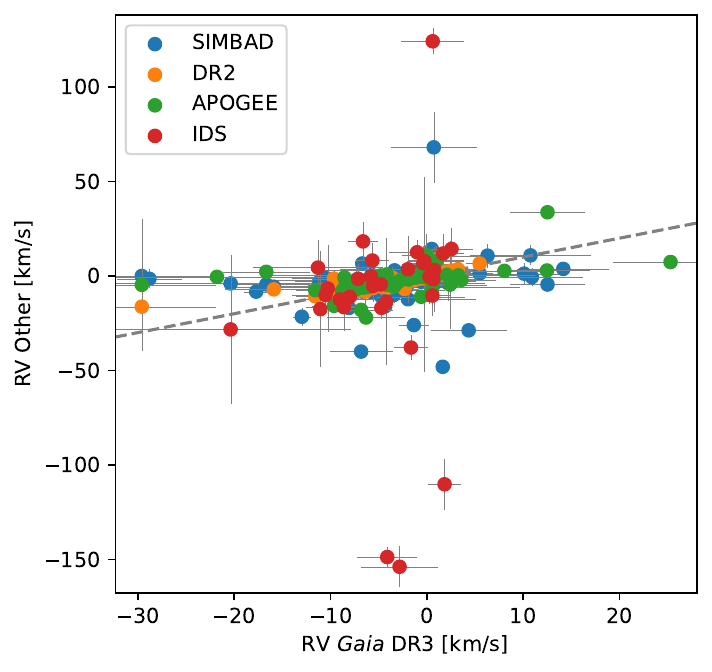}
\vspace{-0.6cm}
     \caption{Comparison of the radial velocities measured here (red dots) and reported in APOGEE (green), SIMBAD (blue) and \textit{Gaia} DR2 (orange) with respect to those of \textit{Gaia} DR3 (horizontal axis). The values shown correspond to the subsets of Coma Ber, Latyshev 2 and Mecayotl 1 candidate members having radial velocities from both \textit{Gaia} DR3 and the other source. The grey dashed line shows the identity relation.}
\label{fig:comparison_rvs}
\end{figure}

With the aim of investigating the quality of the radial velocities measured here, we compare them with those reported in APOGEE, SIMBAD, \textit{Gaia} DR2 and \textit{Gaia} DR3 for the candidate members of Coma Ber, Latyshev 2 and Mecayotl 1, and using as reference the \textit{Gaia} DR3 values due to their precision and large completeness. Figure \ref{fig:comparison_rvs} shows one-to-one scatter plots of the subsets of candidate members having radial velocity measurements in \textit{Gaia} DR3 and each of the other surveys. In the case of our IDS measurements (red dots), the figure shows the radial velocities of our 26 targets having \textit{Gaia} DR3 values as well. As can be observed, our IDS radial velocity measurements compare well with the \textit{Gaia} DR3 values and with those of APOGEE, SIMBAD, and \textit{Gaia} DR2, although with larger uncertainties. Nonetheless, there are four IDS outliers that have similar dispersion as the five SIMBAD outliers. We notice that in all these cases, the \textit{Gaia} DR3 values were used instead.

\section{Radial velocity zero-points}
\label{appendix:rv_zero-points}

We evaluate the consistency of the radial velocity measurements originating from the different input catalogues using as reference the \textit{Gaia} DR3 radial velocities. We compute zero-point offsets for APOGEE, SIMBAD, and IDS using the subsets of our candidate members simultaneously having radial velocities from \textit{Gaia} DR3 and the other survey. The zero-point offset of SIMBAD, APOGEE and IDS are $0.02 \pm 1.46\, \rm{km\,s^{-1}}$, $0.44 \pm 0.81\, \rm{km\,s^{-1}}$, and $0.32 \pm 8.86\, \rm{km\,s^{-1}}$, respectively. We notice that given their uncertainties (computed as standard errors) the latter values are compatible with zero and thus we assume they can be neglected.

\begin{figure}[ht!]
    \centering
     \includegraphics[width=\columnwidth]{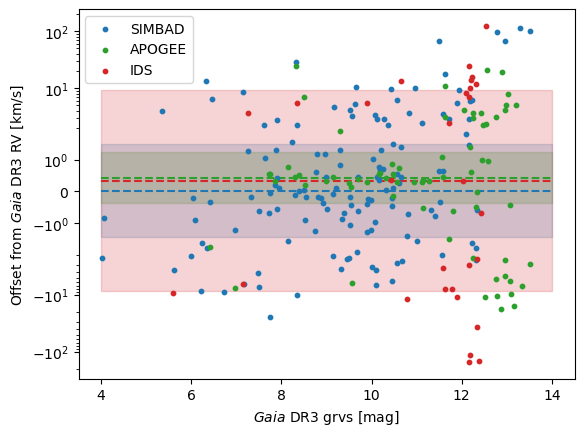}
     \vspace{-0.6cm}
     \caption{APOGEE, SIMBAD, and IDS radial velocities residuals with respect to \textit{Gaia} DR3 as a function of \texttt{grvs\_mag}. The dashed lines and shaded regions depict the median value and its standard error, respectively.}
\label{fig:zero-point_trends}
\end{figure}

\citet[][Sect. 6.1]{2022arXiv220605902K} compared the \textit{Gaia} DR3 radial velocities to those measured by other radial velocity surveys including APOGEE. In their comparison, these authors used two groups of stars: dwarf-turn-off stars and red giants and clump stars and found that the median radial velocity residuals (subtracting the \textit{Gaia} radial velocities) show a trend in magnitude \texttt{grvs\_mag}.  They  recommend correcting the \textit{Gaia} radial velocities of stars cooler than 8500 K and fainter than 11 mag using the relation given by their Eq. 5. Given that the two samples of stars that those authors use are not representative of our main sequence dwarf candidate members, we explored possible trends in the residuals of the APOGEE, SIMBAD, and IDS radial velocities of our candidate members as functions of \texttt{grvs\_mag}. Figure \ref{fig:zero-point_trends} shows the radial velocity residuals of APOGEE, SIMBAD, and IDS with respect to \textit{Gaia} DR3. As can be observed there is no clear trend in magnitude except by an increase of the dispersion for sources fainter than 11 mag. In addition, the figure also shows the median radial velocity residual (dashed lines) and its standard error (shaded regions). As mentioned above, these median values are consistent with zero. Furthermore, they are more than four times smaller than the median value of \textit{Gaia} DR3 radial velocity uncertainties of our candidate members ($1.8\, \rm{km\,s^{-1}}$) and thus, we assume that they can be neglected.

\section{Extinction distributions inferred under a uniform prior: $A_v\in[0,10]$ mag}
\label{appendix:extinction}

\begin{figure}[ht!]
    \centering
     \includegraphics[width=\columnwidth,page=1]{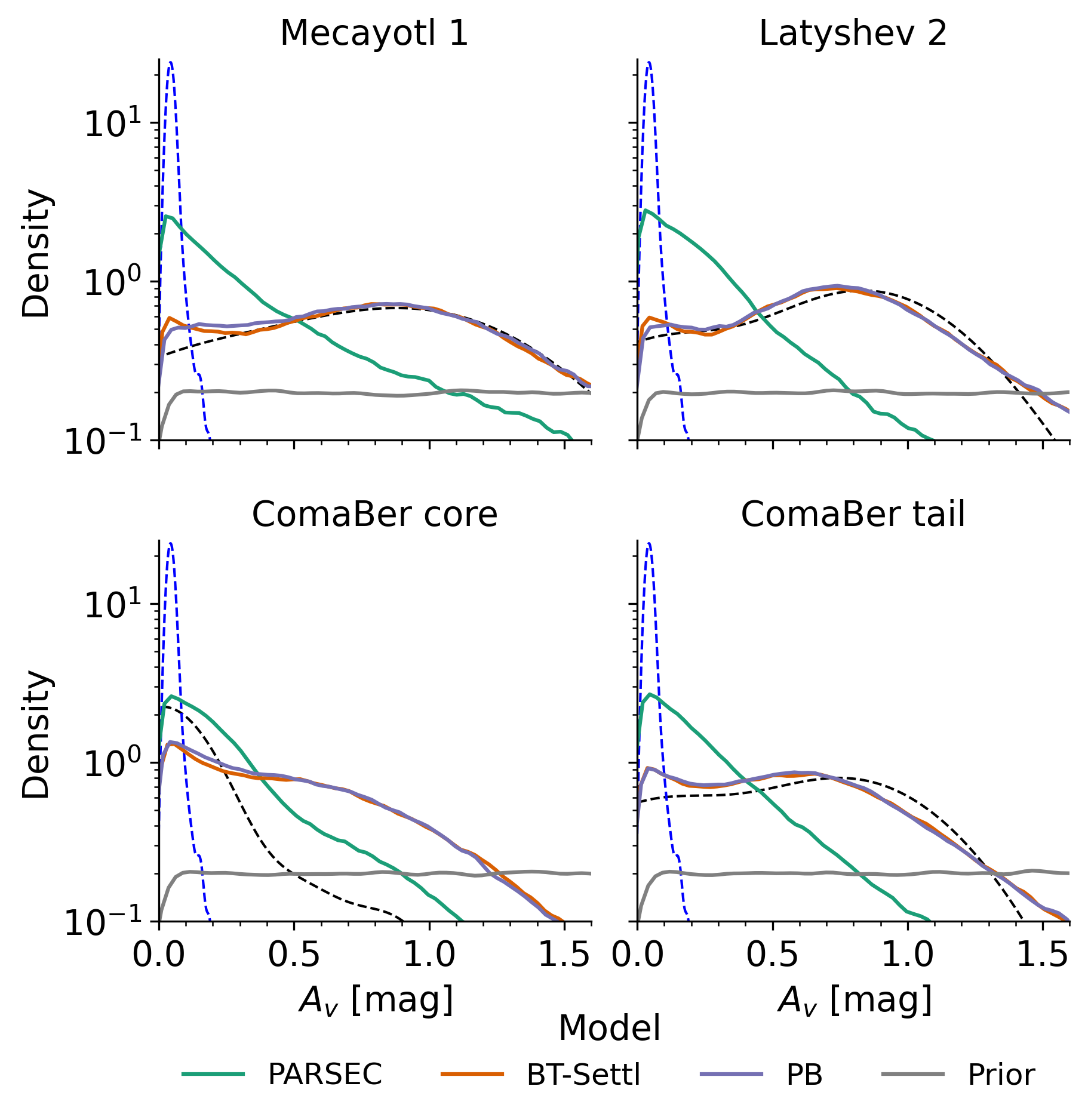}
     \vspace{-0.6cm}
     \caption{Distributions of $A_v$ as inferred using a uniform prior ($A_v\in[0,10]$ mag, gray lines) and different theoretical models (colour coded legend). The black and blue dashed lines show the $A_v$ distributions of our candidate members as reported by \textit{Gaia} DR3 and \citet{2016A&A...596A.109P}, respectively.}
\label{fig:Av}
\end{figure}

In a first attempt to infer the luminosity and mass distributions of our candidate members, we used a wide and uniform extinction prior, with $A_v\in[0,10]$ mag. Figure \ref{fig:Av} shows the $A_v$ posterior distributions resulting from the use of this prior (gray lines) and the PARSEC, BT-Settl, and PB models (colour coded lines). In addition, the black and blue dashed lines show the \textit{Gaia} DR3 and \citet{2016A&A...596A.109P} values reported for our candidate members, respectively. The \textit{Gaia} $A_v$ values were obtained by transforming the \textit{Gaia} DR3 \texttt{azero\_gspphot} to $A_v$ with the \citet{1989ApJ...345..245C} relation assuming $Rv=3.1$. The \citet{2016A&A...596A.109P} extinction values where obtained by querying the reddening values with the \textit{dustmaps} package \citep{2018JOSS....3..695M} at the sky coordinates of our candidate members and transforming them to $A_v$ values assuming $Rv=3.1$.

As can be observed in Fig. \ref{fig:Av}, the PARSEC model results in distributions with modal values $A_v<0.1$ mag for all groups. Instead, the BT-Settl and the joint PB models deliver bimodal distributions with two over-densities, one located at $A_v\!\lesssim\!0.1$ mag that coincides with the mode of the PARSEC model, and another one at $A_v\!\sim\!0.6-0.9$ mag. In the case of the core of Coma Ber, there is only one peak at $A_v\!\lesssim\!0.1$ mag. The disagreement between the observed photometry and the BT-Settl models, which is visible in Fig. \ref{fig:cmds} at the colour interval $G-RP\in[0.8,1.2]$ mag where the BT-Settl isochrone is bluer than the PARSEC one, resulted in larger $A_v$ values for this model and thus for the PB one. Nonetheless, we notice that the PARSEC models have been empirically corrected to reproduce the optical and near-infrared photometric sequences of low-mass stars in metal-poor Globular clusters and the solar-metallicity open clusters M67 and Praesepe \citep{2014MNRAS.444.2525C}.% Therefore, we conclude that the observed discrepancy between the extinction distributions obtained from the BT-Settl and PARSEC models results mainly from the empirical correction applied to the latter.

We further investigated the origin of the previous discrepancy by comparing our inferred $A_v$ distributions with those reported by \textit{Gaia} DR3 and \citet{2016A&A...596A.109P}. As explained above, these latter are also shown in Fig. \ref{fig:Av}. As can be observed, the \textit{Gaia} extinction distributions agree well with those inferred with the BT-Settl and PB models, having similar modal values and shape, except in the core of Coma Ber, where its decay is more pronounced. On the other hand, the extinction distributions delivered by \citet{2016A&A...596A.109P} are contained within $A_v<0.2$ mag and have modal values that agree with those obtained with the PARSEC model: $A_v<0.1$ mag.

It is well known that that stellar models face problems to reproduce the observed photometry of low-mass stars \citep[e.g.,][]{2014MNRAS.444.2525C, 2012ApJ...757...42F}, with the empirical correction applied to the PARSEC models \citep{2014MNRAS.444.2525C} offering a practical solution to this issue. Moreover, it is also known that \textit{Gaia} DR3 extinction values tend to be overestimated for low-mass dwarfs \citep[see Sect. 3.6.1 of][]{2022arXiv220606138A}. Due to the previous well known issues of the stellar models, in Sect. \ref{methods:mass}, we decided to infer stellar masses with an extinction uniform prior fixed to $A_v\in[0.0-0.1]$ mag. We choose this interval because of the following reasons. First, it contains the peak (the mode) values inferred using the PARSEC models with the wider prior ( $A_v\in[0,10]$ mag). Second, it is the the most probable extinction interval according to the \citet{2016A&A...596A.109P} values, which are based on the diffuse thermal dust emission, and therefore are independent of the assumptions and issues of to stellar models.

\onecolumn
\section{List of candidate members}
\label{appendix:list_of_members}

\begin{table*}[!ht]
\caption{Properties of the 634 candidate members.}
\label{table:list_of_members}
\centering
\resizebox{\textwidth}{!}{
\begin{tabular}{|c|c|c|c|c|c|c|c|c|c|c|c|c|c|c|c|c|c|c|c|c|c|c|}
\toprule
         source\_id &          Group & mem\_prob &      ra &    dec &     mean\_X &       sd\_X &     mean\_Y &       sd\_Y &     mean\_Z &       sd\_Z &              mean\_U &                sd\_U &              mean\_V &                sd\_V &              mean\_W &                sd\_W &      lower\_Mass &        map\_Mass &      upper\_Mass & lower\_Av & map\_Av & upper\_Av \\
                  - &              - &         - &     deg &    deg & $\rm{[pc]}$ & $\rm{[pc]}$ & $\rm{[pc]}$ & $\rm{[pc]}$ & $\rm{[pc]}$ & $\rm{[pc]}$ & $\rm{[km\, s^{-1}]}$ & $\rm{[km\, s^{-1}]}$ & $\rm{[km\, s^{-1}]}$ & $\rm{[km\, s^{-1}]}$ & $\rm{[km\, s^{-1}]}$ & $\rm{[km\, s^{-1}]}$ & $\rm{[M_\odot]}$ & $\rm{[M_\odot]}$ & $\rm{[M_\odot]}$ &         - &       - &         - \\
\midrule
1512001425052135680 &    Latyshev\_2 &     1.000 & 212.692 & 52.021 &     -49.460 &       0.166 &     -31.743 &       0.108 &      75.281 &       0.229 &               -3.045 &                0.281 &                7.478 &                0.179 &               -4.883 &                0.424 &            0.269 &            0.278 &            0.290 &     0.021 &   0.096 &     0.087 \\
1405053814507528832 &    Mecayotl\_1 &     1.000 & 239.061 & 53.250 &     -32.918 &       0.418 &     -54.925 &       0.714 &      85.742 &       1.061 &               -2.803 &                0.389 &                7.009 &                0.623 &               -7.537 &                0.982 &            0.132 &            0.136 &            0.139 &     0.018 &   0.051 &     0.088 \\
1559650754307292544 &    Latyshev\_2 &     1.000 & 203.036 & 51.872 &     -52.115 &       1.256 &     -22.226 &       0.637 &      72.393 &       1.742 &               -2.901 &                0.417 &                7.532 &                0.214 &               -4.886 &                0.482 &            0.359 &            0.370 &            0.379 &     0.016 &   0.068 &     0.085 \\
1669197262444734976 &    Mecayotl\_1 &     1.000 & 223.262 & 65.723 &     -31.321 &       0.649 &     -29.474 &       0.634 &      95.329 &       2.042 &               -4.309 &                0.396 &                7.036 &                0.399 &               -7.542 &                1.126 &            0.236 &            0.251 &            0.267 &     0.016 &   0.080 &     0.087 \\
1483407937977056384 & ComaBer\_tails &     1.000 & 210.965 & 36.378 &     -59.826 &       0.384 &     -35.899 &       0.234 &      51.399 &       0.331 &               -1.831 &                0.265 &                3.314 &                0.156 &               -3.891 &                0.224 &            0.158 &            0.161 &            0.165 &     0.019 &   0.079 &     0.087 \\
1459368662422928384 & ComaBer\_tails &     1.000 & 197.925 & 25.963 &     -72.689 &       0.181 &     -23.513 &       0.065 &      37.199 &       0.091 &               -2.447 &                0.083 &                4.004 &                0.030 &               -3.042 &                0.043 &            0.730 &            0.738 &            0.744 &     0.018 &   0.063 &     0.086 \\
1598228528516035200 &    Mecayotl\_1 &     1.000 & 236.477 & 54.692 &     -35.106 &       0.084 &     -52.993 &       0.133 &      89.755 &       0.230 &               -3.462 &                0.376 &                6.844 &                0.568 &               -7.046 &                0.961 &            0.401 &            0.411 &            0.422 &     0.023 &   0.078 &     0.088 \\
4008390933229933184 &  ComaBer\_core &     1.000 & 186.910 & 25.912 &     -75.475 &       0.203 &      -9.146 &       0.036 &      36.938 &       0.111 &               -2.388 &                0.534 &                4.678 &                0.068 &               -3.139 &                0.262 &            0.320 &            1.729 &            1.736 &     0.020 &   0.080 &     0.087 \\
1597061946678694016 &    Mecayotl\_1 &     1.000 & 236.059 & 53.874 &     -35.274 &       0.242 &     -52.416 &       0.363 &      86.551 &       0.544 &               -2.852 &                0.414 &                7.015 &                0.611 &               -7.320 &                1.009 &            0.174 &            0.177 &            0.182 &     0.019 &   0.074 &     0.087 \\
1560767995559564032 &    Latyshev\_2 &     1.000 & 207.493 & 53.246 &     -49.202 &       0.086 &     -25.605 &       0.046 &      74.269 &       0.127 &               -2.648 &                0.054 &                7.626 &                0.031 &               -5.001 &                0.079 &            0.476 &            0.485 &            0.495 &     0.024 &   0.070 &     0.088 \\
\bottomrule
\end{tabular}

}
\tablefoot{{\scriptsize This table will be available in its entirety in machine-readable form at the CDS. For each candidate member we provide its \textit{Gaia} DR3 source id, parent group, sky coordinates, mean and standard deviation of the Cartesian ICRS phase-space coordinates X, Y, Z, U, V, and W, and lower percentiles (2.5\%), maximum-a-posteriori, and upper percentiles (97.5\%) of the estimated mass and extinction in the $V$ band.}}
\end{table*}
\end{appendix}
\end{document}